\documentclass[pdflatex,sn-mathphys-num]{sn-jnl}

\usepackage{algorithm}
\usepackage{algpseudocode}
\usepackage{amsmath}
\usepackage{array}
\usepackage{booktabs}
\usepackage{graphicx}
\usepackage{hyperref}
\usepackage{lipsum}
\usepackage{mathptmx}
\usepackage{multirow}
\usepackage{subfigure}
\usepackage{tabularx}
\usepackage{xspace}
\usepackage{listings}%
\usepackage{xcolor}%
\usepackage{colortbl}
\usepackage{tabulary}

\newcommand{\ie}{\textrm{i.e.,}\@\xspace}

\newcommand{\activity}[1]{\textsl{#1}\@\xspace}

\begin{document}

\title[SecMLOps: A Comprehensive Framework for Integrating Security Throughout the MLOps Lifecycle]{SecMLOps: A Comprehensive Framework for Integrating Security Throughout the Machine Learning Operations Lifecycle}

\author[1,4]{\fnm{Xinrui} \sur{Zhang}}
\email{xinrui.zhang@carleton.ca}

\author[2]{\fnm{Pincan} \sur{Zhao}}
\email{pincanzhao@cmail.carleton.ca}

\author[1]{\fnm{Jason} \sur{Jaskolka}}
\email{jason.jaskolka@carleton.ca}

\author[3]{\fnm{Heng} \sur{Li}}
\email{heng.li@polymtl.ca}

\author[4]{\fnm{Rongxing} \sur{Lu}}
\email{rongxing.lu@queensu.ca}

\affil[1]{\orgdiv{Department of Systems and Computer Engineering}, \orgname{Carleton University}, \orgaddress{\state{ON}, \country{Canada}}}

\affil[2]{\orgdiv{School of Information Technology}, \orgname{Carleton University}, \orgaddress{\state{ON}, \country{Canada}}}

\affil[3]{\orgdiv{ Department of Computer and Software Engineering}, \orgname{Polytechnique Montréal}, \orgaddress{\state{QC}, \country{Canada}}}

\affil[4]{\orgdiv{School of Computing}, \orgname{Queen's University}, \orgaddress{\state{ON}, \country{Canada}}}

\abstract{
Machine Learning (ML) has emerged as a pivotal technology in the operation of large and complex systems, driving advancements in fields such as autonomous vehicles, healthcare diagnostics, and financial fraud detection. Despite its benefits, the deployment of ML models brings significant security challenges, such as adversarial attacks, which can compromise the integrity and reliability of these systems. To address these challenges, this paper builds upon the concept of Secure Machine Learning Operations (SecMLOps), providing a comprehensive framework designed to integrate robust security measures throughout the entire ML operations (MLOps) lifecycle.
SecMLOps builds on the principles of MLOps by embedding security considerations from the initial design phase through to deployment and continuous monitoring. This framework is particularly focused on safeguarding against sophisticated attacks that target various stages of the MLOps lifecycle, thereby enhancing the resilience and trustworthiness of ML applications.
A detailed advanced pedestrian detection system (PDS) use case demonstrates the practical application of SecMLOps in securing critical MLOps. Through extensive empirical evaluations, we highlight the trade-offs between security measures and system performance, providing critical insights into optimizing security without unduly impacting operational efficiency. Our findings underscore the importance of a balanced approach, offering valuable guidance for practitioners on how to achieve an optimal balance between security and performance in ML deployments across various domains.
}

\keywords{Machine learning security, MLOps, SecMLOps}

\maketitle

\section{Introduction}
\label{Sec.I_Introduction}

Machine Learning (ML) has become a cornerstone of modern technology, revolutionizing the way large and complex systems operate across various industries~\cite{sarker2021machine,chen2022security}. As a subset of artificial intelligence, ML enables systems to learn from data, adapt to new information, and make informed decisions without explicit programming. Its importance is particularly evident in applications requiring the processing and analysis of vast amounts of data, such as autonomous vehicles, healthcare diagnostics, financial fraud detection, and industrial automation~\cite{sarker2021machine,paleyes2022challenges}. For instance, in autonomous vehicles, ML algorithms are crucial for real-time object detection and decision-making, ensuring safety and efficiency on the roads~\cite{Liu_2023_CVPR}. In healthcare, ML models assist in diagnosing diseases from medical images and predicting patient outcomes, thereby enhancing the precision and reliability of medical care~\cite{qayyum2020secure}. The financial sector leverages ML for detecting fraudulent transactions and managing risks, significantly improving the security and trustworthiness of financial operations~\cite{sarker2021machine}. 
The scalability and adaptability of ML make it indispensable for managing the complexities of large-scale systems, driving innovation and efficiency across diverse domains.

As ML becomes increasingly integral to large and complex systems, efficiently managing the MLOps lifecycle, particularly in production phases, becomes critical. The rapid adoption of ML necessitates robust methodologies to deploy models at scale and speed to meet evolving business requirements. This includes continuous monitoring of model performance and health, and the ability to update models responsively to maintain their relevance and effectiveness. Moreover, establishing a centralized governance system for ML models ensures better compliance, oversight, and alignment with organizational goals.
Traditional software development methods often inadequately address these needs due to the distinct challenges posed by ML models. These challenges include, but are not limited to, model versioning, ongoing training requirements, and rigorous performance monitoring. Machine learning operations (MLOps)~\cite{kreuzberger2023machine,warnett2024understandability}, inspired by DevOps principles from software engineering, emerges as a critical solution to these issues. MLOps not only ensures that ML models are developed with scientific rigor but also emphasizes the scalability and maintainability of these models in production environments~\cite{alla_beginning_2021, kreuzberger2023machine,warnett2024understandability}.
MLOps enhances the traditional model lifecycle by promoting a seamless integration of development and operational deployment. This integration facilitates improved collaboration among stakeholders, including data scientists, operations teams, and business analysts, ensuring that the models developed are both scientifically valid and practically applicable~\cite{kreuzberger2023machine}. Moreover, MLOps introduces enhanced efficiency through automation of deployment processes and advanced monitoring capabilities, enabling real-time observability and scalability. Such capabilities ensure that ML systems can dynamically adapt to new data and operational conditions without compromising performance or accuracy under normal circumstances.

However, the widespread integration of ML models into critical and large-scale operational environments also introduces significant security challenges. These vulnerabilities range from data breaches to sophisticated adversarial attacks, which not only compromise the integrity and confidentiality of the data but also affect the reliability and performance of the ML models themselves. 
The open and dynamic nature of ML systems, especially those deployed in networked environments like autonomous vehicles, makes them susceptible to various forms of cyberattacks.
Prominent among these are data poisoning (DP) and adversarial examples (AE)~\cite{10057473}, which represent particularly effective and insidious threats at different stages of the MLOps lifecycle. DP~\cite{KASYAP2024121192,10461694} targets the training phase, involving the deliberate manipulation of training data to skew outcomes, thus compromising the model’s integrity and effectiveness. AE, introduced during the inference stage, are subtly modified inputs designed to deceive the model into making incorrect predictions while appearing normal to human observers.
These security challenges highlight the critical need for incorporating robust security measures within MLOps. Ensuring the security of ML models from the onset of development through to deployment is imperative to safeguard against potential attacks and to maintain the trustworthiness and functionality of ML applications in diverse domains. MLOps, therefore, must evolve to include comprehensive security strategies that address these vulnerabilities, enhancing the resilience and dependability of ML systems within operational environments. This integration of security measures not only protects data and models but also ensures that ML systems can dynamically adapt to new threats without compromising their performance or accuracy.

With the increasing concerns over ML security risks, the concept of Secure Machine Learning Operations (SecMLOps)~\cite{10062371} was proposed to extend the MLOps with security considerations. This paradigm advocates for the explicit integration of security measures throughout the entire MLOps lifecycle. By embedding security considerations from the outset, SecMLOps aims to cultivate more secure, reliable, and trustworthy ML-based systems. This holistic security integration not only enhances the resilience of ML deployments but also ensures their alignment with organizational security policies and regulatory requirements, thereby fortifying the foundation of trust and dependability in ML applications across various sectors.

While existing literature provides valuable insights into defending ML models against specific attacks such as DP~\cite{10461694}, AE~\cite{XIONG2023103141, zhao2024attack}, and model inversion attacks~\cite{10080996}, it primarily focuses on countermeasures for isolated incidents rather than embracing a holistic, developmental approach to security across the entire MLOps lifecycle. These studies, although crucial, often do not integrate security as a continuous and integral component throughout all stages of ML development and deployment. Additionally, the majority of current research tends to concentrate on individual security challenges without offering a comprehensive framework that seamlessly incorporates these measures into MLOps practices.

SecMLOps addresses this gap by advocating for a security-first approach in the development and operation of ML systems. By integrating security measures from the design phase through to deployment and maintenance, SecMLOps ensures continuous protection and adaptability of ML systems against emerging threats. This approach not only mitigates risks more effectively but also incorporates a critical trade-off analysis. This analysis helps in balancing security measures with operational efficiency, allowing for informed decision-making based on the security requirements and the potential impact on system performance. Such evaluations are often overlooked in traditional models, where security may be siloed or considered only after system functionalities are developed. Thus, SecMLOps provides a more rounded and sustainable solution to the security challenges in MLOps, promoting the development of ML systems that are not only robust against specific threats but are also resilient and adaptable within their operational environments. This proactive posture enables organizations to understand and manage the trade-offs between security and other factors, ensuring that security considerations are integral to the system design and operation. Our main contributions can be summarized as follows:

\begin{itemize}
    \item \textbf{A Comprehensive Security Paradigm for MLOps}: We present SecMLOps, the first systematic paradigm integrating security throughout the MLOps lifecycle. Built on the PTPGC (People, Technology, Processes, Governance, and Compliance) model, it defines eight specialized security roles, establishes phase-specific security activities, and provides a technology-agnostic approach adaptable to existing infrastructure. Unlike single point solutions, SecMLOps addresses organizational, technical, and operational dimensions to create sustainable ML security practices.
    \item \textbf{Systematic Analysis of ML-Specific Security Threats and Mitigations}: We provide the comprehensive mapping of security threats to distinct MLOps phases, recognizing ML's unique vulnerabilities versus traditional software. Our analysis identifies phase-specific threats including data poisoning, membership inference, adversarial examples, and drift exploitation. We demonstrate the system through detailed STRIDE (Spoofing, Tampering, Repudiation, Information Disclosure, Denial of Service, and Elevation of Privilege) threat analysis with concrete ML-specific controls, enabling targeted security measures rather than generic IT practices.
    \item \textbf{Comprehensive Case Study with Empirical Validation}: We demonstrate SecMLOps' practical application through a detailed pedestrian detection system (PDS) case study, showing how all framework components integrate in practice. Our empirical evaluation validates the framework's effectiveness against multiple attack scenarios, providing practitioners with a concrete implementation template and quantitative evidence of security improvements achievable through systematic security integration.
\end{itemize}

This paper is structured as follows: Section~\ref{Sec.II_Preliminary} provides preliminaries on DevOps, MLOps, and PTPGC to lay the groundwork for understanding SecMLOps. Section~\ref{Sec.III_SecMLOps} introduces the SecMLOps framework, while Section~\ref{Sec.IV_CaseStudy} presents a PDS use case employing SecMLOps. Section~\ref{Sec.V_Evaluation} discusses the implications and findings, and Section~\ref{Sec.VI_Threats} analyzes the threats to validity. Section~\ref{Sec.VII_Conclusion} concludes the paper with a summary of key insights and future directions.

\section{Preliminaries}
\label{Sec.II_Preliminary}
This section explores the foundational concepts of DevOps and MLOps, which form the origins of the SecMLOps approach, and introduces PTPGC as the conceptual framework used to further develop and define SecMLOps. This groundwork is crucial for understanding the subsequent sections.

\subsection{DevOps}
DevOps, initiated by Patrick Debois in 2008 at the Agile Conference, was developed as a response to the inefficiencies and conflicts arising from the traditional separation of development and operations teams in software engineering. This division often led to a dichotomy where developers prioritized frequent updates and operators emphasized system stability, creating systemic inefficiencies~\cite{4599477}.

As a hybrid of agile and lean practices, DevOps integrates cultural philosophies, practices, and tools that enhance an organization’s capacity to deliver applications and services rapidly and reliably~\cite{zhang2025navigating}. This approach fosters a culture of collaboration and continuous improvement, breaking down traditional barriers between departments~\cite{10.1145/3359981, hemon2022conceptualising}. The core principles of DevOps, often summarized in the CAMS model (Culture, Automation, Measurement, and Sharing) have been expanded to include aspects like Security and Feedback, illustrating the evolving nature of these practices~\cite{willis}.

While DevOps has significantly improved efficiency and productivity across various sectors, it is not without challenges. Issues such as organizational restructuring, cultural resistance, and the need for new skills can hinder its implementation~\cite{10.1007/978-3-319-49094-6_44}. Despite these obstacles, the impact of DevOps extends well beyond software development, influencing a wide range of industries by promoting innovative practices and accelerating technological adaptation.

\subsection{MLOps}
Building on the principles of DevOps, MLOps adapts these to meet the unique demands of ML workflows. MLOps has emerged as a critical discipline to facilitate the deployment and maintenance of ML models in production environments. It was developed in response to the challenges associated with operationalizing ML models, which require continuous training, monitoring, testing, and updating to perform effectively in real-world scenarios~\cite{alla_beginning_2021, kreuzberger2023machine}.

By inheriting and expanding on DevOps practices, MLOps integrates key practices such as continuous integration, delivery, and deployment (CI/CD) and expands to continuous training, in facilitating faster development, enhanced quality, and expedited deployment times. This approach not only accelerates the pace of model production but also facilitates vast scalability. It allows for the management and monitoring of numerous models, ensuring their performance remains robust across varied applications. Additionally, MLOps promotes the reproducibility of ML pipelines, fostering a more integrated collaboration across teams and reducing conflicts between data scientists and IT operations.

Beyond the security challenges, MLOps also contends with several critical issues. These include managing the inherent complexity of ML models, ensuring the quality of data used for training and inference, and the absence of standardized maturity models for guiding the development and deployment processes. These factors collectively complicate the implementation and scalability of ML systems in production environments~\cite{tamburri_sustainable_2020}.

\subsection{PTPGC}
\begin{figure}[t!]
    \centering
    \includegraphics[width=0.5\linewidth]{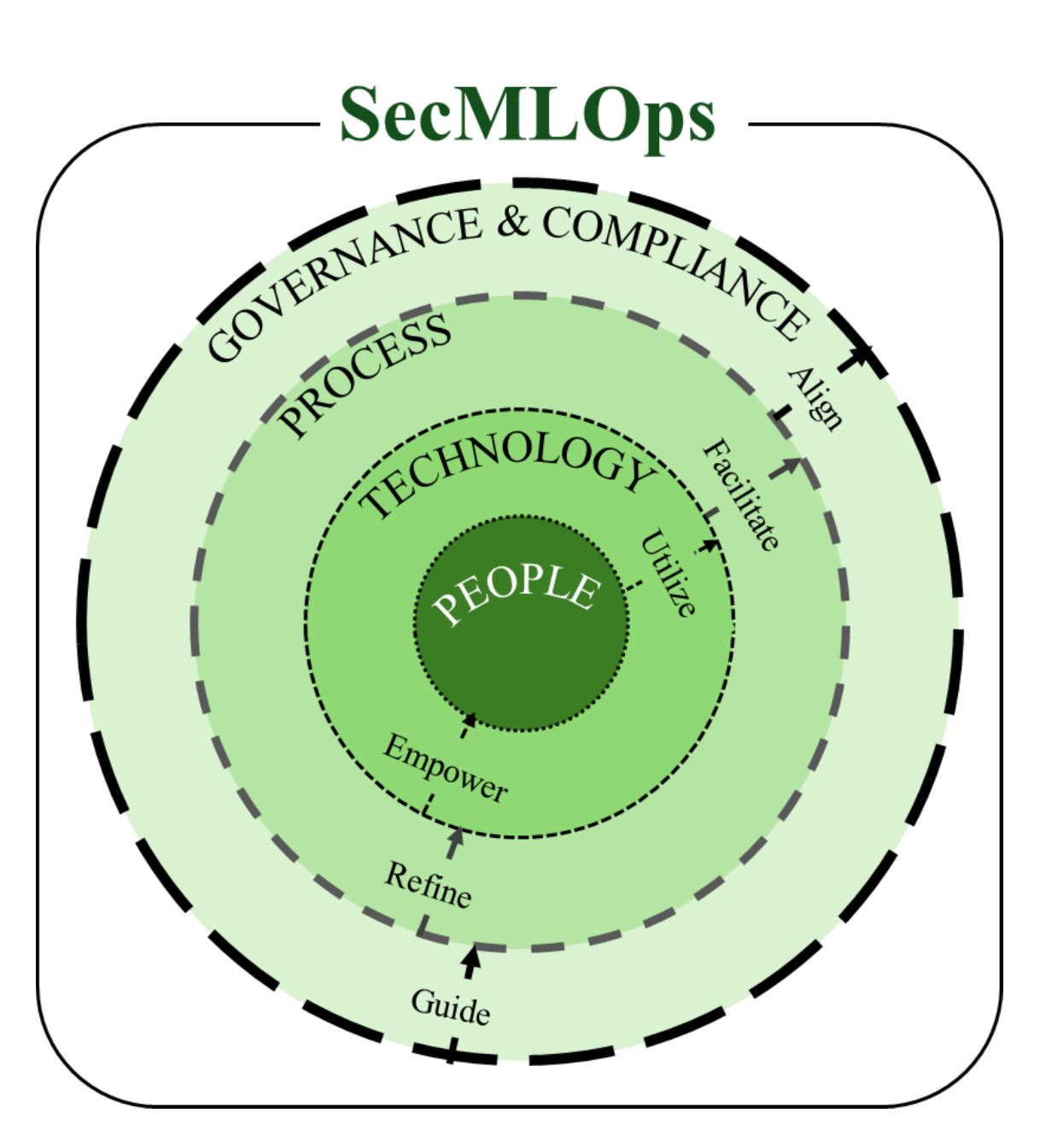}
    \caption{People, Technology, Processes, Governance, and Compliance (PTPGC) framework and the dynamic interaction between them under SecMLOps.}
    \label{fig:ptpgc}
   
\end{figure}

The People, Technology, Processes, Governance, and Compliance (PTPGC) framework, illustrated in Figure~\ref{fig:ptpgc}, builds upon the established People, Processes, and Technologies (PPT) framework, widely used across various IT domains such as product development and customer relationship management~\cite{ELIA2024}. This adaptation extends the original framework to include Governance and Compliance, underscoring their essential roles within DevOps variants and secure operations.

\begin{table*}[pt]
\renewcommand{\arraystretch}{1.3} 
\caption{Summary of key roles and corresponding responsibilities in SecMLOps}
\label{tab:roles and responsibilities}
\centering
\begin{tabular}{>{\raggedright\arraybackslash}p{0.20\textwidth}>{\raggedright\arraybackslash}p{0.35\textwidth}>{\raggedright\arraybackslash}p{0.35\textwidth}}
\toprule
\textbf{Role} & \textbf{Responsibility} & \textbf{Security Knowledge} \\
\midrule
R1. Business Stakeholder & Directs ML system security by formulating strategies and overseeing policy implementation. & Foundational understanding of cybersecurity principles and risk management to guide security strategies within the organization. \\
R2. Solution Architect & Adopts a security-by-design approach in ML systems, ensuring proactive security integration. &  Deep expertise in security architectures and threat modeling to design and implement robust security frameworks. \\
R3. Data Scientist & Specializes in ML model development with a focus on secure and robust ML practices. & Secure data ingestion and data management and well-versed in secure coding practices. \\
R4. Data Engineer & Manages secure data management and data pipeline management. & Adversarial ML, training techniques for robustness and data security and privacy. \\
R5. Software Engineer & Applies secure design patterns and coding practices in software development. & Knowledge of security patterns and secure programming practices. \\
R6. DevOps Engineer & Integrates security in SecMLOps, focusing on CI/CD and monitoring. & SecDevOps knowledge and vulnerability management within development pipelines. \\
R7. MLOps Engineer & Oversees cross-functional management of ML operations and assets. & Comprehensive security and privacy knowledge on IT and ML. \\
R8. SecMLOps Engineer & Manages operational security, performs incident responses, and conducts security assessments and testing.  & Expertise in security testing, operational security monitoring, and incident handling within ML environments. \\
\bottomrule
\end{tabular}
\end{table*}

Within the PTPGC framework, the \textit{people} component involves individuals and teams who carry out tasks, emphasizing necessary roles and expertise for effective performance. The \textit{technology} component covers essential tools and platforms that support specific operational functions. The \textit{process} aspect details the methodologies and workflows through which these tasks are executed, while the \textit{governance and compliance} component addresses the regulatory standards and guidelines that define and restrict the operational framework, ensuring adherence to legal and ethical norms.

The framework facilitates a dynamic and cyclic interaction among its components, creating a robust environment for the implementation of SecMLOps. It starts with people who utilize technology to enhance efficiency and capabilities. This technology, in turn, facilitates processes through streamlined tools and solutions. These processes are aligned with governance and compliance to ensure they meet regulatory and ethical standards. Governance and compliance also guide processes for continuous improvement and alignment. Refined by these processes, technology enhances its tools based on operational feedback, which then empowers people by providing them with advanced tools and data for more informed decision-making and innovation. Each arrow in the diagram (Figure~\ref{fig:ptpgc}) symbolizes these interactions, illustrating the continuous feedback loop that drives the efficacy and compliance of SecMLOps.

In the SecMLOps context, the PTPGC framework provides a comprehensive structure for integrating security throughout the MLOps lifecycle, supporting the development of robust, compliant ML systems. This ensures that the security considerations are woven seamlessly into every aspect of ML deployment and maintenance.

Having established the foundational concepts of DevOps, MLOps, and the PTPGC framework, we now present how these elements converge in the SecMLOps framework. The following section demonstrates how SecMLOps extends traditional MLOps by systematically integrating security considerations through each component of the PTPGC model, creating a unified approach to secure ML system development and deployment.

\section{SecMLOps}
\label{Sec.III_SecMLOps}
SecMLOps integrates rigorous security practices throughout the MLOps lifecycle, addressing the increasing need for robust cybersecurity measures in automated systems~\cite{10062371}. This section discusses the pivotal role of SecMLOps in enhancing security from initial design to continuous deployment and monitoring, extending from roles in MLOps~\cite{kreuzberger2023machine}. It elaborates on the framework's structure, its essential components, and the processes that ensure comprehensive security across all phases of ML system development.

\subsection{Necessity of SecMLOps}
The development of SecMLOps arises from the growing recognition of security vulnerabilities in ML systems and the limitations of current approaches that treat security as an afterthought rather than a foundational component. Several factors motivate the need for a comprehensive security framework:
First, the increasing deployment of ML in critical systems has elevated the potential consequences of security breaches from mere inconvenience to life-threatening scenarios \cite{sarker2021machine, Liu_2023_CVPR}. As ML systems gain decision-making authority in sensitive domains, the security implications become proportionately more significant.
Second, research has demonstrated that ML models are vulnerable to sophisticated attacks at various stages of their lifecycle \cite{10057473, 10461694, XIONG2023103141}. These vulnerabilities cannot be adequately addressed by isolated security measures or post-deployment patches; they require a systematic approach that integrates security throughout the entire MLOps workflow.
Third, regulatory frameworks such as GDPR \cite{voigt2017eu}, HIPAA \cite{act1996health}, and industry standards like ISO/IEC 27001 \cite{disterer2013iso} increasingly demand robust security and privacy safeguards for automated systems, creating a compliance imperative for organizations deploying ML solutions.

\textcolor{red}{The SecMLOps framework presented in this paper was developed through a multi-stage process as a paradigmatic methodology rather than a prescriptive implementation. Initially, we conducted a comprehensive review of existing literature on ML security, MLOps practices, and DevSecOps principles, recognizing that effective ML security requires context-specific adaptations rather than one-size-fits-all solutions. This review synthesized current knowledge on threats, vulnerabilities, and defense mechanisms relevant to ML systems while acknowledging that each deployment context, from medical diagnosis to autonomous driving, demands unique security patterns tailored to its specific threat model and operational requirements~\cite{ELIA2024}.}

With the critical need for SecMLOps established, we now detail how the framework operationalizes security through its primary components. The following subsections examine each element of the PTPGC model showing how they work in concert to address the security challenges identified above.

\subsection{Primary Components of SecMLOps}
\label{Sub.Components}
In this section, we delve into the primary components that constitute the SecMLOps framework, focusing on people, technology, processes and governance and compliance. Each component is crucial in building and maintaining secure ML systems, ensuring that every aspect of the lifecycle is safeguarded against potential security threats.

\subsubsection{People}
\textit{People} are at the core of SecMLOps, serving as the essential agents whose actions directly influence and enhance system security. It is their expertise and decision-making that fundamentally secure and shape the systems. Each role, in the context of SecMLOps, contributes uniquely to the security lifecycle of ML systems. Table~\ref{tab:roles and responsibilities} concludes the key roles and their responsibilities in SecMLOps.

\textbf{R1. Business Stakeholder} (\textit{similar role}: Product Owner, Project Manager): 
Directs ML system security by formulating high-level strategies and security policies within business domains, often consulting external experts to clarify security objectives and requirements~\cite{4359475}, particularly in sensitive areas like healthcare~\cite{qayyum2020secure} and cybersecurity~\cite{chen2022security}. They also oversee the implementation of these policies across ML development and operations.

\textbf{R2. Solution Architect} (\textit{similar role}: IT Architect). Adopts a security-by-design approach, integrating security throughout the design, technology selection, and evaluation of ML systems. Collaboration with external experts helps identify attack surfaces and conduct threat modeling, risk analysis, and mitigation strategies~\cite{LEZZI201897}.

\textbf{R3. Data Scientist} (\textit{similar roles}: ML Specialist, ML Developer). 
Specializes in adversarial ML and secure training techniques~\cite{9798870} to design ML algorithms that fulfill robustness and privacy requirements set forth by the Solution Architect(\ie role R2).

\textbf{R4. Data Engineer} (\textit{similar role}: DataOps Engineer). Understands potential attacks during data ingestion and feature engineering, whether specialized in security or acting as internal security experts. Guided by the Solution Architect's design (\ie role R2), they implement appropriate defenses and mitigations to ensure secure data handling and produce the desired datasets~\cite{LEZZI201897}.

\textbf{R5. Software Engineer}. Software engineers specialized in security or the internal security experts are aware of, and follow, secure coding practices and guidelines~\cite{10418104} to lay a foundation for secure and well-engineered ML-based systems. They may need to refactor the code into a more sustainable manner for the purpose of security management.

\textbf{R6. DevOps Engineer}. 
Integrates security into ML operations through safe CI/CD automation, workflow orchestration, and monitoring, applying SecDevOps principles~\cite{ZHAO2024112063}.

\textbf{R7. MLOps Engineer} (\textit{similar role}: ML Engineer).
Synchronizes the work of data scientists, data engineers, software engineers, and DevOps engineers(\ie roles R3, R4, R5 and R6)  ensuring operational coherence and efficiency. They oversee the ML infrastructure, manage automated ML workflow pipelines, handle model deployment to production, and monitor both the model and the ML infrastructure, updating the system as necessary.

\textbf{R8. SecMLOps Engineer}. Ensures that security considerations are dynamically integrated and managed throughout the MLOps lifecycle. This role involves ensuring regulatory compliance, conducting security assessments, monitoring system security, leading incident responses, and promoting security awareness. 

\subsubsection{Technology}

\textit{Technology} plays a pivotal role in SecMLOps, underpinning the framework with robust tools and platforms that ensure security throughout the MLOps lifecycle. In the context of the PTPGC framework, we explore various technological implementations that are essential for securing ML systems. The key technologies and corresponding security features are summarized in Table~\ref{tab:technology_ptpgc}.

\begin{table*}[!t]
\renewcommand{\arraystretch}{1.3} 
\centering
\caption{Summary of technologies, security features, and tools in SecMLOps}
\label{tab:technology_ptpgc}
\begin{tabular}{>{\raggedright\arraybackslash}p{0.20\textwidth}>{\raggedright\arraybackslash}p{0.5\textwidth}>{\raggedright\arraybackslash}p{0.25\textwidth}}
\toprule
\textbf{Technology} & \textbf{Security Features} & \textbf{Tools/Platform Examples} \\
\midrule
Secure Data Storage and Access & Provides robust encryption protocols and data access controls to safeguard sensitive information. & AWS KMS, Azure Key Vault, Google Cloud KMS \\
Secure Model Training and Deployment & Enables secure model training and data analysis without exposing raw data, maintaining data privacy across distributed environments. & TensorFlow Privacy, PySyft\\
Automated Security Testing and Compliance & Automates security scanning and compliance checks to ensure continuous protection and adherence to regulatory standards. & SonarQube, OWASP ZAP\\
Real-time Monitoring and Anomaly Detection & Offers real-time observability and anomaly detection to quickly identify and respond to security threats. & Splunk, Datadog \\
Advanced Encryption and Anonymization & Allows computation on encrypted data and removes personally identifiable information to maintain data privacy. & IBM Homomorphic Encryption, ARX Data Anonymization Tool \\
Integration Tools for SecMLOps & Incorporates security testing and evaluations into the continuous integration and deployment pipelines, ensuring ongoing security assurance. & Jenkins, GitLab CI/CD \\
Security Configuration and Management Tools & Facilitates the automated setup and configuration of security settings, reducing human error and ensuring consistent security practices across deployments. & Terraform, Ansible \\
\bottomrule
\end{tabular}
\end{table*}

\paragraph{Secure Data Storage and Access}
Ensuring the security of data, both at rest and in transit, is critical. Technologies such as encrypted databases and secure cloud storage solutions, including AWS KMS\footnote{\url{https://aws.amazon.com/kms/}}, Azure Key Vault\footnote{\url{https://azure.microsoft.com/en-us/products/key-vault}} and Google Cloud KMS\footnote{\url{https://cloud.google.com/security/products/security-key-management}} provide robust encryption protocols like AES and TLS. These technologies safeguard sensitive data and regulate access through advanced data access control mechanisms, thus preventing unauthorized data breaches.

\paragraph{Secure Model Training and Deployment}
The deployment phase of ML models is susceptible to multiple security risks. Utilizing privacy-preserving technologies such as differential privacy and federated learning, alongside secure multi-party computation, can mitigate these risks. These technologies help maintain the confidentiality and integrity of data during model training and deployment. Additionally, container security technologies like TensorFlow Privacy\footnote{\url{https://www.tensorflow.org/responsible_ai/privacy/guide}} and PySyft\footnote{\url{https://blog.openmined.org/tag/pysyft/}} enhance model security by isolating environments and maintaining strict access controls.

\paragraph{Automated Security Testing and Compliance}
Automated tools play an indispensable role in continuous security assessment and compliance monitoring. Tools such as SonarQube\footnote{\url{https://www.sonarsource.com/products/sonarqube/}} for static code analysis and OWASP ZAP\footnote{\url{https://www.zaproxy.org}} for vulnerability scanning automate the detection of potential security flaws. Similarly, compliance management software ensures that ML models adhere to legal and regulatory standards, facilitating continuous compliance throughout the lifecycle.

\paragraph{Real-time Monitoring and Anomaly Detection}
Security information and event management systems and ML-based anomaly detection tools are crucial for monitoring operations in real-time including Splunk\footnote{\url{https://www.splunk.com}} and Datadog\footnote{\url{https://www.datadoghq.com}}. These technologies detect and respond to potential security threats, leveraging advanced analytics to identify unusual patterns that may signify a security incident, thereby enabling timely mitigation strategies.

\paragraph{Advanced Encryption and Anonymization Techniques}
To further secure data while enabling complex analyses, technologies such as IBM Homomorphic Encryption\footnote{\url{https://www.ibm.com/topics/homomorphic-encryption}} allow operations on encrypted data, preserving privacy without decryption. ARX Data Anonymization Tool\footnote{\url{https://arx.deidentifier.org}}, which removes personally identifiable information, ensure data privacy by transforming sensitive data into a format where the identity of the individual cannot be inferred.

\paragraph{Integration Tools for SecMLOps}
The integration of security into ML operations is facilitated by CI/CD tools such as Jenkins\footnote{\url{https://www.jenkins.io}} and GitLab CI/CD\footnote{\url{https://docs.gitlab.com/ee/ci/}}, which can be augmented with security plugins to incorporate security assessments directly into the deployment pipelines. This ensures that security is not an afterthought, but a continuous consideration integrated into every phase of the development and deployment process.

\paragraph{Security Configuration and Management Tools}
Infrastructure as code (IaC) tools like Terraform\footnote{\url{https://www.terraform.io}} and Ansible\footnote{\url{https://www.ansible.com}} play a crucial role in automating the configuration and management of security settings. By codifying security parameters, these tools ensure consistent application of security policies across different environments, reducing the potential for human error and enhancing the overall security posture.\\

Each of these technological components contributes to the robustness of the SecMLOps framework, ensuring that security measures are seamlessly integrated and continuously maintained throughout the ML development and deployment lifecycle.

\begin{figure*}[t]
    \centering
    \includegraphics[width=\linewidth]{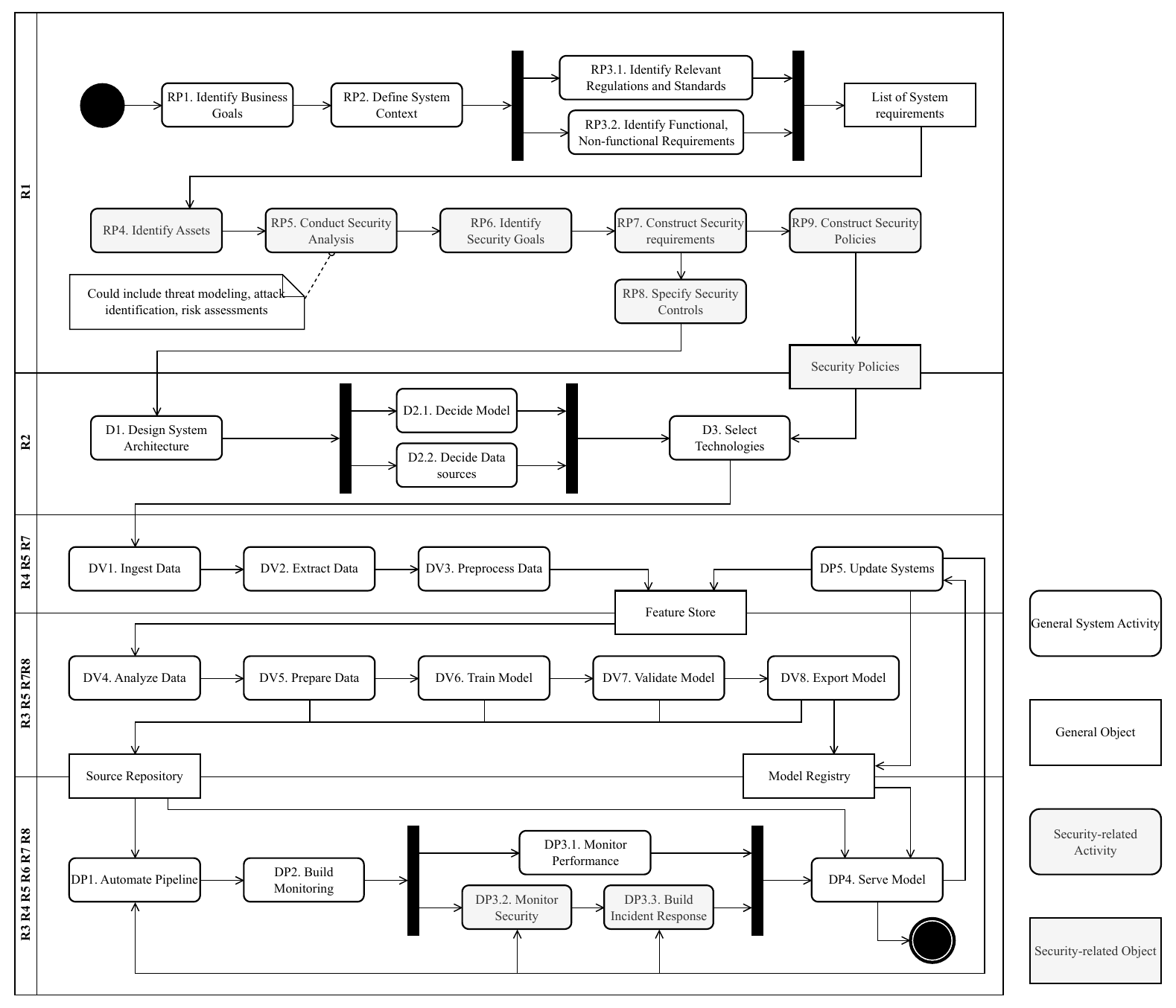}
    \caption{Comprehensive Process Diagram of SecMLOps. This diagram showcases the integrated security activities structured within the SecMLOps framework. Each activity is labeled according to its respective stage to illustrate the sequential progress and detailed tasks undertaken in each phase. Swimlanes represent the roles of actors responsible for these activities as defined in Section~\ref{Sub.Components}, ensuring a clear visualization of role-specific tasks throughout the SecMLOps lifecycle.}
    \label{fig:processdiagram}
    
\end{figure*}

\subsubsection{Process}
\label{subsubsec3.2process}
The \textit{Process} of SecMLOps formalizes how security practices are operationalized and continuously reinforced throughout the MLOps lifecycle. It focuses on transforming security from a one-time validation step into a continuous, automated, and evidence-driven cycle embedded in every pipeline execution.

Reproducibility and provenance tracking are fundamental to secure ML operations~\cite{kreuzberger2023machine, tamburri_sustainable_2020, wazir2023mlops}. Each dataset, feature transformation, training run, and model artifact must be traceable to its source and associated configurations to enable forensic accountability in the event of a security incident. Practices such as \textit{datasheets for datasets} and \textit{model cards} have been proposed to standardize documentation and transparency of data and models, facilitating auditable lineage and reducing technical debt in ML pipelines~\cite{gebru2018datasheets,mitchell2019modelcards}. Provenance tracking not only supports compliance but also helps detect data tampering and unauthorized model changes during retraining or deployment~\cite{paleyes2022challenges}.

Automation is critical to ensuring consistency and scalability of security practices. Security checks should be integrated into continuous integration and continuous deployment (CI/CD) pipelines~\cite{10.1145/3359981}, including data validation, schema conformity, privacy leakage detection, adversarial robustness evaluation, and compliance enforcement gates. Empirical studies demonstrate that automated testing and monitoring significantly reduce hidden technical debt and operational vulnerabilities in production ML systems~\cite{sculley2015hidden,breck2017themltest}. Automation minimizes human error while providing repeatable, verifiable assurance that every deployed model meets predefined security baselines.

Continuous monitoring must extend beyond performance metrics to include indicators of data and concept drift, feature integrity, and model behavior stability. ML models are uniquely sensitive to changes in data distributions, which can degrade performance or mask adversarial manipulation. Research on drift adaptation~\cite{gama2014survey} emphasizes that both statistical and adversarial drifts must be monitored and mitigated through redundant detection mechanisms, secure baselines, and controlled retraining protocols~\cite{korycki2023adversarial}. By integrating drift-aware monitoring into operational workflows, SecMLOps ensures that evolving threats are detected and addressed before they compromise system integrity.

SecMLOps institutionalizes a continuous risk management and feedback loop. Detected anomalies trigger automated incident responses, root-cause analysis using provenance data, and iterative refinement of models and policies. Immutable feature stores and versioned model registries support rollback and recovery, ensuring resilience against data corruption and poisoning attacks. This cyclical process transforms security from a reactive control into a proactive and adaptive capability that evolves with the operational environment.

\subsubsection{Governance and Compliance}
In the \textit{Governance and Compliance} of SecMLOps, it is crucial to distinguish between external influences and internal mechanisms. External factors encompass the regulatory and industry standards that dictate foundational security requirements and ethical guidelines for ML systems. Internally, organizations must develop robust frameworks and processes that ensure these external standards are effectively implemented and consistently adhered to within their operational practices.

Externally, the governance of SecMLOps is primarily shaped by regulatory requirements and industry standards that dictate comprehensive guidelines for the secure and ethical management of ML systems. Regulatory frameworks such as the General Data Protection Regulation (GDPR)~\cite{voigt2017eu} in the European Union and the Health Insurance Portability and Accountability Act (HIPAA)~\cite{act1996health} in the United States set strict protocols for data privacy and security, necessitating adherence to high standards of operational integrity. Furthermore, industry-specific guidelines, such as those from the International Organization for Standardization (ISO), particularly ISO/IEC 27001~\cite{disterer2013iso}, provide a framework for information security management that imposes additional compliance obligations influencing strategic oversight and policy formulation within organizations engaged in ML operations.

Internally, organizations operationalize these external mandates through rigorous governance structures and compliance mechanisms, ensuring consistent adherence across all facets of ML development and deployment. This internalization process includes the development of detailed policies and procedures that govern data handling, model training, and deployment, supported by regular audits and real-time monitoring systems to enforce compliance and manage risks effectively. Additionally, internal compliance is enhanced through the maintenance of comprehensive documentation and robust reporting systems, which not only facilitate regulatory audits but also promote an organizational culture of transparency and accountability. These practices ensure that the governance and compliance frameworks are not merely reactive but are proactive measures that integrate security and ethical considerations into the core operational processes of SecMLOps.

Having detailed the individual components and their roles, we now demonstrate how these elements integrate into a cohesive operational framework. Figure~\ref{fig:processdiagram} illustrates this integration through a comprehensive process diagram that maps the interaction between the eight roles across four stages, showing the practical workflow of SecMLOps implementation.

\subsection{The SecMLOps Framework}
SecMLOps represents an advanced methodological framework designed to integrate security seamlessly into every stage of the MLOps life cycle, as shown in a process diagram (Figure~\ref{fig:processdiagram}). The framework integrates the components described in Section~\ref{Sub.Components}.

In the Research and Planning stage (denoted by activities labelled with RP in Figure~\ref{fig:processdiagram}), business stakeholders or product owners (R1) initiate the development by identifying core business goals (\activity{RP1}) and defining the system's operational context (\activity{RP2}). This crucial phase aligns the project with organizational objectives and specific security needs. Concurrently, R1 usually collaborates with external security experts to navigate the complex regulatory landscape, addressing relevant laws, standards, and security frameworks (\activity{RP3.1}). This partnership is vital for establishing comprehensive system requirements that encompass both functional and non-functional aspects (\activity{RP3.2}), including stringent security policies and compliance mandates. Key processes such as asset identification (\activity{RP4}), security analysis (\activity{RP5}), security goal setting (\activity{RP6}), and the formulation of security requirements (\activity{RP7}), security controls (\activity{RP8}), and policies (\activity{RP9}) are meticulously developed. Security analysis includes but not limit to threat modeling,  attack identification, risk assessment. These activities lay the groundwork for integrating security deeply into the business strategy and system architecture, ensuring readiness for the subsequent design stage where specific security controls are specified for system architecture and established security policies guide further development.

The SecMLOps framework systematically considers distinct security threats that emerge at each phase of the MLOps lifecycle, recognizing that ML systems face fundamentally different vulnerabilities than traditional software systems~\cite{papernot2018sok}. During the Research and Planning phase, the primary security concerns center on the integrity of initial datasets and the trustworthiness of pretrained models. Data poisoning attacks represent a critical threat at this stage, where adversaries may inject malicious samples into training datasets to compromise model behavior. Our framework addresses this through mandatory data provenance tracking (Activity RP4) and comprehensive threat modeling using STRIDE methodology (Activity RP5), ensuring that potential poisoning vectors are identified before training begins.

The Development phase introduces vulnerabilities specific to the model training process itself. Beyond traditional code vulnerabilities, ML systems face unique risks such as membership inference attacks that can extract information about training data, and model extraction attempts through repeated querying during validation~\cite{gebru2018datasheets}. The framework mandates secure training environments with restricted access (Activities DV6-DV7), implements differential privacy techniques to protect training data confidentiality, and establishes validation protocols that limit information leakage through model responses. Particularly critical is the protection of gradient information during distributed training, where federated learning scenarios may expose model updates to adversarial participants.

During the Deployment phase, ML systems become vulnerable to adversarial examples and evasion attacks that don't exist in conventional software. An attacker might craft inputs that appear benign to human observers but cause misclassification in production models. Our framework addresses these threats through mandatory adversarial robustness testing (Activity DP1) before deployment, implementing input validation mechanisms that detect statistical anomalies indicative of adversarial perturbations, and establishing secure model serving infrastructure that prevents direct access to model weights. The deployment phase also requires specific controls against model inversion attacks, where adversaries attempt to reconstruct training data from deployed models.

The continuous monitoring phase faces the unique ML challenge of concept drift and model degradation, which can be exploited by adversaries through slow, deliberate data distribution shifts. Unlike traditional software that maintains consistent behavior, ML models may silently fail as input distributions change over time~\cite{vzliobaite2015overview}. SecMLOps addresses this through continuous monitoring systems (Activity DP3.2) that track not just performance metrics but also data distribution statistics, prediction confidence patterns, and feature importance stability. When drift is detected, the framework triggers secure retraining protocols that validate new data against poisoning attempts while maintaining model versioning for rollback capabilities. 

In the Design stage (denoted by activities labelled with D in Figure~\ref{fig:processdiagram}), the solution architect (R2) takes a central role, shaping the system architecture with an emphasis on security (\activity{D1}). This stage is directly influenced by the previously established security policies, guiding the selection of secure technologies (\activity{D3}). R2 ensures that each aspect of the system's design adheres to these policies, fostering a secure and robust architecture with required security controls. The process involves critical decision-making regarding model (\activity{D2.1}) and data source selection (\activity{D2.2}), which are fundamental to constructing a resilient ML system. Each decision is made with a comprehensive understanding of how architectural choices impact the overall security and functionality of the system, ensuring that the architecture not only supports but enhances security measures.

In the Development stage (denoted by activities labelled with DV in Figure~\ref{fig:processdiagram}), the focus shifts to implementing the secure system design. This phase involves collaboration among data scientists or ML developers (R3), data engineers (R4), and software engineers (R5), and is coordinated by the MLOps Engineer (R7). R4 manages the data ingestion (\activity{DV1}), extraction (\activity{DV2}), and preprocessing of data to the feature store (\activity{DV3}), ensuring that each step complies with security controls to protect data integrity and confidentiality. Following this, R3 undertakes the data analysis (\activity{DV4}), data preparation (\activity{DV5}), model training (\activity{DV6}), model validation (\activity{DV7}), and model export (\activity{DV8}), embedding comprehensive security measures throughout these phases to mitigate vulnerabilities effectively. They cooperate with software engineer (R5) and DevOps engineer (R6) to automate the feature engineering and ML model into orchestrated pipeline (\activity{DP1}) under the secure coding principles as the CI/CD component in source repository. The Model Registry is critical at this stage, as it archives the model's architecture, status, and logs, along with performance metrics that are essential for ongoing security monitoring and the formulation of responsive actions. When the model advances to the Deployment stage (denoted by activities labelled with DP in Figure~\ref{fig:processdiagram}) and is actively served (\activity{DP4}), it is crucial to build monitoring (\activity{DP2}) for both performance (\activity{DP3.1}) and security (\activity{DP3.2}), as well as to establish incident response procedures (\activity{DP3.3}). It is also important to routinely update the systems (\activity{DP5}) to respond to new security challenges or anomalies, such as adversarial attacks. R7 plays a key role in operations to maintain a robust feedback loop that potentially revisits and updates the feature store, feature engineering pipeline, model registry, and orchestrated experimentation pipeline. Concurrently, the SecMLOps Engineer (R8) focuses on the security aspects, leading the assessment and integration of new security measures. They are responsible for revising monitoring strategies and refining incident response plans in response to new security threats. This role ensures that all security updates align with regulatory compliance and the system's security framework, enhancing the system's resilience against evolving threats. This comprehensive approach not only maintains the system's integrity but also enhances its adaptability, securing its functionality against future security challenges.

To demonstrate the practical application of the SecMLOps framework presented above, we now present a comprehensive use case using a state-of-the-art PDS. This case study illustrates how each component of the PTPGC model and the defined roles collaborate in practice, moving from abstract framework to concrete implementation. Each subsection that follows maps directly to the activities in our process diagram (Figure~\ref{fig:processdiagram}), providing a step-by-step demonstration of SecMLOps in action.

\section{Use Case: Pedestrian Detection Systems}
\label{Sec.IV_CaseStudy}
This section presents a comprehensive use case that demonstrates the practical application of the SecMLOps framework within the context of the Vision-Language semantic self-supervision for context-aware Pedestrian Detection (VLPD) system~\cite{Liu_2023_CVPR}. This section provides a detailed examination of the VLPD system, focusing on its architecture, key operational assets, compliance regulations, security analysis, security goals, requirements, controls, policy frameworks, and strategies for monitoring security and incident response. Each subsection corresponds to specific security activities outlined in the SecMLOps process diagram (Figure~\ref{fig:processdiagram}), namely \activity{RP2, RP5, RP3.1, RP6, RP7-9, RP10,} and \activity{DP3.2-3.3} respectively. By structuring the use case in this manner, we provide a clear and comprehensive overview of how security is seamlessly integrated and managed throughout the VLPD system, in alignment with the SecMLOps framework. This use case serves as a practical illustration of how SecMLOps can be applied to real-world ML systems, enhancing their security, reliability, and compliance.

Table~\ref{tab:use_case_mapping} demonstrates the systematic application of the SecMLOps framework to the VLPD implementation. Each subsection of the use case directly implements specific activities from the framework's process diagram (Figure~\ref{fig:processdiagram}), with clear role assignments and PTPGC component utilization. This mapping illustrates how the theoretical framework translates into practical implementation steps, with each phase building upon the previous to create a comprehensive security integration throughout the MLOps lifecycle.

\begin{table*}[ht]
\renewcommand{\arraystretch}{1.3}
\caption{Mapping of VLPD Use Case Implementation to SecMLOps Framework Components}
\label{tab:use_case_mapping}
\centering
\begin{tabular}{>{\raggedright\arraybackslash}p{0.18\textwidth}>{\raggedright\arraybackslash}p{0.25\textwidth}>{\raggedright\arraybackslash}p{0.25\textwidth}>{\raggedright\arraybackslash}p{0.22\textwidth}}
\toprule
\textbf{Use Case Section} & \textbf{Framework Activity} & \textbf{Primary Roles} & \textbf{PTPGC Components} \\
\midrule
4.1 System Overview & RP2: Define operational context & R1 (Business Stakeholder), R2 (Solution Architect) & People, Process \\
4.2 Assets Identification & RP4: Asset identification & R2 (Solution Architect), R8 (SecMLOps Engineer) & Process, Technology \\
4.3 Compliance and Regulations & RP3.1: Address regulations and standards & R1 (Business Stakeholder), R8 (SecMLOps Engineer) & Governance, Compliance \\
4.4 Security Analysis & RP5: Security analysis (threat modeling, risk assessment) & R2 (Solution Architect), R8 (SecMLOps Engineer), R3 (Data Scientist) & Process, Technology \\
4.5 Security Goals, Requirements and Controls & RP6: Security goal setting, RP7: Security requirements, RP8: Security controls & R2 (Solution Architect), R8 (SecMLOps Engineer) & Process, Technology, Governance \\
4.6 Security Policy & RP9: Security policies formulation & R1 (Business Stakeholder), R8 (SecMLOps Engineer) & Governance, Compliance \\
4.7 Monitor Security and Incident Response & DP3.2: Security monitoring, DP3.3: Incident response & R8 (SecMLOps Engineer), R7 (MLOps Engineer), R6 (DevOps Engineer) & Technology, Process \\
\bottomrule
\end{tabular}
\end{table*}

\subsection{System Overview}
\begin{figure*}[t!]
    \centering
    \includegraphics[width=\linewidth]{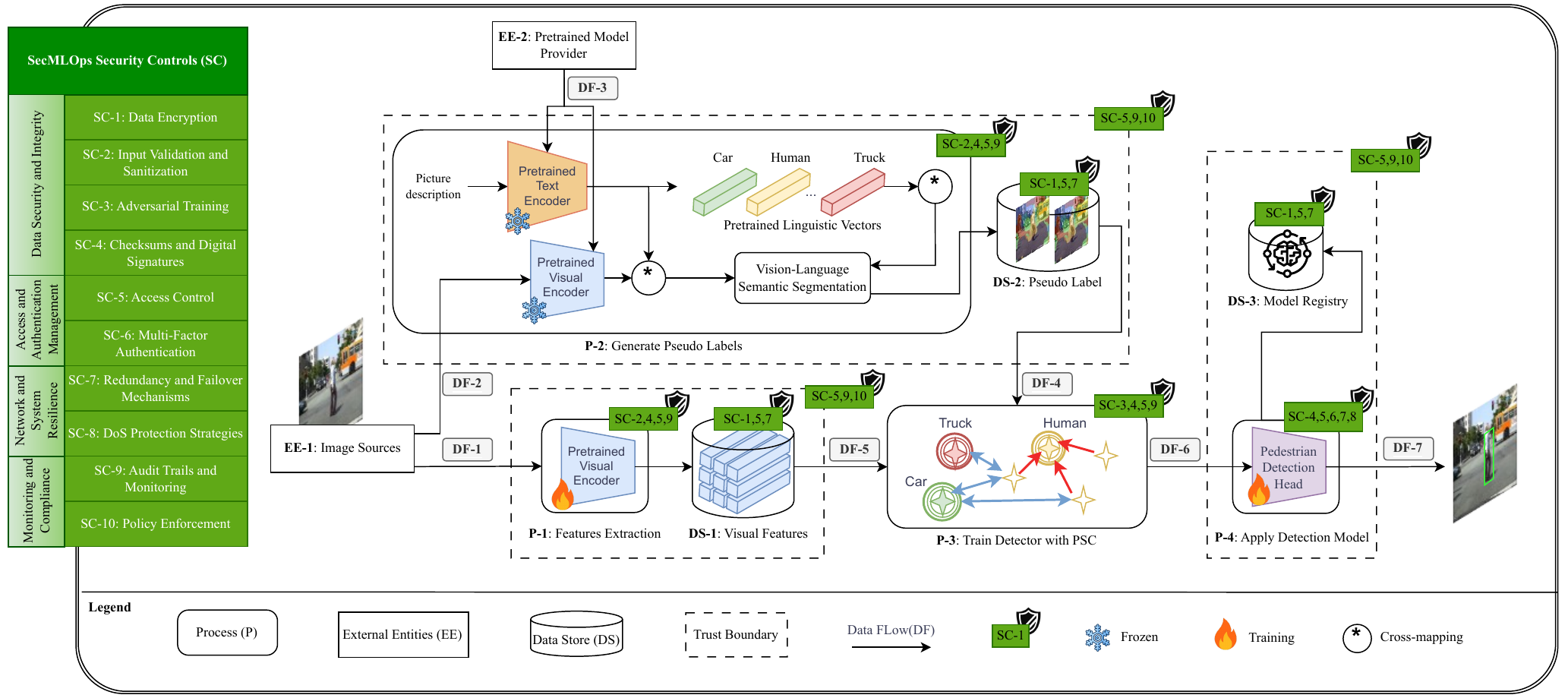}
    \caption{Data Flow Diagram of SecMLOps-enabled VLPD System, integrating security controls within MLOps lifecycle to ensure system's robustness and security.}
    \label{fig:vlpd}
\end{figure*}

The VLPD system (shown in Figure~\ref{fig:vlpd}) represents a sophisticated, state-of-the-art approach that effectively navigates the complexities of real-world environments.The VLPD system implementation follows the SecMLOps framework's four-stage process (RP, D, DV, DP) with clear role assignments. The Business Stakeholder (R1) initiated the project by identifying the need for secure pedestrian detection in autonomous driving contexts, while the Solution Architect (R2) designed the system architecture incorporating security-by-design principles from the framework's Design stage.

The VLPD system employs the Vision-Language Semantic (VLS) segmentation method, which utilizes pretrained vision-language models for effective cross-modal mapping. This mapping generates pseudo labels that bridge visual features with corresponding linguistic vectors, representing semantic classes enriched by the model's linguistic insights. This mechanism enables the visual encoder to directly recognize and segment diverse semantic classes from images, thereby bypassing the need for manual annotation.  Subsequently, the Prototypical Semantic Contrastive (PSC) learning method capitalizes on these rich semantic contexts to refine the system's ability to discern pedestrians from other elements in a scene through class-specific prototypes. 

\subsection{Assets Identification}
Following the Research and Planning stage of our SecMLOps framework, we begin with asset identification (Activity RP4 in Figure~\ref{fig:processdiagram}), a crucial step that the Solution Architect (R2) undertakes in collaboration with Business Stakeholders (R1). We conducted a systematic inventory using the three-tier classification defined in Section~\ref{subsubsec3.2process}: data assets, computational models, and operational algorithms. The Solution Architect (R2) led structured sessions using the framework's asset categorization template, which guided the team to identify not just obvious assets like training data, but also critical components such as the cross-modal mapping algorithm that might otherwise be overlooked.

In the VLPD system, the assets are fundamental to its operational integrity and effectiveness in complex real-world environments. This subsection corresponds to the activity labeled as \activity{RP5} in Figure~\ref{fig:processdiagram}, which involves identifying and categorizing the critical assets within the system that require protection. These assets are broadly categorized into data assets, computational models, and operational algorithms, each playing a pivotal role in the system's architecture.

Data assets primarily include raw image data, which constitutes the visual input that the system processes. Alongside this, visual features and pseudo labels, generated through the system's advanced processing capabilities, transform raw data into a structured format that is essential for further semantic analysis and classification. These data assets are fundamental as they form the basis for all subsequent processing and decision-making tasks within the system.

Computational models are central to the VLPD's ability to interpret and analyze visual data. This category includes the pretrained visual and text encoders, which are crucial for converting visual input into detailed feature maps and generating linguistic vectors that encapsulate semantic information relevant to the visuals. Another significant model, the pedestrian detection head, is specifically optimized to identify and localize pedestrians within the data, leveraging the refined semantic classifications provided by earlier processing stages.

Operational algorithms such as the cross-modal mapping algorithm and the PSC framework form the backbone of the system's processing capabilities. The cross-modal mapping algorithm is indispensable for linking visual features with linguistic vectors to produce pseudo labels, which are crucial for the semantic segmentation of images. The PSC framework enhances the detection process by improving feature discrimination, ensuring that the system can effectively distinguish pedestrians from other scene elements.

\subsection{Compliance and Regulations}
In the context of VLPD system, governance and compliance are integral to ensuring that the system adheres to established legal and industry-specific standards. This subsection corresponds to the activity labeled as \activity{RP3.1} in Figure~\ref{fig:processdiagram}, which involves identifying and addressing the external regulatory requirements and standards applicable to the system.

Externally, the system is required to comply with stringent data protection regulations such as GDPR~\cite{voigt2017eu} in the European Union. This regulation mandates specific measures for the protection of personal data, crucial for systems like the VLPD that process visual data in public spaces, necessitating robust anonymization protocols and secure data handling practices.

Moreover, the VLPD system must align with automotive safety standards such as ISO 26262~\cite{ISO26262}, which addresses the safety of electronic and electrical systems within road vehicles, and ISO/PAS 21448~\cite{ISOPAS21448}, known as SOTIF (Safety Of The Intended Functionality). These standards are critical for ensuring that the system functions reliably and safely under various operational conditions, guiding rigorous development and validation processes specific to autonomous driving technologies~\cite{zhao2025intelligent}.

Internally, the VLPD's governance structure includes strict adherence to these external standards through the enforcement of detailed internal policies. This governance ensures that operations ranging from data collection to model deployment are conducted under controlled conditions. Auditing processes are regularly implemented to monitor compliance, and advanced cybersecurity measures such as end-to-end encryption and secure data storage are utilized to protect data integrity and confidentiality. These internal governance mechanisms are vital for maintaining regulatory compliance and ensuring that all aspects of system operation are secure and controlled.

\subsection{Security Analysis}
\label{Subsec.IV_analysis}

With assets identified and compliance requirements established, we proceed to security analysis (Activity RP5), where the SecMLOps Engineer (R8) works with the Solution Architect (R2) to systematically evaluate threats using the STRIDE methodology introduced in our framework.

Given the critical nature of the application in fields like autonomous driving and urban surveillance, ensuring the robustness of VLPD against potential security threats is paramount. This subsection corresponds to the activity labeled as \activity{RP6} in Figure~\ref{fig:processdiagram}, which involves systematically evaluating the security implications of the VLPD system and identifying potential threats, risks, and impacts.
This analysis adopts the STRIDE methodology~\cite{ABUABED2023103391} to systematically evaluate the security implications of VLPD systems, specifically focusing on identifying potential threats, risks, and impacts associated with data flow elements and interactions including external entities, processes, data flows, and data stores shown in Figure~\ref{fig:vlpd}. The primary goal of this security analysis is to achieve a thorough understanding of the system's security threats, risks, and impacts. 
This analysis forms a crucial part of the SecMLOps framework, laying the groundwork for establishing well-defined security goals and implementing effective controls to protect the system.

The STRIDE threat analysis began with a discussion where the SecMLOps Engineer (R8) facilitated sessions with the Solution Architect (R2) and Data Scientist (R3). The Solution Architect mapped the system's data flow components while the Data Scientist identified ML-specific vulnerabilities in the model pipeline. For each identified threat, the team followed the framework's risk assessment process: (1) the SecMLOps Engineer evaluated threat likelihood based on known attack patterns, (2) the Solution Architect assessed potential system impact, and (3) the MLOps Engineer (R7) estimated operational consequences. This collaborative process, guided by the framework's RP5 activity template, produced the comprehensive threat matrix shown in Table~\ref{tab:stride-matrix}.

\begin{table}[ht!]
\centering
\caption{STRIDE threat and impact matrix for VLPD systems. Each cell indicates the likelihood of a threat occurring (H: High, M: Medium, L: Low, N: Not applicable) and is color-coded to show the severity of the impact(Green: Low, Yellow: Medium, Red: Severe).}
\label{tab:stride-matrix}
\begin{tabular}{|l|l|l|l|l|l|l|}
\hline
              & \textbf{S}                & \textbf{T}                & \textbf{R}                & \textbf{I}                & \textbf{D}                & \textbf{E}                \\ \hline
\textbf{EE-1} & \cellcolor[HTML]{FF6347}H & \cellcolor[HTML]{FF6347}H & \cellcolor[HTML]{98FB98}L & \cellcolor[HTML]{98FB98}L & \cellcolor[HTML]{FFFF00}M & \cellcolor[HTML]{98FB98}L \\ \hline
\textbf{EE-2} & \cellcolor[HTML]{FFFF00}M & \cellcolor[HTML]{FFFF00}M & \cellcolor[HTML]{98FB98}L & \cellcolor[HTML]{98FB98}L & \cellcolor[HTML]{98FB98}L & \cellcolor[HTML]{98FB98}L \\ \hline
\textbf{P-1}  & \cellcolor[HTML]{98FB98}L & \cellcolor[HTML]{FFFF00}M & \cellcolor[HTML]{FFFF00}L & \cellcolor[HTML]{FFFF00}M & \cellcolor[HTML]{FF6347}M & \cellcolor[HTML]{FF6347}M \\ \hline
\textbf{P-2}  & \cellcolor[HTML]{98FB98}L & \cellcolor[HTML]{FFFF00}M & \cellcolor[HTML]{FFFF00}L & \cellcolor[HTML]{98FB98}M & \cellcolor[HTML]{FF6347}M & \cellcolor[HTML]{FF6347}M \\ \hline
\textbf{P-3}  & \cellcolor[HTML]{98FB98}L & \cellcolor[HTML]{FF6347}M & \cellcolor[HTML]{FFFF00}M & \cellcolor[HTML]{FFFF00}M & \cellcolor[HTML]{FF6347}M & \cellcolor[HTML]{FF6347}M \\ \hline
\textbf{P-4}  & \cellcolor[HTML]{98FB98}L & \cellcolor[HTML]{FF6347}M & \cellcolor[HTML]{FFFF00}M & \cellcolor[HTML]{98FB98}M & \cellcolor[HTML]{FF6347}H & \cellcolor[HTML]{FF6347}M \\ \hline
\textbf{DF-1} & \cellcolor[HTML]{FFFF00}H & \cellcolor[HTML]{FF6347}H & \cellcolor[HTML]{98FB98}L & \cellcolor[HTML]{98FB98}L & \cellcolor[HTML]{FF6347}H & \cellcolor[HTML]{98FB98}L \\ \hline
\textbf{DF-2} & \cellcolor[HTML]{98FB98}L & \cellcolor[HTML]{FFFF00}M & \cellcolor[HTML]{98FB98}L & \cellcolor[HTML]{98FB98}L & \cellcolor[HTML]{FFFF00}M & \cellcolor[HTML]{98FB98}L \\ \hline
\textbf{DF-3} & \cellcolor[HTML]{98FB98}L & \cellcolor[HTML]{FFFF00}M & \cellcolor[HTML]{98FB98}L & \cellcolor[HTML]{98FB98}L & \cellcolor[HTML]{FFFF00}M & \cellcolor[HTML]{98FB98}L \\ \hline
\textbf{DF-4} & \cellcolor[HTML]{98FB98}L & \cellcolor[HTML]{FFFF00}M & \cellcolor[HTML]{98FB98}L & \cellcolor[HTML]{98FB98}L & \cellcolor[HTML]{FFFF00}M & \cellcolor[HTML]{98FB98}L \\ \hline
\textbf{DF-5} & \cellcolor[HTML]{98FB98}L & \cellcolor[HTML]{FFFF00}M & \cellcolor[HTML]{98FB98}L & \cellcolor[HTML]{98FB98}L & \cellcolor[HTML]{FFFF00}M & \cellcolor[HTML]{98FB98}L \\ \hline
\textbf{DF-6} & \cellcolor[HTML]{98FB98}L & \cellcolor[HTML]{FFFF00}M & \cellcolor[HTML]{98FB98}L & \cellcolor[HTML]{98FB98}L & \cellcolor[HTML]{FFFF00}M & \cellcolor[HTML]{98FB98}L \\ \hline
\textbf{DF-7} & \cellcolor[HTML]{98FB98}L & \cellcolor[HTML]{FFFF00}M & \cellcolor[HTML]{98FB98}L & \cellcolor[HTML]{FFFF00}M & \cellcolor[HTML]{FFFF00}M & \cellcolor[HTML]{98FB98}L \\ \hline
\textbf{DS-1} & \cellcolor[HTML]{98FB98}L & \cellcolor[HTML]{FFFF00}M & \cellcolor[HTML]{FFFF00}M & \cellcolor[HTML]{FFFF00}M & \cellcolor[HTML]{98FB98}L & \cellcolor[HTML]{98FB98}L \\ \hline
\textbf{DS-2} & \cellcolor[HTML]{98FB98}L & \cellcolor[HTML]{FFFF00}M & \cellcolor[HTML]{FFFF00}M & \cellcolor[HTML]{FFFF00}M & \cellcolor[HTML]{98FB98}L & \cellcolor[HTML]{98FB98}L \\ \hline
\textbf{DS-3} & \cellcolor[HTML]{98FB98}L & \cellcolor[HTML]{FFFF00}M & \cellcolor[HTML]{FFFF00}M & \cellcolor[HTML]{FFFF00}M & \cellcolor[HTML]{98FB98}L & \cellcolor[HTML]{98FB98}L \\ \hline
\end{tabular}%
\end{table}

The STRIDE Threat and Impact Matrix systematically categorizes the probability and potential impact of various security threats across different system components. In Table~\ref{tab:stride-matrix}, each cell within the matrix corresponds to a specific combination of system element (ranging from External Entities, Processes, Data Flows, to Data Stores) and a type of STRIDE threat. The likelihood of each threat occurring is indicated by a letter ('H' for High, 'M' for Medium, and 'N' for Low) providing a qualitative measure of vulnerability. Concurrently, the background color of each cell illustrates the severity of the impact should the threat materialize, categorized as follows:
\begin{itemize}
    \item Green (Low Impact): Inconveniences or minor disruptions that do not affect the core functionality of the system or lead to any significant damages.
    \item Yellow (Medium Impact): Disruptions that affect system performance or data integrity but do not cause direct physical harm or significant financial loss.
    \item Red (High Impact): Situations that could lead to severe consequences such as human injury, substantial financial loss, or significant damages to the system's infrastructure and reputation.
\end{itemize}

This matrix not only enables a quick visual evaluation of potential security vulnerabilities but also assists in strategically prioritizing mitigation efforts by considering both the likelihood and severity of the identified threats. The following will dive into a detailed analysis of each threat type.

 \textit{Spoofing}. In the VLPD system, spoofing targets key external entities: EE-1, which provides raw images, and EE-2, which supplies pretrained models. Attackers may manipulate EE-1 by injecting poisoned data or AE~~\cite{10057473}, leading directly to flaws in pedestrian detection during the model's training or inference phases. Such spoofing compromises the system’s ability to accurately classify pedestrians, raising the risk of incorrect pedestrian detection or misclassification. Similarly, spoofing EE-2 could cause the system to adopt compromised models with embedded backdoors, enabling attackers to manipulate outputs during operational deployment. These compromised inputs and models affect DF-1, escalating errors through the system’s feature extraction process and degrading the overall decision-making accuracy and reliability, potentially leading to critical failures in pedestrian detection.
 
 \textit{Tampering}. 
In the VLPD system, tampering primarily affects EE-1 and EE-2, where unauthorized alterations of input data and models pose significant threats. Tampering with EE-1's image data can lead to inaccuracies in initial pedestrian detection during both training and operational phases. Similarly, compromised models from EE-2 can introduce biases, causing systemic errors in the processing algorithms. This malicious alteration propagates through DF-1 to the P-1, further affecting P-2. Such distortions in early processing stages amplify through the system, degrading the accuracy of pedestrian detection and classification, culminating in a heightened risk of erroneous decisions in critical applications like autonomous driving, potentially leading to unsafe vehicle behaviors and accidents.

 \textit{Repudiation}.
The VLPD system's processes (P-1 through P-4) are vulnerable to repudiation threats, where unauthorized modifications or deletions could go unchecked. If actions within these processes are not properly logged, attackers could alter detection algorithms or manipulate data outputs without detection. This lack of accountability can severely disrupt the integrity of pedestrian detection, leading to incorrect classifications or missed detections. Such compromises in data accuracy directly impact subsequent processes and overall system performance, potentially resulting in flawed decision-making and compromising the safety of autonomous vehicle operations dependent on this system.

 \textit{Information Disclosure}.
The threat of information disclosure arises mainly from its handling of proprietary model and image data. While the exposure of image data through External Entities EE-1 and EE-2 is deemed low risk and has minimal impact on system functionality, it could potentially affect pedestrian privacy. More critical are Data Stores (DS-1, DS-2, DS-3) and core Processes (P-1 through P-4), where the risk of exposing operational details or ML models is medium. If data from these components were disclosed, it could reveal detailed insights into the system's operational mechanics or ML models, posing a threat to the system's confidentiality. 
Such disclosure could enable model stealing, membership or model inference attack, leading to broader security implications that could undermine the effectiveness and competitive advantage of the VLPD system.

\textit{Denial of Service}.
Denial of Service (DoS)~\cite{guo2023distributed} attacks primarily target the essential data flow and processing components, disrupting the system's ability to function correctly. An attack on DF-1, which serves as the principal gateway for the ingress of raw image data, could severely impact the system by halting the initial data input, which is crucial for all subsequent processing stages. Such an interruption would directly affect processes like P-1, where the absence of incoming data prevents the extraction of meaningful features necessary for further analysis. This disruption cascades through to P-2 and P-3, stalling the generation of pseudo labels and the training of detection models, respectively. The halted flow in these critical pathways not only degrades the system's performance but also interrupts real-time processing essential for applications like autonomous driving, leading to operational paralysis and potential safety hazards.

\textit{Elevation of Privilege}.
Elevation of Privilege attack targets crucial processes such as P-1 through P-4, which handle sensitive operations like feature extraction and model application. If an attacker gains elevated access within these areas, they can manipulate system algorithms, access restricted functionalities or extract proprietary data, compromising the system's integrity and confidentiality of sensitive data. This unauthorized access can lead to altered algorithm parameters or the disruption of operational logic, significantly impacting the system's reliability.

The STRIDE analysis reveals phase-specific vulnerabilities unique to ML systems. During data ingestion (DF-1), the system faces high risk of poisoning attacks where adversaries inject malicious pedestrian labels to cause systematic misclassification. This differs fundamentally from traditional data integrity threats as even small poisoning percentages (5-10\%) can significantly degrade model performance, as demonstrated in our evaluation in Section~\ref{Sec.V_Evaluation}. The training phase (P-3) exhibits medium vulnerability to model extraction through careful observation of training metrics and validation outputs, while the deployment phase (P-4) faces high risk from adversarial examples specifically crafted to evade pedestrian detection. 

\subsection{Security Goals, Requirements and Controls}

The security control selection followed the framework's iterative process. The Solution Architect (R2) initially proposed technical controls based on the threat analysis. These were reviewed by the SecMLOps Engineer (R8) for completeness and refined through consultations with the Data Engineer (R4) for data pipeline controls and the Software Engineer (R5) for implementation feasibility. For instance, the decision to implement adversarial training (SC-3) emerged from a specific discussion where the Data Scientist (R3) demonstrated vulnerability to FGSM attacks, leading the team to prioritize robustness measures in the training pipeline.

To ensure security for the VLPD system, the following security goals are formulated within the context of the CIAAAA (Confidentiality, Integrity, Availability, Authentication, Authorization, and Accountability) security objectives. This subsection corresponds to the activities labeled as \activity{RP7}, \activity{RP8}, and \activity{RP9} in Figure~\ref{fig:processdiagram}, which involve establishing security goals, defining precise security requirements, and proposing corresponding security controls based on the vulnerabilities and potential threats identified through the security analysis.
Each goal is prominently outlined in its own subsection, specifically tailored to address the vulnerabilities and potential threats identified through the security analysis of the system. In each subsection, we discuss the goals and establish precise security requirements, then we propose the corresponding security controls which are comprehensively illustrated in Figure~\ref{fig:vlpd}.
This ensures that each security aspect is tightly integrated and effectively managed to protect the system against diverse security challenges~\cite{Zhang2022aa}.

\subsubsection{\textbf{Confidentiality of Proprietary Algorithms and Data} }
Protecting the confidentiality of ML models and the data they process is paramount, especially given the system’s usage in sensitive applications like autonomous driving. Unauthorized disclosure could lead to model stealing or reverse engineering, undermining competitive advantages and operational security. To address these challenges, we identify the following security requirements: (1) ensuring all data, including model details and training data, remain inaccessible to unauthorized entities, and (2) protecting data integrity and privacy during transfer and storage. To meet these requirements, we propose a set of targeted security controls that are detailed below:

\begin{itemize}
    \item \textit{SC-1. Data Encryption.} Employ state-of-the-art encryption technologies to safeguard all data at rest and in transit within the VLPD system. Utilize cryptographic protocols such as AES-256 and TLS to ensure that model details and proprietary algorithms remain inaccessible to unauthorized parties~\cite{lanza2022memristive}. This measure prevents attackers from gaining insights into the model architecture or training data, which could facilitate reverse engineering or model replication efforts.
    \item \textit{SC-2. Access Controls.} Implement stringent access control mechanisms, such as role-based access control (RBAC) \cite{laverdiere2021rbac}, to restrict access to model configurations and training datasets. Access should be granted based on the principle of least privilege, ensuring that only personnel with a legitimate need can view or interact with sensitive model details. Regular audits of access logs will help identify and respond to unauthorized access attempts, further securing proprietary information.
\end{itemize}

\subsubsection{\textbf{Integrity of Data Flow and Processing} }
Ensuring the integrity of data throughout its lifecycle in the VLPD system is crucial. This goal specifically addresses risks such as data tampering, spoofing, and the influence of AE, which could lead to incorrect pedestrian detection and potentially hazardous outcomes in real-world applications. To effectively mitigate these risks, we have identified two essential security requirements: (1) maintaining data accuracy and consistency across all processing stages, and (2) enhancing model robustness to reduce its susceptibility to adversarial manipulations in all operational environments. The following security controls are proposed to address these requirements:

\begin{itemize}
    \item \textit{SC-3. Adversarial Training.} Integrate AE into the training process to prepare the model to recognize and resist adversarial inputs during operational deployment. This proactive strategy enhances the model's resilience by exposing it to a variety of attack vectors under controlled conditions, thereby improving its ability to maintain integrity under adversarial conditions.
    \item \textit{SC-4. Input Validation and Sanitization.} Develop robust input validation mechanisms to detect and reject corrupted or malicious data entries before they are processed by the system. Techniques such as anomaly detection can identify outliers or unusual patterns that may indicate tampering or adversarial interference.
    \item \textit{SC-5. Checksums and Digital Signatures.} Utilize checksums and digital signatures to verify the integrity of data as it moves through the processing stages. These measures ensure that any alterations to the data, whether due to technical glitches or malicious interference, are detected promptly, allowing for immediate remediation.
\end{itemize}

\subsubsection{\textbf{Availability of Real-Time Processing Capabilities}} Continuous system availability is critical for the reliability of autonomous operations and urban surveillance systems. This goal mitigates against DoS attacks, which could disrupt the processing of real-time data, leading to failures in detecting and responding to live environment changes. To ensure resilience, the security requirement is defined as follows: (1) maintaining the capability to continue operations seamlessly, even in the event of component failures or external attacks. To meet this requirement, we propose the following security controls:
\begin{itemize}
    \item \textit{SC-6. Redundancy and Failover Mechanisms.} Design the system architecture with built-in redundancy for critical components and data pathways. Implement failover mechanisms to ensure that backup systems can take over seamlessly in the event of a hardware failure or a cyber-attack.
    \item \textit{SC-7. DoS Protection Strategies.} Deploy advanced network security solutions, including firewalls and intrusion prevention systems, to defend against DoS attacks. Regularly update these systems to guard against emerging threats and ensure they are configured to maintain data throughput and processing capabilities under attack conditions.
\end{itemize}

\subsubsection{\textbf{Authentication of System Interactions}} 
Strong authentication mechanisms are crucial to confirm the identities of interacting entities, preventing unauthorized access. This goal directly counters potential spoofing and elevation of privilege threats by ensuring that only legitimate operations and modifications are executed within the system. A key security requirement has been identified: verifying the identity of all users and processes accessing the system using robust, compromise-resistant authentication mechanisms. To meet this requirement, the following security control is proposed:

\begin{itemize}
    \item \textit{SC-8.Multi-Factor Authentication (MFA).} Enforce MFA for all user interactions with the system, combining something the user knows (password) with something the user has (security token) and something the user is (biometric verification)~\cite{sinigaglia2020survey}. This layered approach significantly reduces the risk of unauthorized access due to compromised credentials.
\end{itemize}

\subsubsection{\textbf{Authorization to Enforce Access Controls}} 
Proper authorization ensures that even authenticated users or processes are only able to perform actions within their permitted scope. This goal is particularly vital in preventing unauthorized changes or access to system configurations and sensitive data, which could compromise system operations. To effectively enforce this goal, we establish the following security requirements: (1) Implement dynamic access control mechanisms that adjust permissions based on real-time analysis of user activity and context, ensuring users have appropriate access levels at all times, and (2) Regularly audit and review authorization policies and permissions to ensure they align with current security policies and the principle of least privilege. The security controls to achieve these requirements include:

\begin{itemize}
    \item \textit{SC-9. Policy Enforcement.} Implement comprehensive policies that define and enforce what actions each user role can perform within the system under what situations. Regularly review and update these policies to adapt to new security challenges or changes in organizational structure~\cite{FLOWERDAY2016169}.
\end{itemize}

\subsubsection{\textbf{Accountability and Non-Repudiation of Actions}} Implementing mechanisms that log and monitor system activities ensures that all modifications can be attributed to a verified source. This goal supports non-repudiation by making it possible to trace every action back to its origin, crucial for auditing and responding to incidents, particularly in repudiation threats. To achieve this, the following security requirements are established: (1) Ensuring all system activities are sufficiently captured to allow for complete traceability of actions, and (2) Facilitating the timely detection of anomalies and suspicious behaviors within system logs. 
\begin{itemize}
    \item \textit{SC-10. Audit Trails and Monitoring.} Maintain detailed logs of all system activities and regularly review these logs to detect and respond to suspicious behavior. Implement automated monitoring tools that can alert administrators to potentially malicious activities, ensuring accountability and facilitating rapid response to incidents.
\end{itemize}

\subsection{Security Policy}
The implementation of ML-specific security policies is crucial for protecting the VLPD system against the unique threats identified in our STRIDE analysis. This subsection corresponds to the activity labeled as RP10 in Figure~\ref{fig:processdiagram}. Unlike traditional IT systems, ML systems face distinctive challenges including model manipulation, data poisoning, adversarial examples, and model extraction attacks. Our security policy framework addresses these ML-specific vulnerabilities through targeted policies that govern the entire ML lifecycle.

The foundation of our ML security policy begins with \textit{\textbf{Model Integrity and Provenance Requirements}}. Every model deployed in the VLPD system must maintain a cryptographically signed record of its training lineage, including dataset versions, hyperparameters, and training infrastructure specifications~\cite{schelter2018challenges}. This policy mandates that any model update or retraining must be traceable through an immutable audit log, with each version requiring approval from both the Data Scientist (R3) and SecMLOps Engineer (R8). Furthermore, all pretrained models from external sources must undergo security validation through our defined verification pipeline before integration, addressing the spoofing threats identified for EE-2 in our STRIDE analysis.

To combat data poisoning threats that scored high in our threat matrix, we establish strict \textit{\textbf{Data Validation and Sanitization Protocols}}. All training data must pass through a multi-stage validation process that includes: statistical anomaly detection to identify distributional shifts that might indicate poisoning attempts, cross-validation against known clean datasets to detect systematic manipulations, and automated screening for adversarial patterns using the detection algorithms specified in our security controls~\cite{paudice2018detection}. The policy requires maintaining a ``golden dataset" of verified clean samples, representing at least 10\% of training data~\cite{steinhardt2017certified}, which serves as a reference for detecting poisoning attempts. Any deviation exceeding defined thresholds triggers automatic quarantine and manual review by the Data Engineer (R4) and SecMLOps Engineer (R8).

Our policy framework mandates \textit{\textbf{Mandatory Adversarial Robustness Testing}} as part of the model development lifecycle. Before any model advances to production, it must demonstrate resilience against a set of adversarial attacks~\cite{carlini2017towards}, including those identified in our evaluation (FGSM, DeepFool, and combined attack scenarios). The policy specifies minimum robustness thresholds: models must maintain at least 80\% of baseline performance under adversarial conditions with $\epsilon \leq 0.03$ for FGSM attacks and $\xi \leq 0.03$ for DeepFool attacks. These thresholds directly correspond to our empirical findings in Section~\ref{Sec.V_Evaluation}, where we demonstrated that properly defended models can maintain acceptable performance under these attack strengths.

The policy establishes \textit{\textbf{Continuous Monitoring and Drift Detection Requirements}} specifically tailored to ML systems. Beyond traditional performance metrics, the system must monitor for concept drift, data distribution shifts, and anomalous prediction patterns that might indicate ongoing attacks~\cite{korycki2023adversarial}. The monitoring system must maintain separate baselines for different operational contexts, such as urban and highway, and trigger alerts when the model's behavior deviates significantly from established patterns. This includes implementing the security monitoring activities (DP3.2) with specific ML-focused metrics: prediction confidence distributions, feature importance stability, and adversarial detection rates~\cite{kreuzberger2023machine}.

For \textit{\textbf{Model Access and Deployment Control}}, the policy defines strict boundaries on model exposure and interaction. Production models must be deployed with rate limiting to prevent extraction attacks, with API access restricted to authenticated services only. The policy prohibits direct model weight access in deployment environments and mandates the use of trusted execution environments for inference operations on sensitive data~\cite{tramer2016stealing}. Model updates follow a staged deployment process with mandatory performance validation at each stage.

The policy also addresses \textit{\textbf{Incident Response Procedures Specific to ML Threats}}. When potential model compromise is detected, the policy mandates immediate model rollback capabilities to the last verified secure version. The incident response team, led by the SecMLOps Engineer (R8), must maintain pre-computed "safe mode" models that prioritize security over performance for emergency deployment~\cite{shankar2017no,kumar2020adversarial}. The policy requires post-incident analysis to include not just system logs but also model behavior analysis, data provenance verification, and potential retraining with cleaned datasets.

Finally, our policy framework includes \textit{\textbf{Compliance and Audit Requirements}} that acknowledge the unique challenges of ML systems. Regular security audits must include adversarial robustness testing, data pipeline integrity verification, and model behavior analysis across different demographic groups to ensure both security and fairness~\cite{zhang2024enhancing}. The policy requires maintaining comprehensive documentation of all security-relevant decisions, including trade-off analyses between model performance and security measures, with quarterly reviews to adjust security parameters based on evolving threat landscapes.

These ML-specific policies, derived directly from our STRIDE analysis and empirical evaluation, provide actionable governance that addresses the unique security challenges of ML systems while maintaining operational efficiency. Regular review and updates of these policies ensure continued effectiveness against emerging ML-specific threats.

\subsection{Monitor Security and Incident Response}
Effective monitoring and incident response are crucial components of maintaining the security posture of VLPD systems. This subsection corresponds to the activities labeled as \activity{DP3.2} and \activity{DP3.3} in Figure~\ref{fig:processdiagram}. Building upon the redundancy and failover mechanisms discussed in SC-6, this section focuses on proactive monitoring and reactive incident handling processes.
For VLPD systems, monitoring must be comprehensive and adaptive, involving:
\begin{itemize}
\item \textit{Performance Metrics Monitoring:} Continuously track processing speeds and detection accuracy to identify deviations that may indicate system malfunctions or attempts to compromise system integrity.
\item \textit{Traffic Surveillance:} Employ network traffic analysis tools to detect unusual patterns or spikes that could signify intrusion attempts or unauthorized data transfers.
\item \textit{Activity Logs:} Implement extensive logging of all user activities, system changes, and data transactions. These logs are vital for forensic analysis and understanding the context of security incidents.
\item \textit{Automated Alert Systems:} Set up real-time alerts for abnormal activities based on predefined criteria or thresholds. This includes alerts for potential security breaches or system performance issues.
\end{itemize}
The incident response plan should outline:
\begin{itemize}
\item \textit{Defined Roles and Immediate Actions:} Establish clear responsibilities within the incident response team to ensure swift actions. This includes immediate containment and mitigation strategies to limit the impact of the breach.
\item \textit{Incident Analysis and Classification:} Develop protocols for the rapid classification and escalation of incidents based on their severity and potential impact on operations.
\item \textit{Evidence Preservation and Analysis:} Outline procedures for securely collecting and analyzing evidence post-incident to determine the breach's cause and to formulate prevention strategies for future threats.
\item \textit{Stakeholder Communication:} Prepare communication plans for timely reporting to relevant stakeholders and, if necessary, to regulatory bodies or the public to manage transparency and regulatory compliance.
\end{itemize}
Regularly conduct simulation exercises and review incidents to refine the monitoring and response strategies. These activities are crucial for adapting to new threats and improving the resilience of VLPD systems against future attacks.

\section{Evaluation}
\label{Sec.V_Evaluation}
This section shows the implementation of SecMLOps within the context of the most prevalent security threats, specifically focusing on DP and AE. The analysis is designed to assess the robustness of the SecMLOps framework under these common attack scenarios, highlighting the effectiveness of the implemented security measures and discussing potential trade-offs.

\subsection{Experimental Setup}
The structured approach used to assess the effectiveness of SecMLOps in pedestrian detection is first described, detailing the use of the CityPersons dataset, adversarial attacks, defense mechanisms, implementation settings, and evaluation metrics.
\subsubsection{Dataset} 
CityPersons~\cite{Shanshan2017CVPR}  is a widely recognized and challenging dataset for pedestrian detection, developed from the CityScapes dataset for autonomous driving. The dataset includes 2,975 images for training and 500 images for validation, captured across 27 cities in Germany. Each pedestrian is annotated with bounding boxes and occlusion levels, and dense groups or misleading regions, such as reflections, are marked as ignore regions. Its comprehensive annotations and diverse settings contribute to its popularity and extensive use in research. 

\subsubsection{Adversarial Attacks}
In this experiment, a DP attack at the training stage, and two state-of-the-art attacks for generating AE in the deployment stage are used and briefly introduced as follows:
\begin{itemize}
    \item DP~\cite{10461694} attacks aim to mislead the model into detecting non-pedestrian objects, such as trees or streetlights, as pedestrians. Given a pedestrian label $y$=[$ y_{\text{center}}$, $y_{\text{height}}$, $y_{\text{offset}}$], where $y_{\text{center}}$ represents the center of the pedestrian, it is sufficient to modify only $y_{\text{center}}$. Dense-pixel features [$f_{\text{center}}, f_{\text{height}}, f_{\text{offset}}$] are identified using edge detection and clustering techniques as the adversarially identified center. In this attack strategy, each pedestrian label's center may be randomly flipped to an adversarial center $f_{\text{center}}$ based on probability $\gamma$, which determines the percentage of the dataset likely affected by this flipping.:
    \begin{equation}
        y_{\text{center}}^{\text{adv}} = \begin{cases} 
        f_{\text{center}} & \text{with } \gamma \\
        y_{\text{center}} & \text{otherwise}
        \end{cases} 
    \end{equation}
    
    \item Fast Gradient Signed Method (FGSM)~\cite{goodfellow2014explaining} is a single-step attack that generates adversarial samples. In our context, input image is perturbed in a way that maximizes the classification loss $\mathcal{L}_{cls}(s, y ; w)$ for the pedestrian center. Mathematically, it perturbs the pedestrian center coordinates  using the sign of the gradient $y_{\text{center}}$, where $\epsilon$ is the adversarial perturbation budget and $w$ denotes the model parameters.
    \begin{equation}
        s^{a d v}=s^{\text {clean }}+\epsilon \cdot \operatorname{sign}\left(\nabla \mathcal{L}_{cls}\left(s^{\text {clean }}, y ; w\right)\right) 
    \end{equation}
    
    \item DeepFool (DF)~\cite{Moosavi-Dezfooli_2016_CVPR} is an untargeted attack that iteratively computes minimal adversarial perturbations under an $\mathcal{L}_{2}$ constraint. By approximating the decision boundary as hyperplanar, DF efficiently misclassifies images with subtle changes that traverse the nearest decision boundary, making these perturbations less perceptible to human observers. The performance is influenced by three key factors: the number of iterations, the perturbation magnitude, and the overshoot parameter. In our context, this attacks effectively exploits overshoot parameter \(\xi\), guided by $\mathbf{x}_{adv} = \mathbf{x} + (1 + \xi) \cdot \mathbf{r}$ where \(\mathbf{x}\) is the original input and \(\mathbf{r}\) is the perturbation vector calculated by DF, to subtly to misalign the detected pedestrian center \( y_{\text{center}} \), thereby testing the robustness of pedestrian localization.
\end{itemize}

\subsubsection{Defense Strategies}
In our SecMLOps framework for secure pedestrian detection, we employ a multi-layered defense strategy that combines various techniques to protect the VLPD system against attacks. 
Based on the security analysis in Section~\ref{Subsec.IV_analysis} that identified high-risk threats such as data poisoning attacks targeting the training phase (DF-1, EE-1) and adversarial examples affecting the inference process (P-3, P-4), we selected defense mechanisms that directly implement the security requirements and controls. These defenses focus on protecting data integrity (SC-3, SC-4) and ensuring model robustness against adversarial manipulations, aligning with the "\textit{Integrity of Data Flow and Processing}" security goal. Our defense approach combines preprocessing techniques, in-training protections, and post-training refinements to establish comprehensive protection throughout the MLOps lifecycle.

\begin{itemize}
    \item CutMix~\cite{yun2019cutmix} is a data augmentation technique applied during the preprocessing stage of the SecMLOps workflow, implementing security control SC-4 (Input Validation and Sanitization) by diversifying the training data. In the context of the VLPD system, CutMix helps reduce the impact of potentially poisoned samples by cutting and pasting patches from different images and their corresponding labels to create new training samples. The probability $P$ of applying CutMix to a given batch of images is controlled by a hyperparameter, which determines the frequency of augmentation. This approach addresses the integrity requirement to maintain data accuracy across processing stages by reducing the model's susceptibility to data poisoning attacks identified as high-risk for external entities (EE-1) in our security analysis.
    
    \item Early Stopping~\cite{bai2021understanding} is a widely adopted in-training defense mechanism against DP, for achieving security control SC-4. It was designed to prevent overfitting and unnecessary training iterations. In the SecMLOps framework, early stopping is implemented by monitoring the model's performance on a validation set during training. If the validation performance does not improve for a specified number of epochs, known as the patience $p$, the training process is halted.  Thus, while early stopping improves general model performance, careful tuning of the patience parameter is required to avoid compromising the robustness of our secure PDS against adversarial threats.
    
    \item Adversarial Training~\cite{9798870} directly implements security control SC-3 (Adversarial Training), enhancing model robustness by including adversarially perturbed examples in the training set. This defense specifically addresses the vulnerability to adversarial examples identified for processes P-3 and P-4.
    By leveraging methods such as projected gradient descent (PGD)~\cite{9798870}, adversarial training generates examples that maximize classification loss within a constrained perturbation budget $\epsilon^{AT}$, as shown in Algorithm~\ref{alg:AdversarialTranining}. This approach fulfills the security requirement to enhance model robustness against adversarial manipulations in the operational environments, directly supporting the security goal of integrity.

\begin{algorithm}
\caption{Adversarial Training using PGD for Pedestrian Detection}
\label{alg:AdversarialTranining}
\begin{algorithmic}[1]
\State \textbf{Input:} Training data $\mathcal{D} = \{(x_i, y_i)\}_{i=1}^N$, loss function $\mathcal{L}$, number of epochs $E$, batch size $B$, learning rate $\eta$, perturbation limit $\epsilon^{AT}$, step size $\alpha$, number of PGD iterations $K$
\State \textbf{Output:} Robust model parameters $w$
\State Initialize model parameters $w$
\For{$epoch = 1$ to $E$}
    \For{each batch $\{(x_i, y_i)\}$ of $B$ from $\mathcal{D}$}
        \For{each $(x, y)$ in batch}
            \State $x^{adv}_0 \leftarrow x$  
            \For{$k = 1$ to $K$}
                \State $\nabla_x \leftarrow \nabla_x \mathcal{L}(x^{adv}_{k-1}, y; w)$ 
                \State $x^{adv}_k \leftarrow \text{clip}(x^{adv}_{k-1} + \alpha \cdot \text{sign}(\nabla_x), x-\epsilon, x+\epsilon)$ 
            \EndFor
            \State $x^{adv} \leftarrow x^{adv}_K$ 
        \EndFor
        \State $w \leftarrow w - \eta \nabla_w \frac{1}{B} \sum_{i=1}^B \mathcal{L}(x^{adv}_i, y_i; w)$ 
    \EndFor
\EndFor
\end{algorithmic}
\end{algorithm}
    
    \item Model Distillation~\cite{mullapudi2019online} serves as a post-training defense technique that implements aspects of security controls SC-3 transferring knowledge from a larger model to a smaller one, creating a more robust and clean representation of the data. In our SecMLOps framework for pedestrian detection, model distillation is applied after model training refines the pre-trained model, by tuning the smoothing factor $\alpha$ to determine how much influence the current model’s (student's) weights have on the teacher model’s weights during the knowledge transfer process. By distilling knowledge from the larger model to a smaller one, the VLPD system benefits from improved robustness against adversarial attempts while maintaining efficient deployment capabilities. This technique provides an additional layer of security, complementing the in-training defenses and enhancing the overall resilience of the VLPD system within the SecMLOps workflow.

\end{itemize}

\subsubsection{Implementation Settings}
Note that our experiments are developed to evaluate the effectiveness of a first-class performance PDS applying SecMLOps. Based on this motivation, we reproduce VLPD systems with the same settings~\cite{Liu_2023_CVPR} and apply defenses that conclude from Section~\ref{Sec.III_SecMLOps}. That is, a mini-batch contains 8 images with the learning rate of $2 \times 10^{-4}$ for training. The size of training images is 640x1280 and the size of tests remains as the original data without resizing.

We use CleverHans\footnote{\url{https://github.com/cleverhans-lab/cleverhans}} and Adversarial Robustness Toolbox (ART)\footnote{\url{https://github.com/Trusted-AI/adversarial-robustness-toolbox}} to build the attacks and defenses for different scenarios. Experiments are all on a single NVIDIA Tesla A100 GPU.

\subsubsection{Evaluation Metrics}
The most popular performance metric utilized for PDS is the log-average miss rate (laMR)~\cite{5975165}, which is computed by averaging the miss rate across nine false positive per image (FPPI) rates. These rates are evenly spaced in log-space within the range of $10^{-2}$ to $10^{0}$. 
The evaluation follows the reasonable setup as defined in~\cite{5975165}. This setup involves using only pedestrians with a height greater than 50 pixels and an occlusion level less than 0.35 for both training and evaluation. While the mean average precision (mAP) metric is commonly used in generic object detection, the laMR metric is preferred in specific applications like autonomous driving. This preference is due to the practical need to maintain an upper limit on the number of false positives per image.

The following definitions employ false positive $(F P)$, false negative $(F N)$, true positive $(T P)$ and false positive $(F P)$ counts at various evaluation confidence thresholds $c$. The miss rate $(MR)$ at a given confidence threshold $c$ is defined as:

\begin{equation}
\operatorname{MR}(c)=\frac{\mathrm{FN}(c)}{\mathrm{TP}(c)+\mathrm{FN}(c)},
\end{equation}

The FPPI at a given confidence threshold $c$ is defined as:

\begin{equation}
\operatorname{FPPI}(c)=\frac{\operatorname{FP}(c)}{N},
\end{equation}

where $N$ represents the number of images in the evaluation set. These measures are combined to calculate the log-average miss rate ($\operatorname{laMR}$), where $f$ values are equally spaced in the interval $f \in \mathcal{F}=\left[10^{-2}, 10^{0}\right]$:

\begin{equation}
\label{equ:lamr}
\operatorname{laMR}=\exp (\frac{1}{9} \sum_{f \in \mathcal{F}} \log (\underset{\operatorname{MRPI}(c) \leq f}{\operatorname{argmax}}(\operatorname{FPPI}(c))))).
\end{equation}

Various evaluation configurations have been suggested in prior research to assess the performance of PDS across different scenarios. This study employs the evaluation configurations introduced with~\cite{5975165}. These configurations categorize pedestrian detections into four overlapping subgroups according to the visibility ratio and pixel height of each detection: \textit{Reasonable}, \textit{Small}, \textit{Heavy occlusion}, and \textit{All}. Table~\ref{tab:evaluation_settings} provides a detailed description of the evaluation settings used in this study.

\begin{table}[ht!]
\caption{Evaluation settings for pedestrian datasets}
\label{tab:evaluation_settings}
\centering
\begin{tabular}{|ccc|}
\hline
\textbf{Setting} & \textbf{Visibility Ratio} & \textbf{Height Pixel} \\
\hline
Reasonable & [0.65, $\infty$] & [50, $\infty$] \\
\hline
Small & [0.65, $\infty$] & [50, 75] \\
\hline
Heavy & [0.25, 0.65] & [50, $\infty$] \\
\hline
All & [0.2, $\infty$] & [20, $\infty$] \\
\hline
\end{tabular}
\end{table}

\subsection{Results and Analysis}
\subsubsection{Performance under Attack Scenarios}
\begin{table}[ht!]
\centering
\caption{laMR performance comparison under various conditions with the attack parameters DP probability 10\%, FGSM budget $\epsilon$ = 0.03, DF overshoot $\xi$ = 0.03 and defense parameters including: CM probability $P=0.5$, ES patience $p=15$, AT perturbation budget $\epsilon^{AT}$ = 0.02, MD smoothing factor $\alpha = 0.5$}
\label{tab:performance_comparison}
\small 
\begin{tabularx}{0.7\columnwidth}{@{}lXcccc@{}}
\toprule
Attack Type & Strategy & Reasonable & Small & Heavy & All \\ \midrule
Baseline & No defense & 9.9 & 11.3 & 39.8 & 35.8 \\
         & SecMLOps       & 11.4 & 14.0 & 41.2 & 38.3 \\ \midrule
DP & No defense & 21.7 & 23.3 & 48.2 & 44.1 \\
          & SecMLOps       & 12.8 & 18.4 & 44.7 & 41.6 \\ \midrule
FGSM      & No defense & 35.6 & 42.8 & 61.4 & 56.0 \\
          & SecMLOps       & 14.9 & 20.5 & 45.2 & 42.3 \\ \midrule
DF  & No defense & 38.4 & 40.7 & 56.8 & 54.1 \\
          & SecMLOps       & 18.1 & 21.8 & 48.3 & 45.4 \\ \midrule
DP+FGSM & No defense & 46.2 & 38.9 & 58.7 & 57.6 \\
              & SecMLOps       & 17.6 & 21.2 & 47.1 & 44.3 \\ \midrule
DP+DF & No defense & 52.5 & 54.0 & 63.3 & 61.2 \\
                   & SecMLOps       & 18.6 & 22.3 & 49.2 & 46.8 \\ \midrule
FGSM+DF & No defense & 61.4 & 64.6 & 75.1 & 72.5 \\
              & SecMLOps       & 22.8 & 28.5 & 52.1 & 49.9 \\ \bottomrule
\end{tabularx}
\end{table}
The results in Table \ref{tab:performance_comparison} reveal crucial insights into our SecMLOps framework's effectiveness across various attack scenarios. In the baseline condition without attacks, we observe a slight performance degradation when our defense strategy is applied, with the laMR increasing from 9.9\% to 11.4\% in the \textit{Reasonable} setting. This marginal decrease is an expected consequence of implementing robust security measures, reflecting the well-known trade-off between security and performance in adversarial machine learning. However, the true value of our approach becomes evident under attack scenarios. Across all attack types - DP, FGSM, DF, and their combinations - our strategy consistently outperforms the undefended baseline. For instance, under the FGSM attack, our method achieves an laMR of 14.9\% in the \textit{Reasonable} setting, compared to 35.6\% for the undefended model. This improvement is even more pronounced in combined attack scenarios, such as DP+FGSM, where our method maintains an laMR of 17.6\% while the undefended model deteriorates to 46.2\%. The effectiveness of our framework can be attributed to its integration of multiple defense mechanisms and emphasis on continuous monitoring and adaptation. In conclusion, while our approach introduces a minor performance cost under normal conditions, it provides substantial benefits in terms of model robustness and reliability against adversarial attacks, a trade-off often necessary and justified in security-critical applications.

\begin{figure}[pt]
    \centering
    \subfigure[]{
        \begin{minipage}{3.1cm}
            \label{fig.dp.1}
            \includegraphics[width=3.1cm]{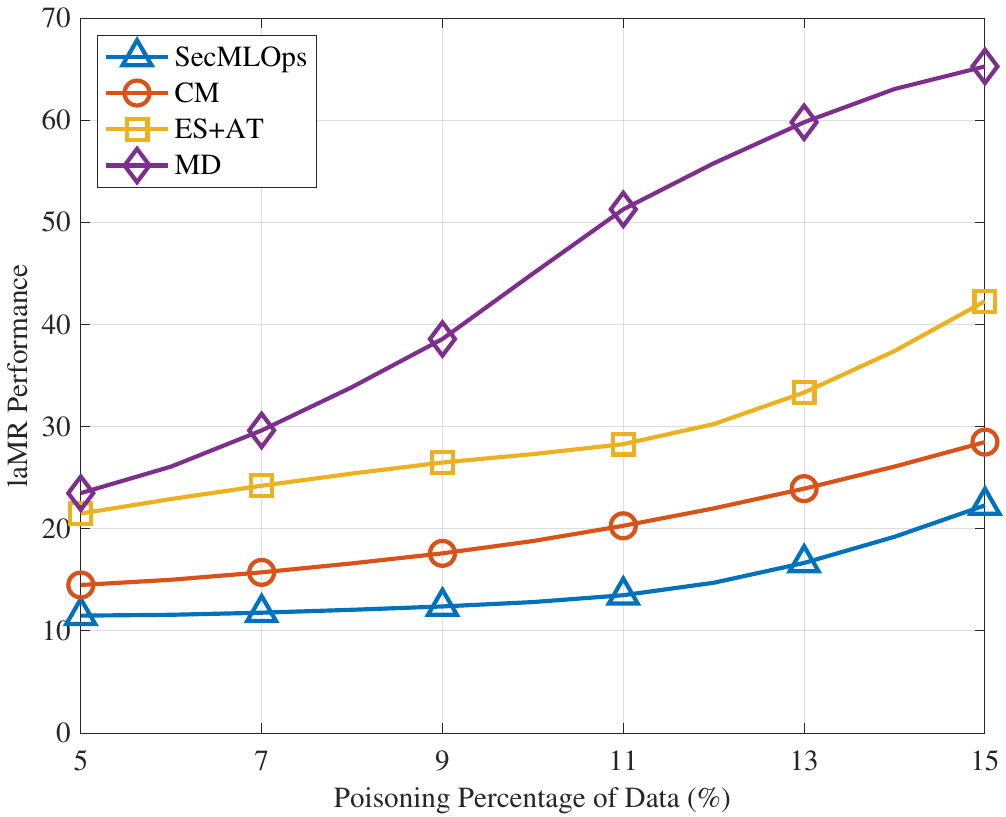}
        \end{minipage}}
    \subfigure[]{
        \begin{minipage}{3.1cm}
            \label{fig.dp.2}
            \includegraphics[width=3.1cm]{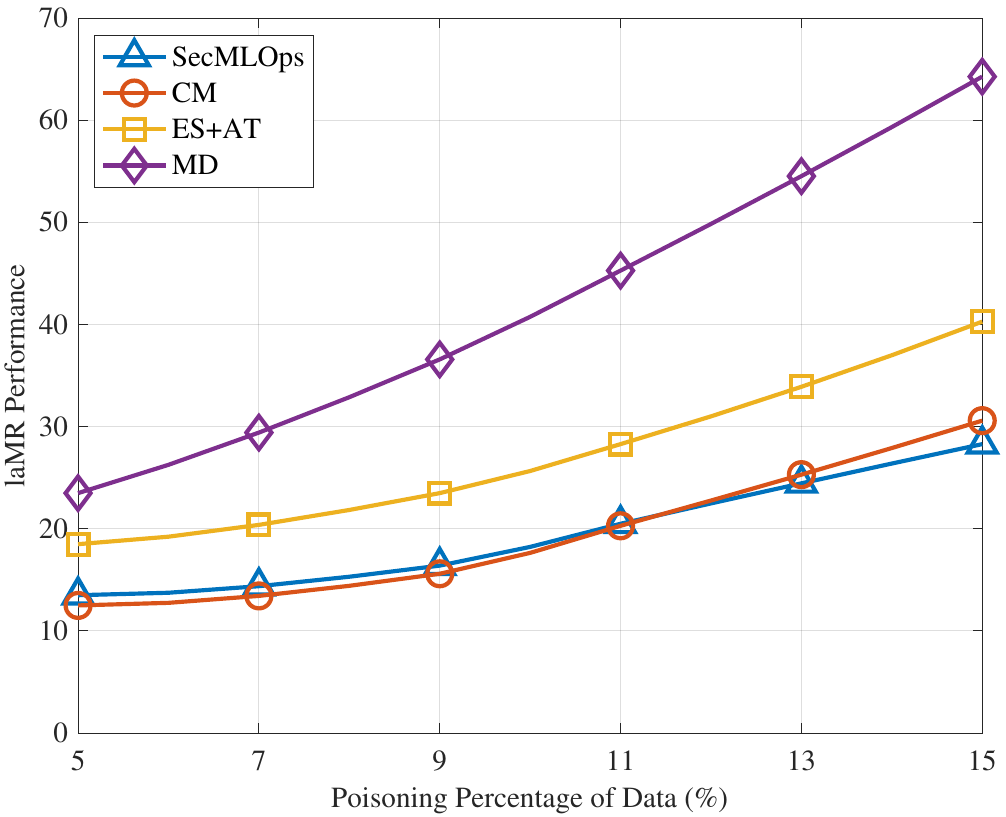}
        \end{minipage}}
    \subfigure[]{
        \begin{minipage}{3.1cm}
            \label{fig.dp.3}
            \includegraphics[width=3.1cm]{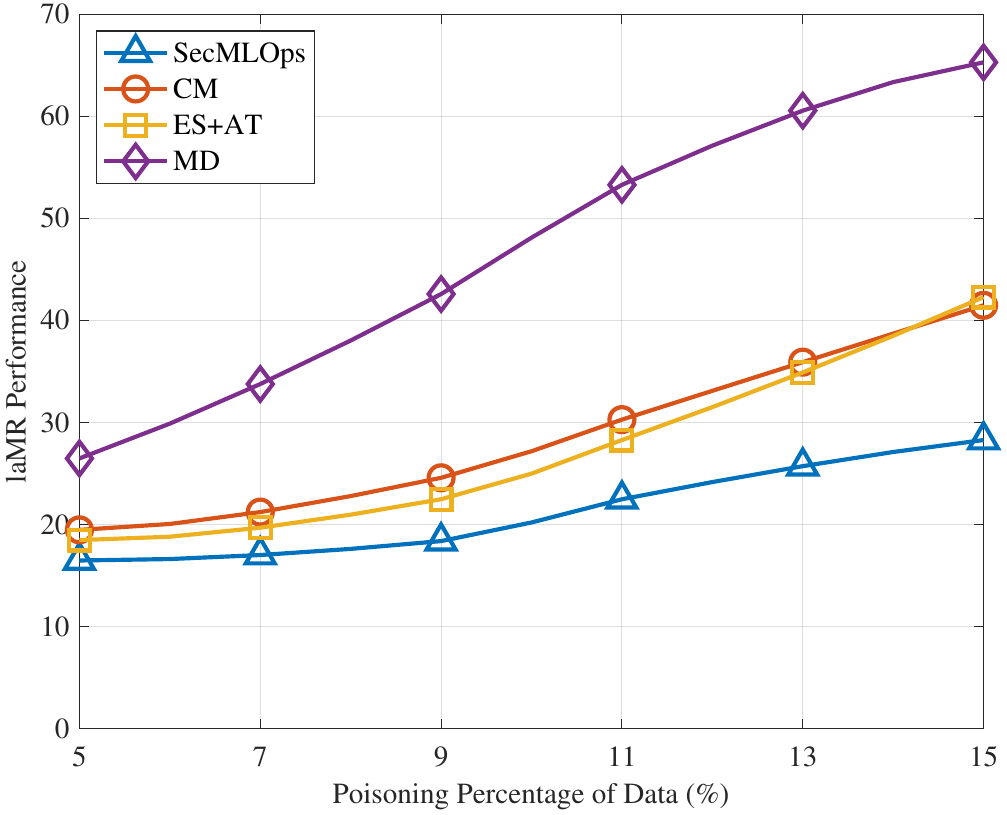}
        \end{minipage}}
    \subfigure[]{
        \begin{minipage}{3.1cm}
            \label{fig.dp.4}
            \includegraphics[width=3.1cm]{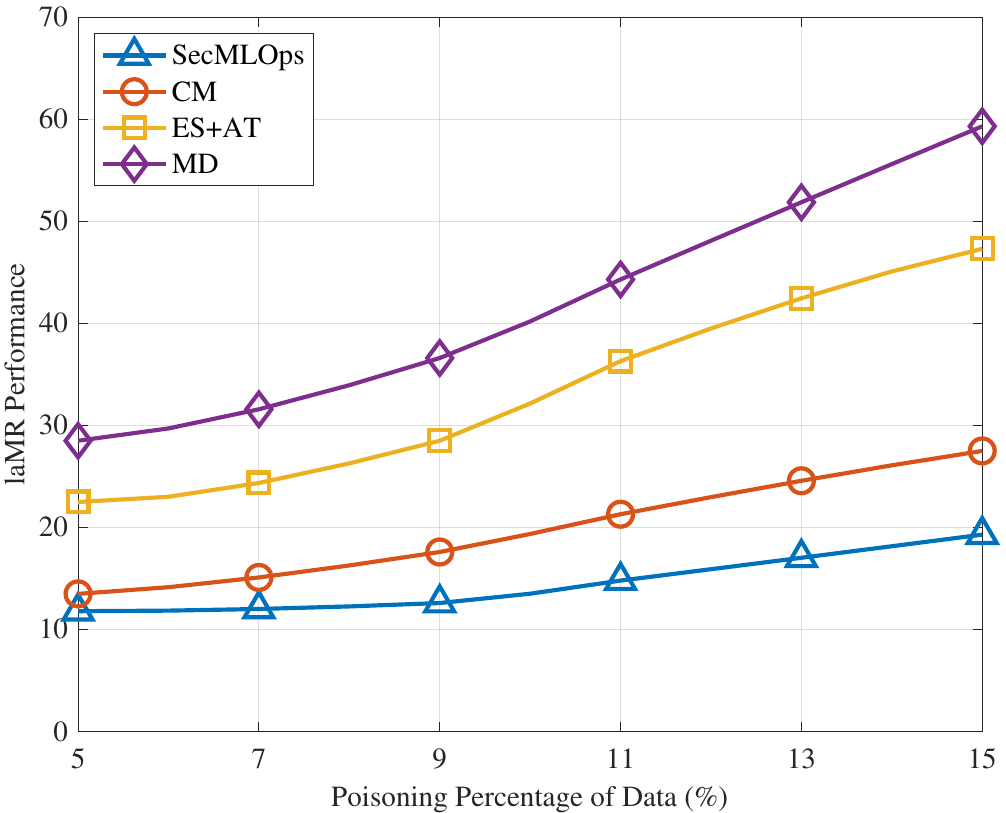}
        \end{minipage}}
    \subfigure[]{
        \begin{minipage}{3.1cm}
            \label{fig.FGSM.1}
            \includegraphics[width=3.1cm]{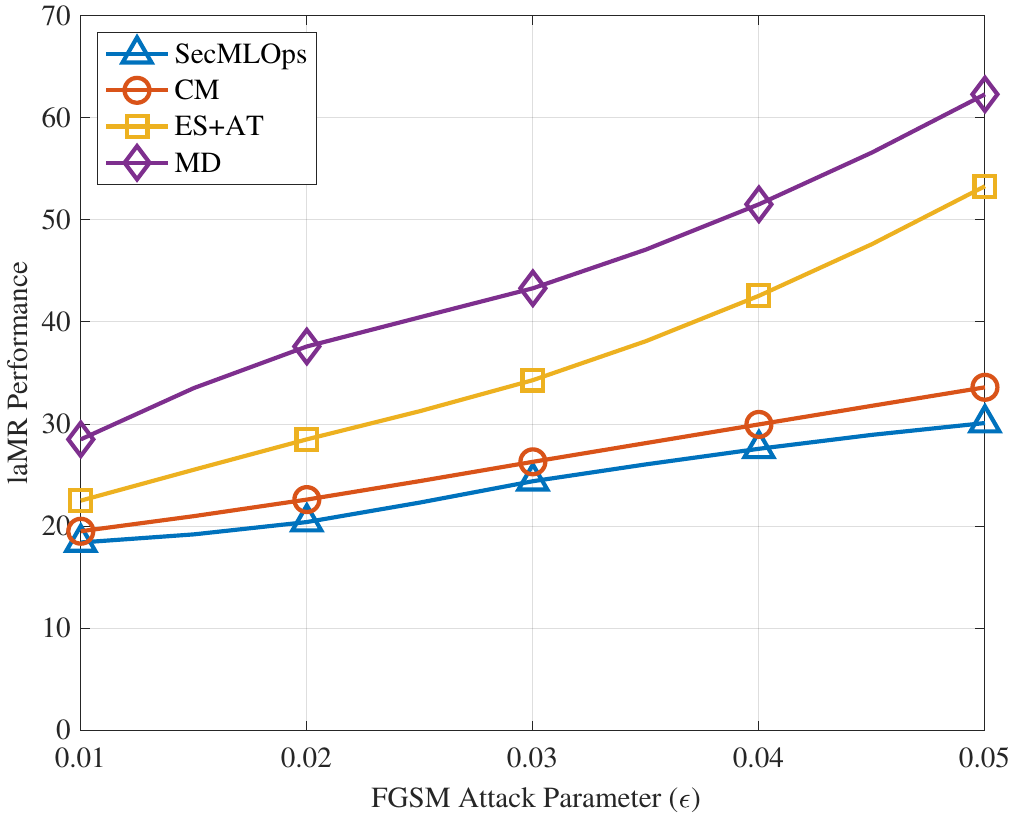}
        \end{minipage}}
    \subfigure[]{
        \begin{minipage}{3.1cm}
            \label{fig.FGSM.2}
            \includegraphics[width=3.1cm]{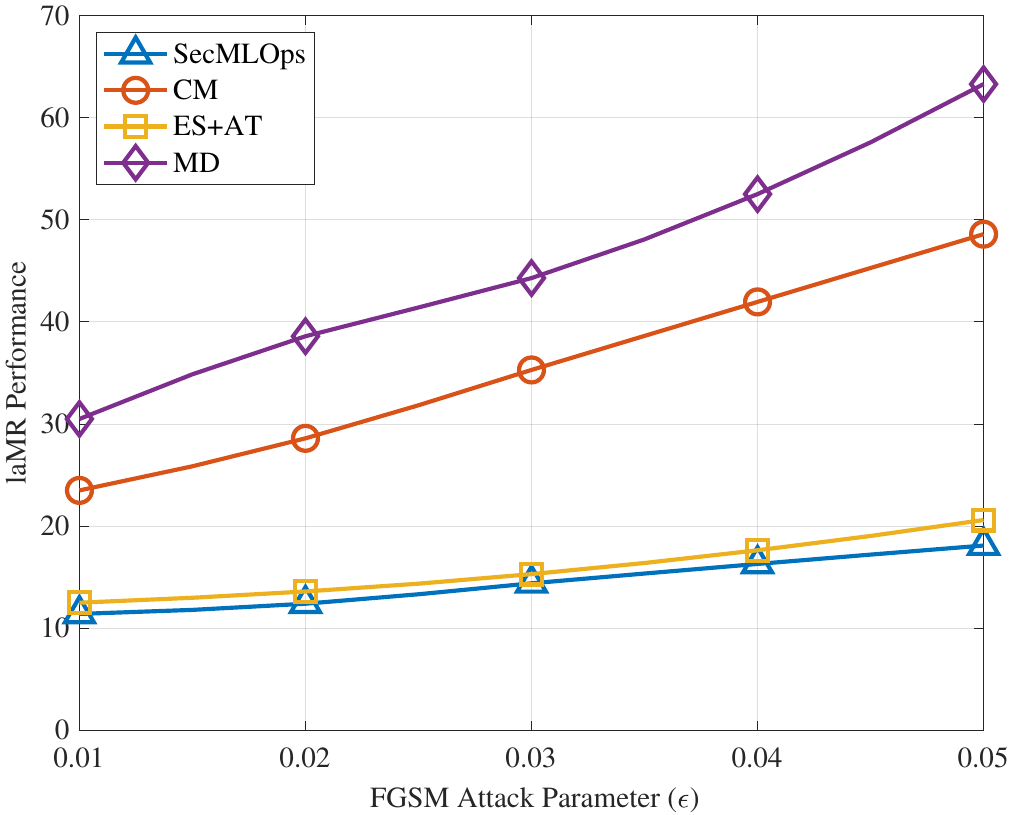}
        \end{minipage}}
    \subfigure[]{
        \begin{minipage}{3.1cm}
            \label{fig.FGSM.3}
            \includegraphics[width=3.1cm]{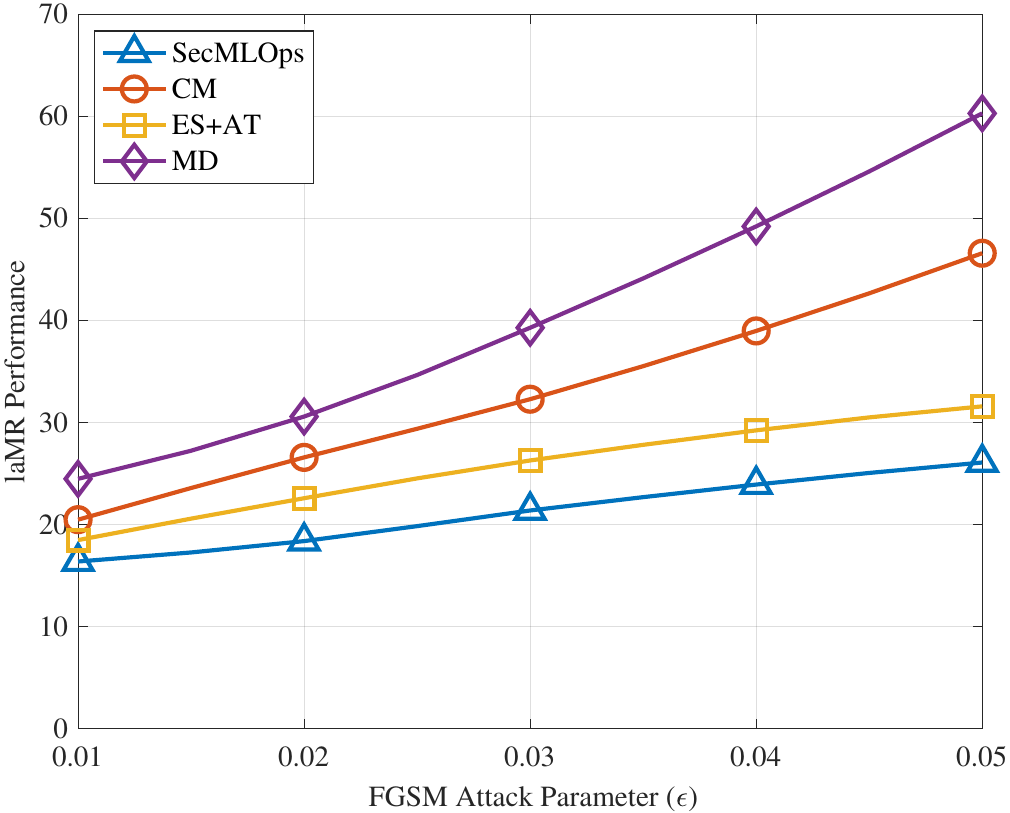}
        \end{minipage}}
    \subfigure[]{
        \begin{minipage}{3.1cm}
            \label{fig.FGSM1.4}
            \includegraphics[width=3.1cm]{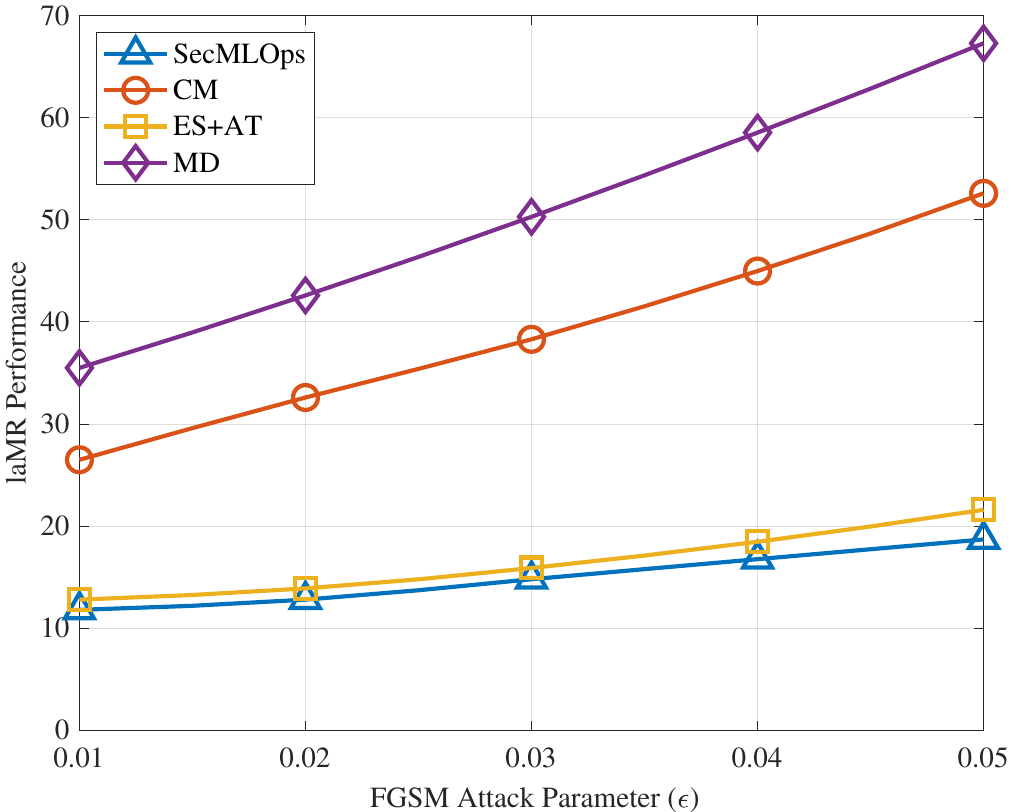}
        \end{minipage}}
    \subfigure[]{
        \begin{minipage}{3.1cm}
            \label{fig.DF.1}
            \includegraphics[width=3.1cm]{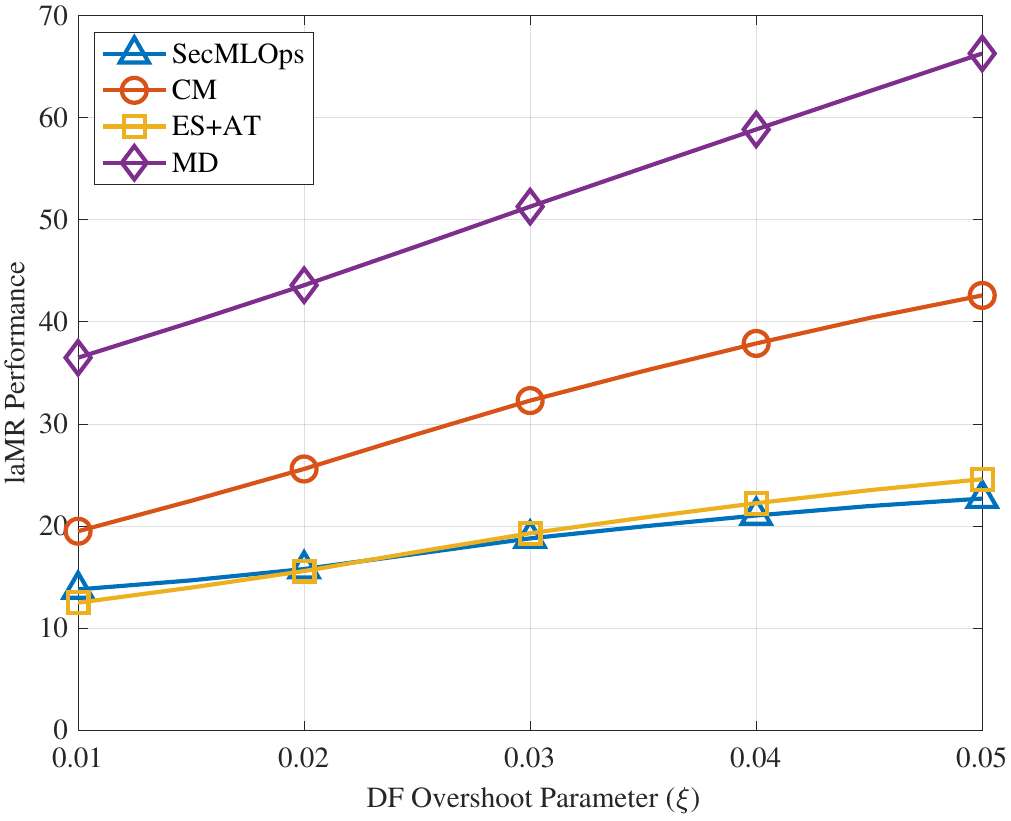}
        \end{minipage}}
    \subfigure[]{
        \begin{minipage}{3.1cm}
            \label{fig.DF.2}
            \includegraphics[width=3.1cm]{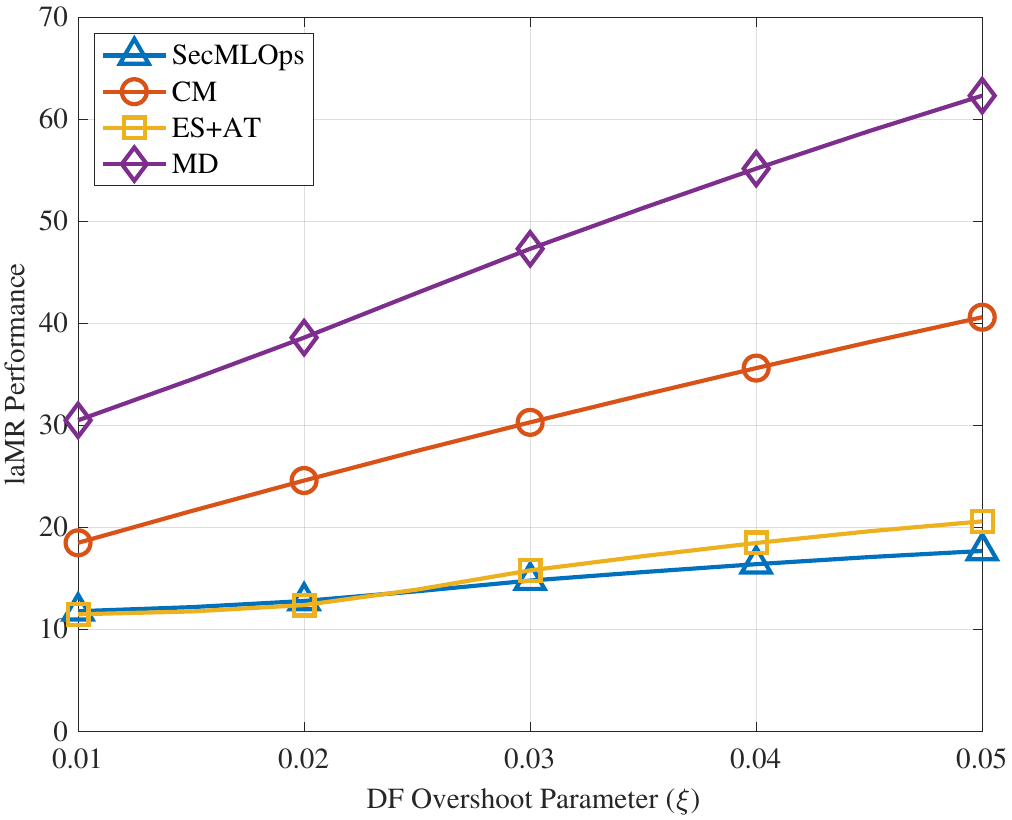}
        \end{minipage}}
    \subfigure[]{
        \begin{minipage}{3.1cm}
            \label{fig.DF.3}
            \includegraphics[width=3.1cm]{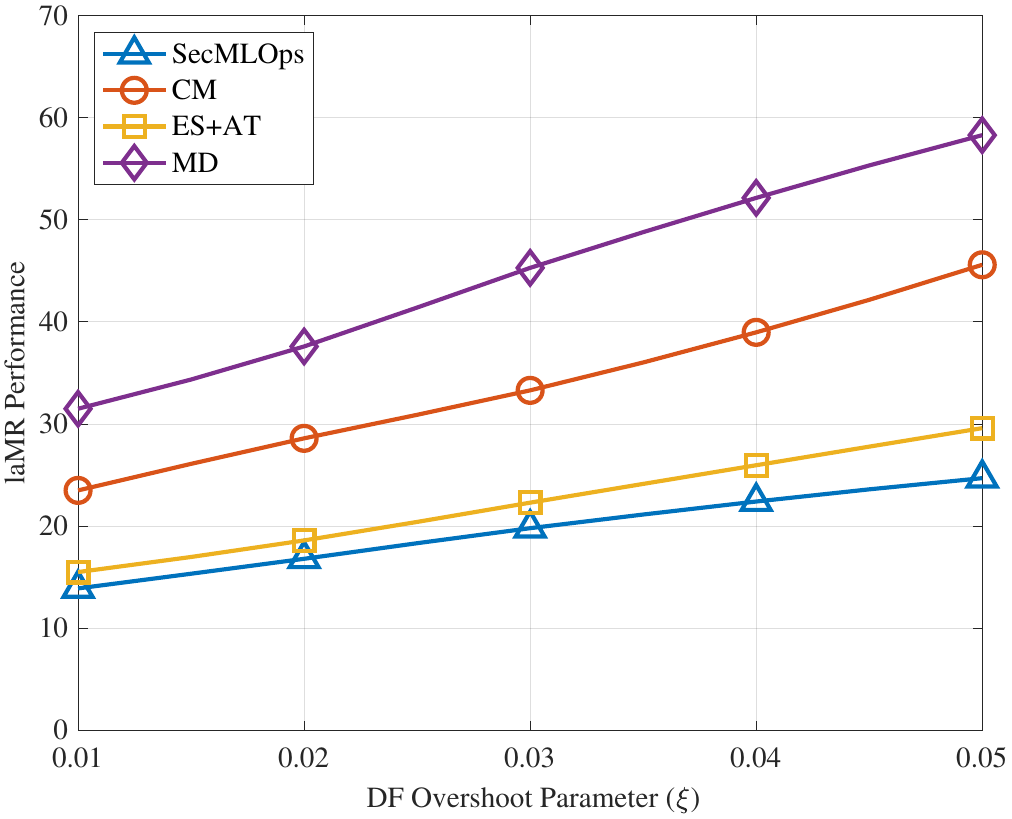}
        \end{minipage}}
    \subfigure[]{
        \begin{minipage}{3.1cm}
            \label{fig.DF.4}
            \includegraphics[width=3.1cm]{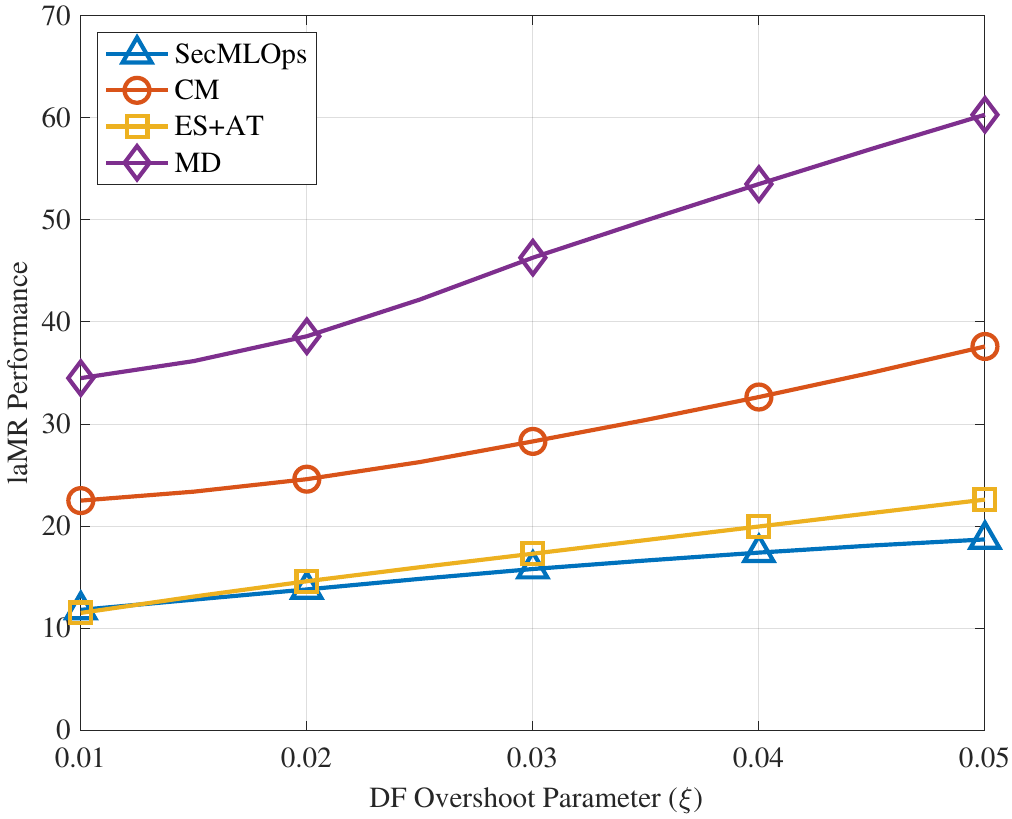}
        \end{minipage}}
    \caption{Performance comparison of SecMLOps against baseline methods (CM, ES+AT, MD) under various attack scenarios. (a-d) DP, (e-h) FGSM, (i-l) DF. Lower laMR indicates better performance. The defense parameters for the first column figures include: CM probability $P=0.5$, ES patience $p=10$, AT perturbation budget $\epsilon^{AT}$ = 0.01, MD smoothing factor $\alpha = 0.2$; The defense parameters for the second column figures include: $P=0.2$, $p=20$, $\epsilon^{AT}$ = 0.02, $\alpha = 0.2$; The defense parameters for the third column figures include: $P=0.2$, $p=10$, $\epsilon^{AT}$ = 0.01, $\alpha = 0.4$; The defense parameters for the fourth column figures include: $P=0.5$, $p=20$, $\epsilon^{AT}$ = 0.02, $\alpha = 0.2$.}
      \label{fig:individualattack}

\end{figure}

\begin{figure}[pt]
    \centering
    \subfigure[]{
        \begin{minipage}{3.1cm}
            \label{fig.DPandFGSM.1}
            \includegraphics[width=3.1cm]{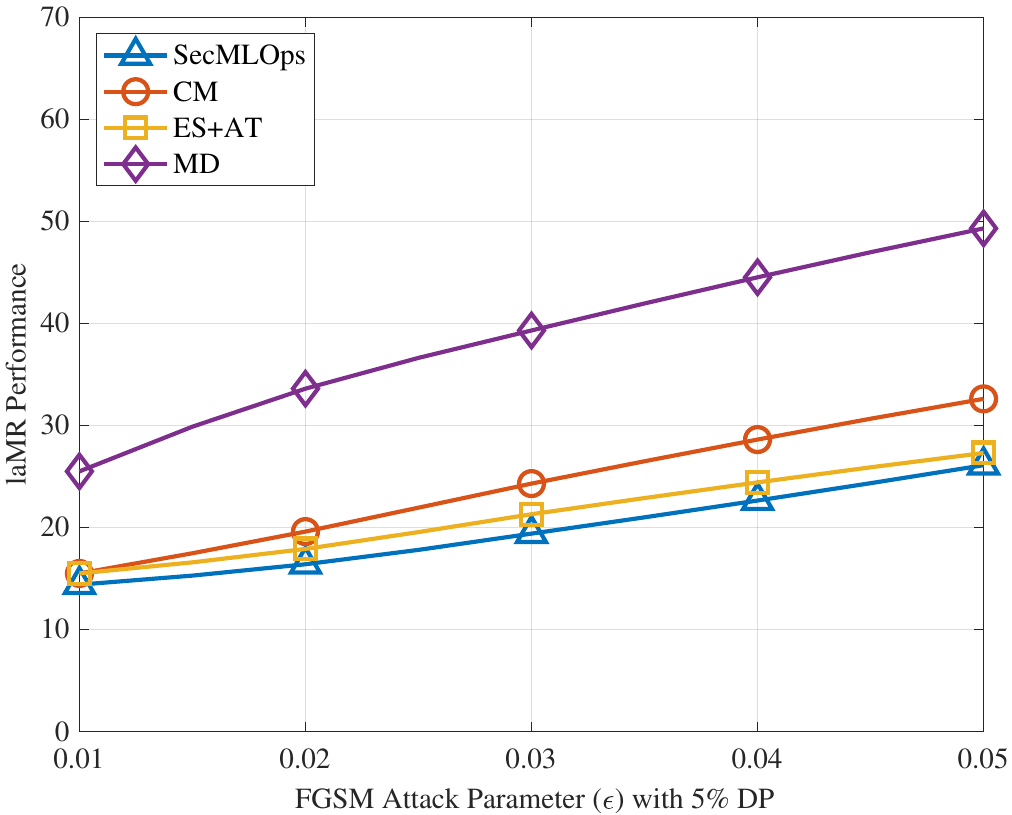}
        \end{minipage}}
    \subfigure[]{
        \begin{minipage}{3.1cm}
            \label{fig.DPandFGSM.2}
            \includegraphics[width=3.1cm]{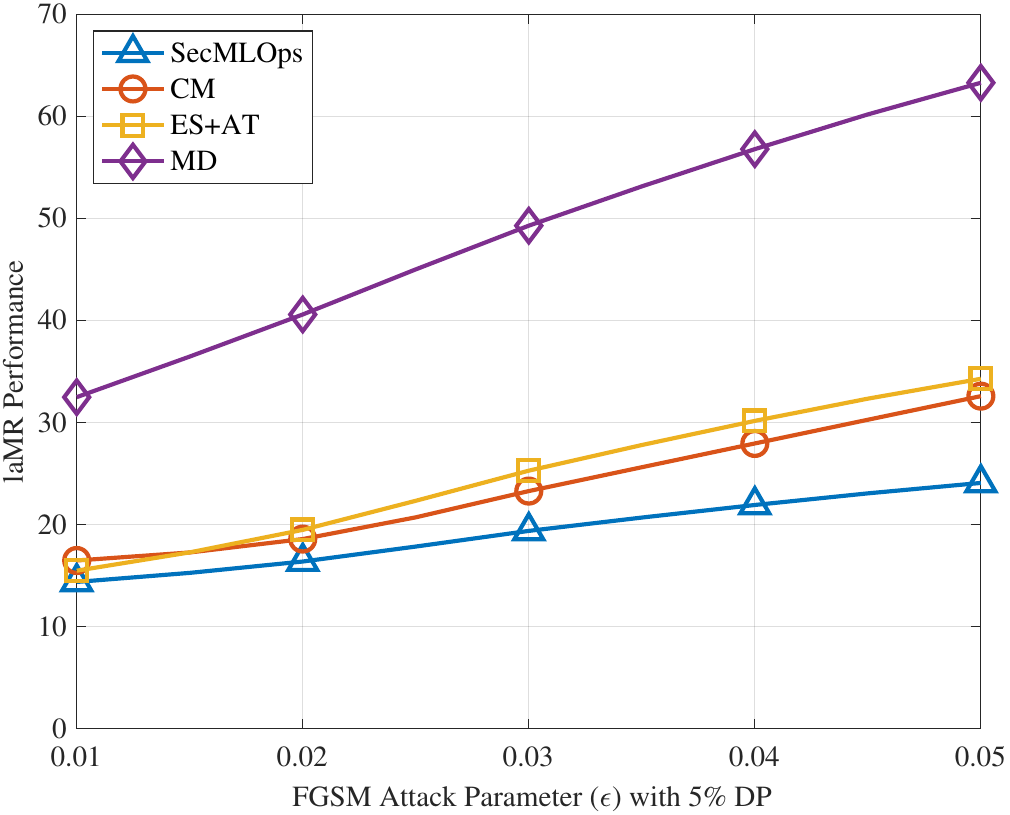}
        \end{minipage}}
    \subfigure[]{
        \begin{minipage}{3.1cm}
            \label{fig.DPandFGSM.3}
            \includegraphics[width=3.1cm]{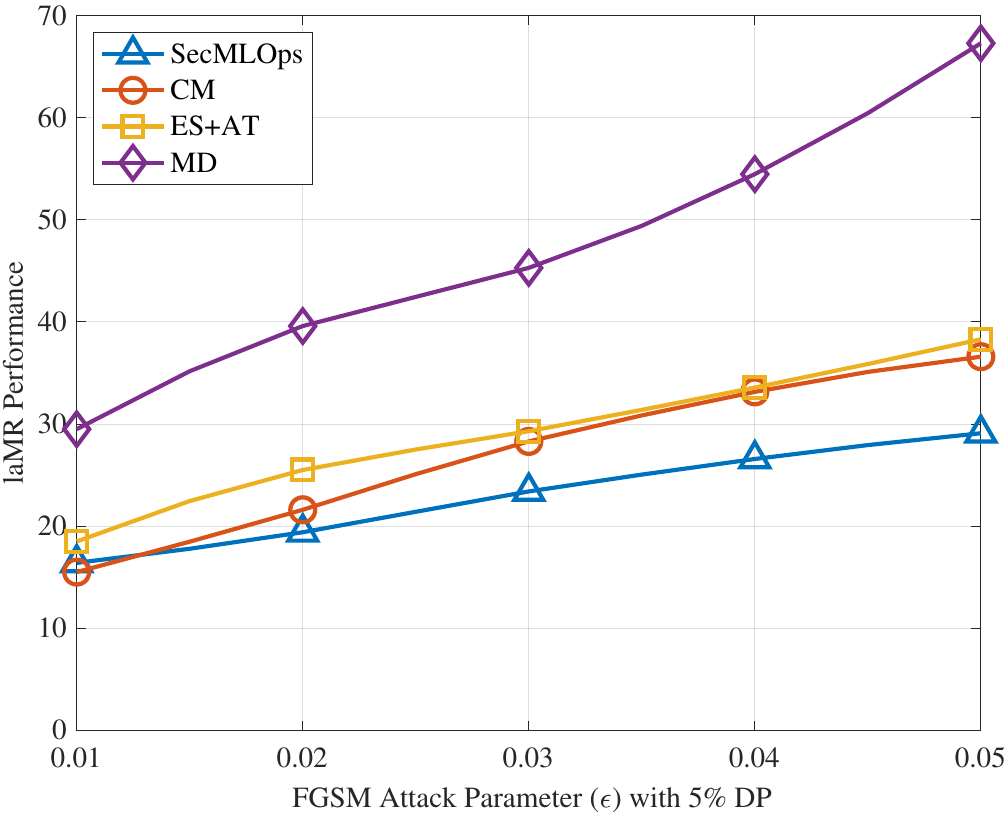}
        \end{minipage}}
    \subfigure[]{
        \begin{minipage}{3.1cm}
            \label{fig.DPandFGSM.4}
            \includegraphics[width=3.1cm]{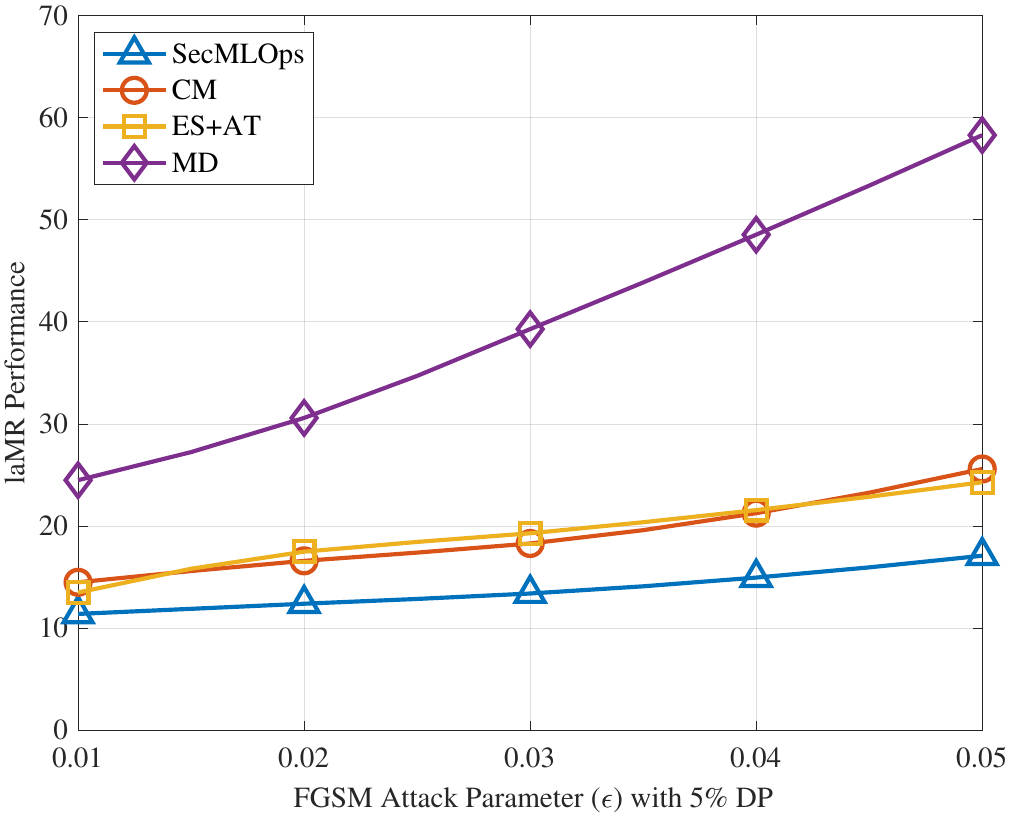}
        \end{minipage}}
    \subfigure[]{
        \begin{minipage}{3.1cm}
            \label{fig.DPandDF.1}
            \includegraphics[width=3.1cm]{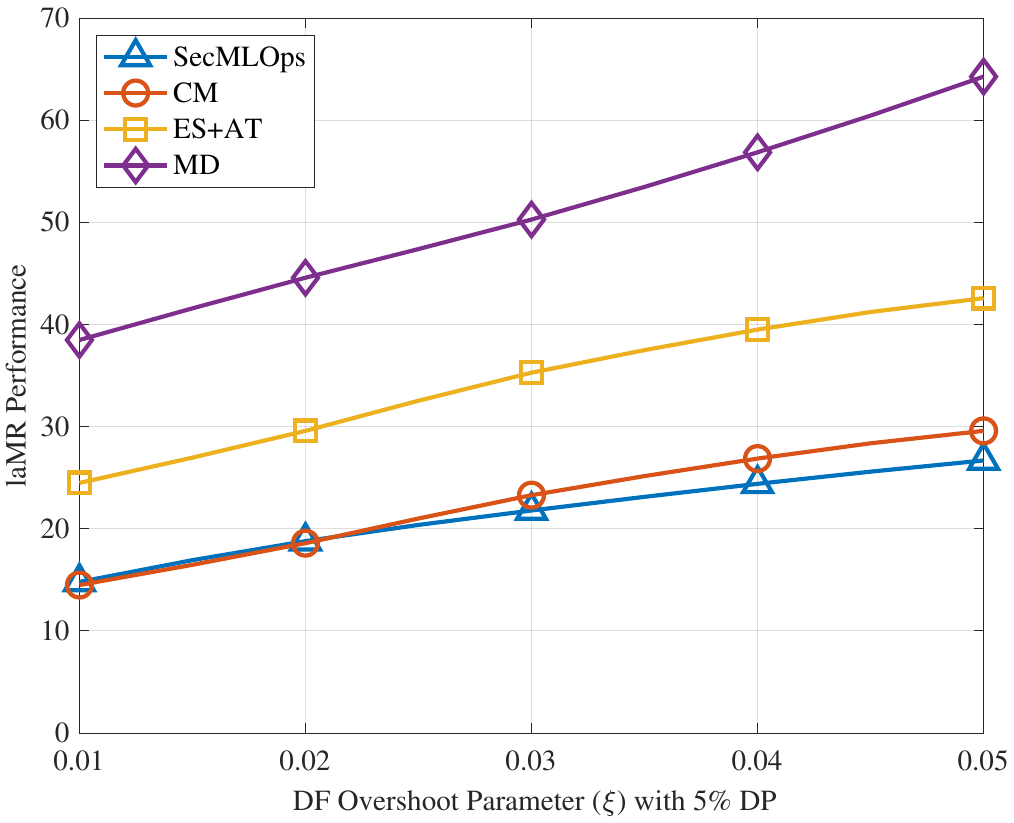}
        \end{minipage}}
    \subfigure[]{
        \begin{minipage}{3.1cm}
            \label{fig.DPandDF.2}
            \includegraphics[width=3.1cm]{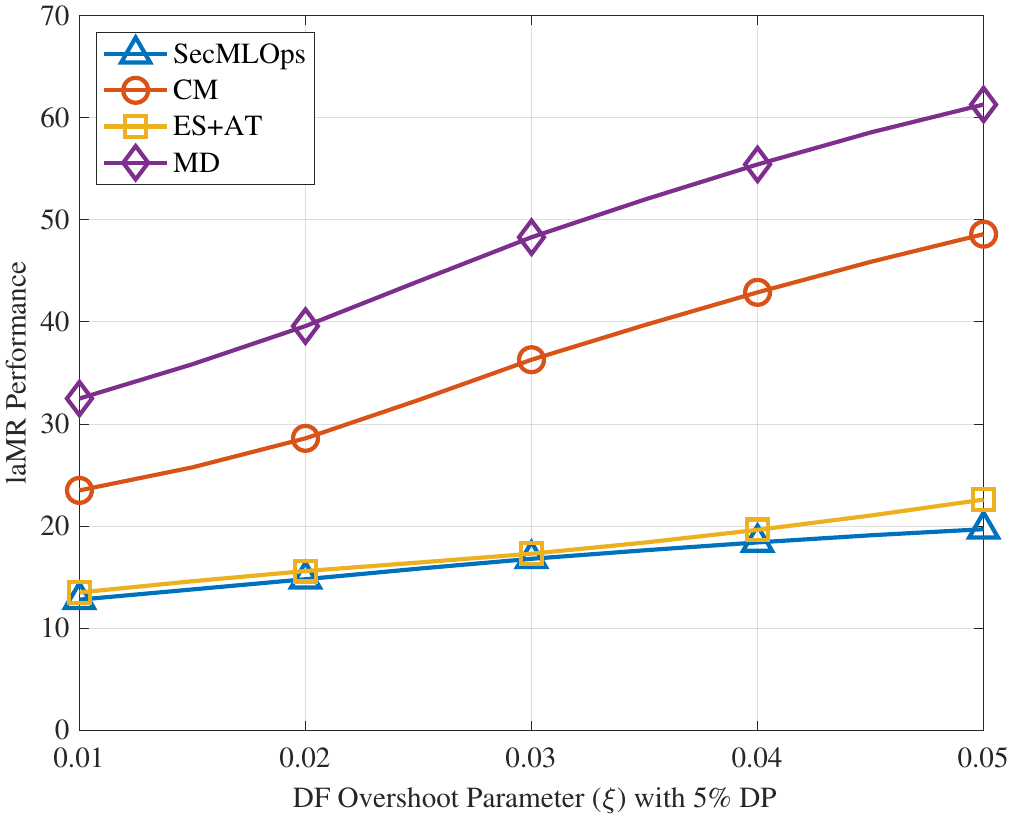}
        \end{minipage}}
    \subfigure[]{
        \begin{minipage}{3.1cm}
            \label{fig.DPandDF.3}
            \includegraphics[width=3.1cm]{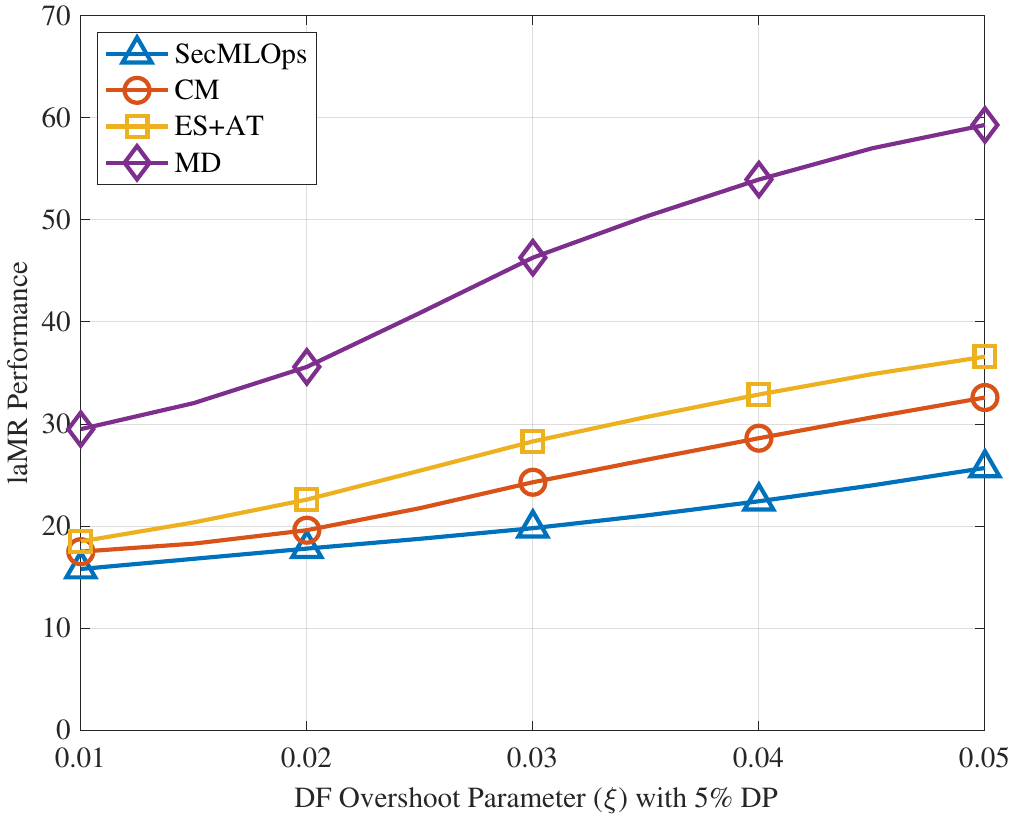}
        \end{minipage}}
    \subfigure[]{
        \begin{minipage}{3.1cm}
            \label{fig.DPandDF.4}
            \includegraphics[width=3.1cm]{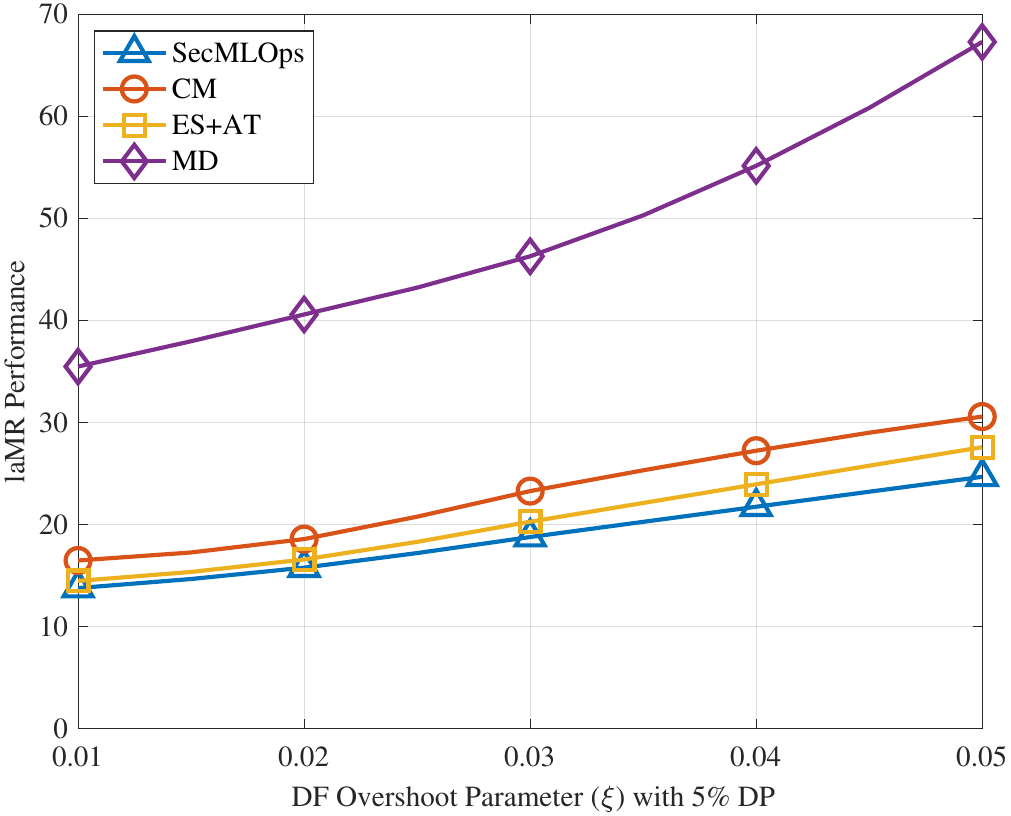}
        \end{minipage}}
    \subfigure[]{
        \begin{minipage}{3.1cm}
            \label{fig.FGSMandDF.1}
            \includegraphics[width=3.1cm]{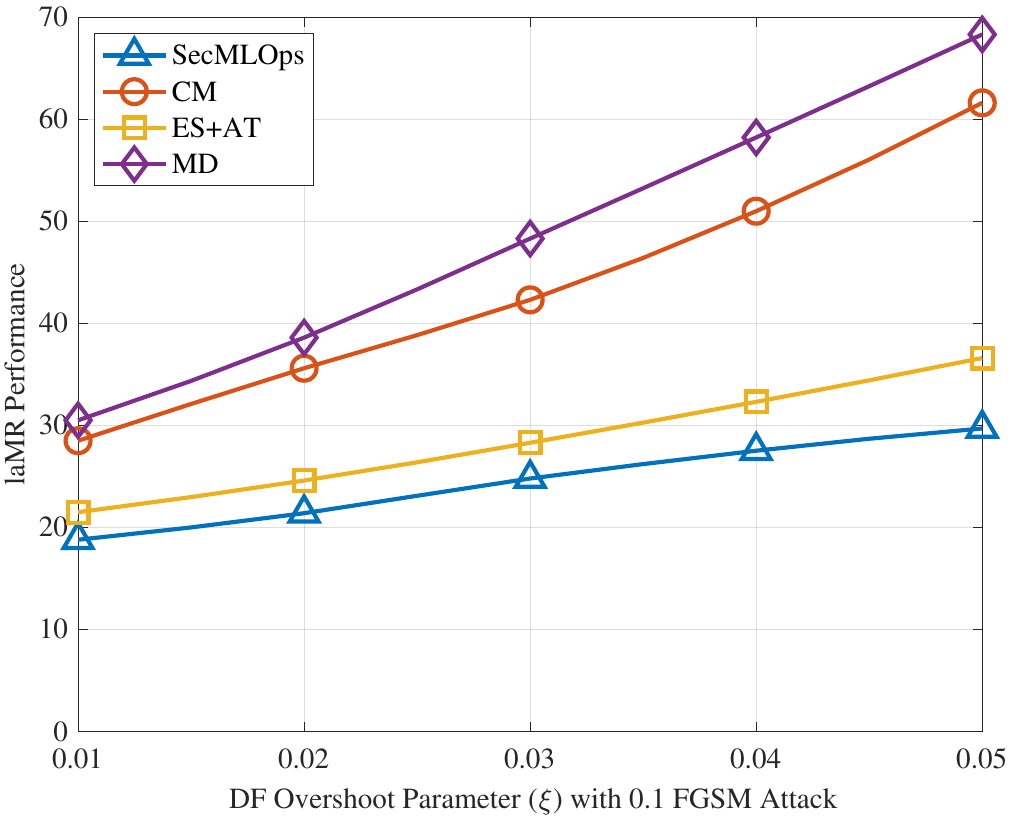}
        \end{minipage}}
    \subfigure[]{
        \begin{minipage}{3.1cm}
            \label{fig.FGSMandDF.2}
            \includegraphics[width=3.1cm]{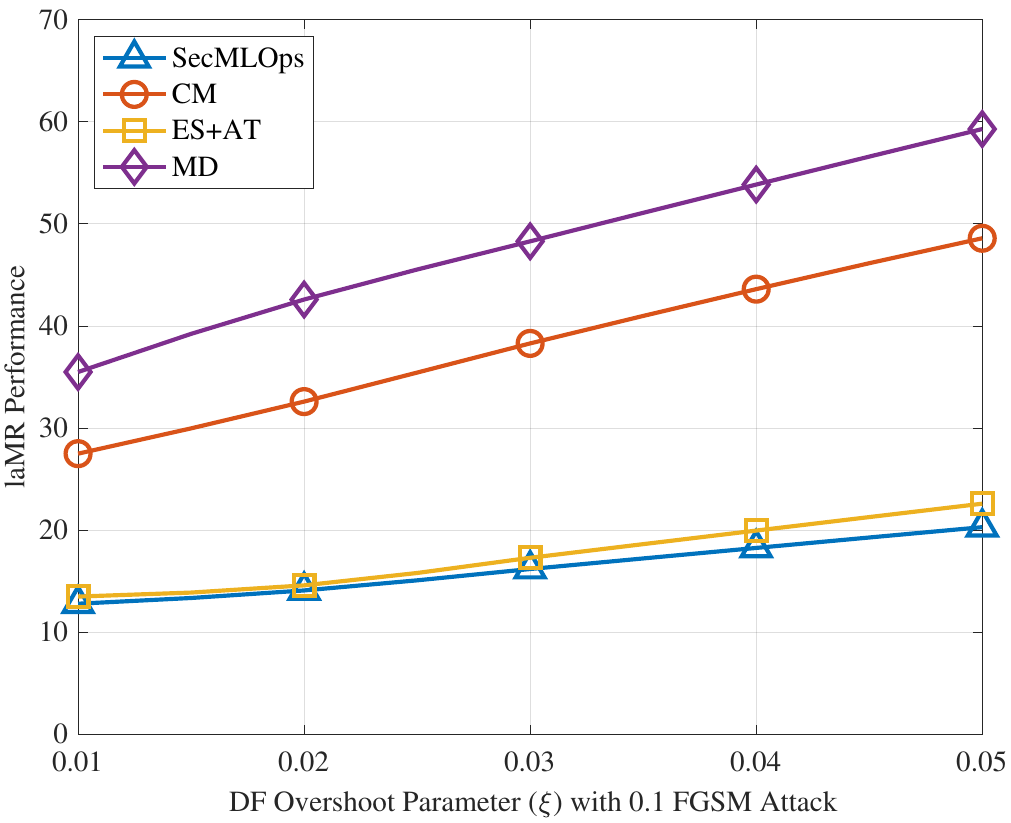}
        \end{minipage}}
    \subfigure[]{
        \begin{minipage}{3.1cm}
            \label{fig.FGSMandDF.3}
            \includegraphics[width=3.1cm]{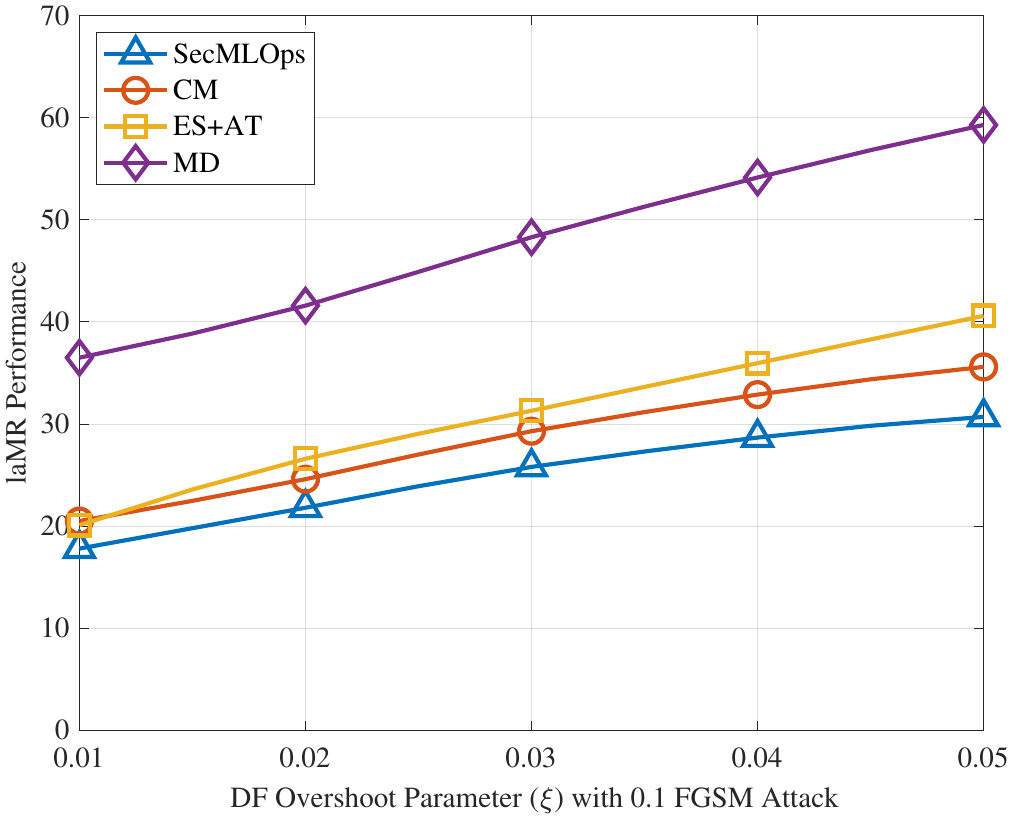}
        \end{minipage}}
    \subfigure[]{
        \begin{minipage}{3.1cm}
            \label{fig.FGSMandDF.4}
            \includegraphics[width=3.1cm]{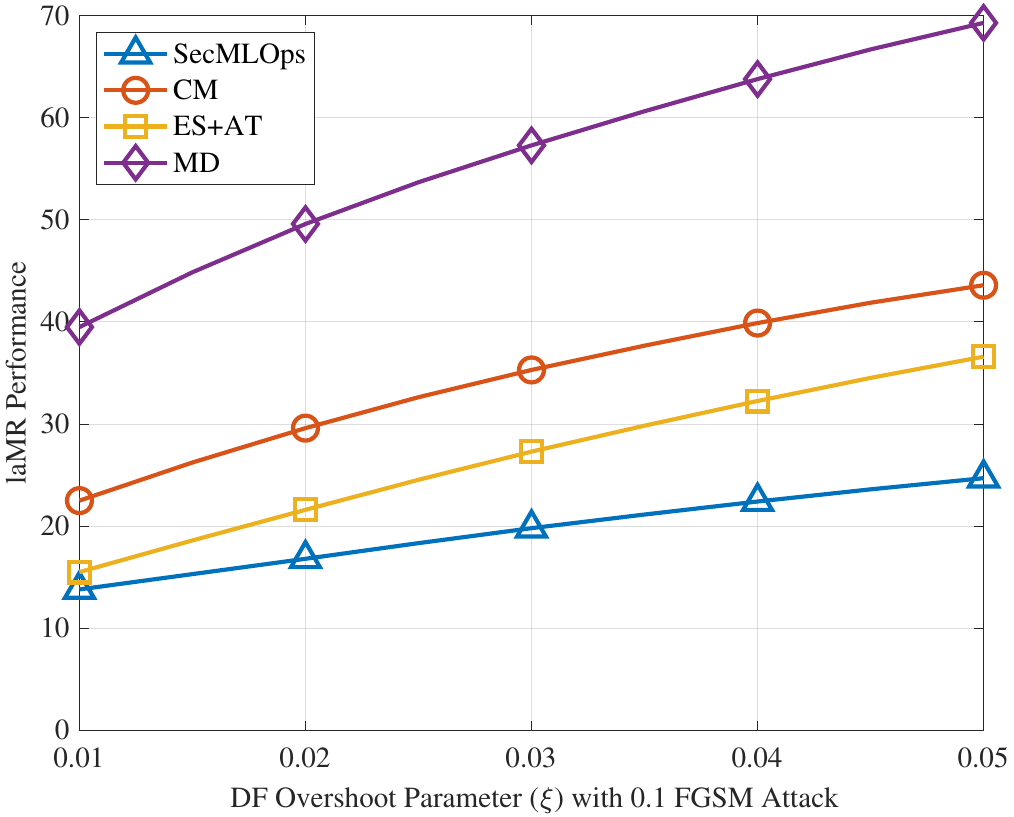}
        \end{minipage}}
    \subfigure[]{
        \begin{minipage}{3.1cm}
            \label{fig.All.1}
            \includegraphics[width=3.1cm]{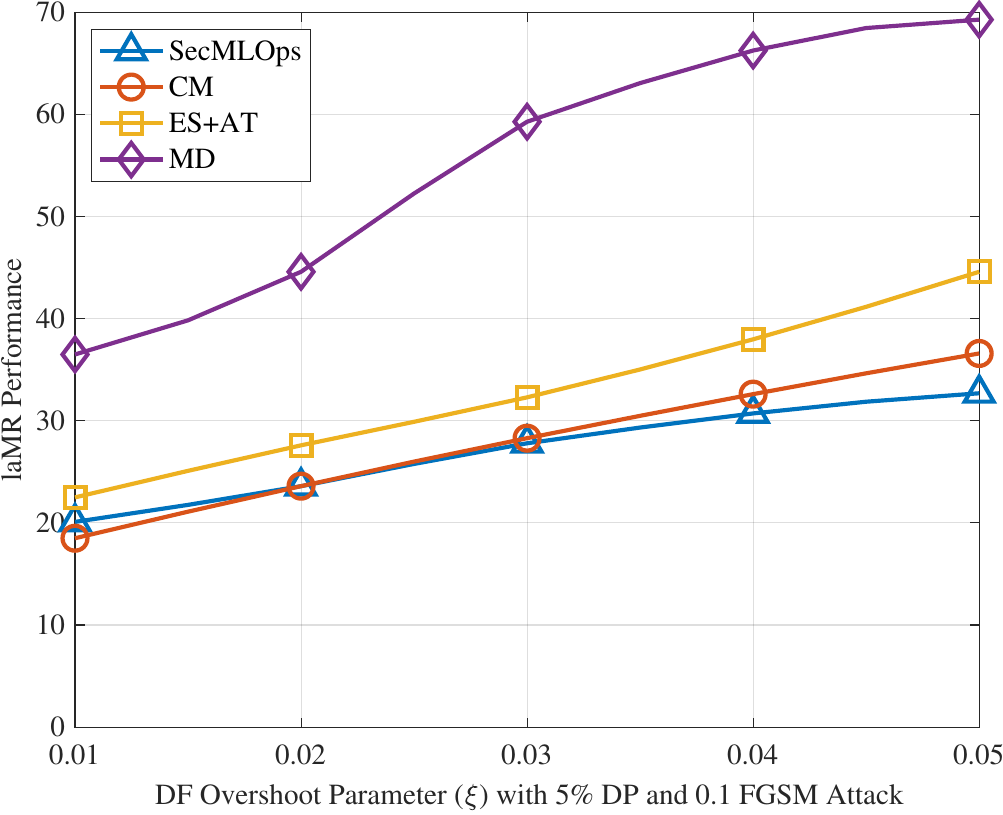}
        \end{minipage}}
    \subfigure[]{
        \begin{minipage}{3.1cm}
            \label{fig.All.2}
            \includegraphics[width=3.1cm]{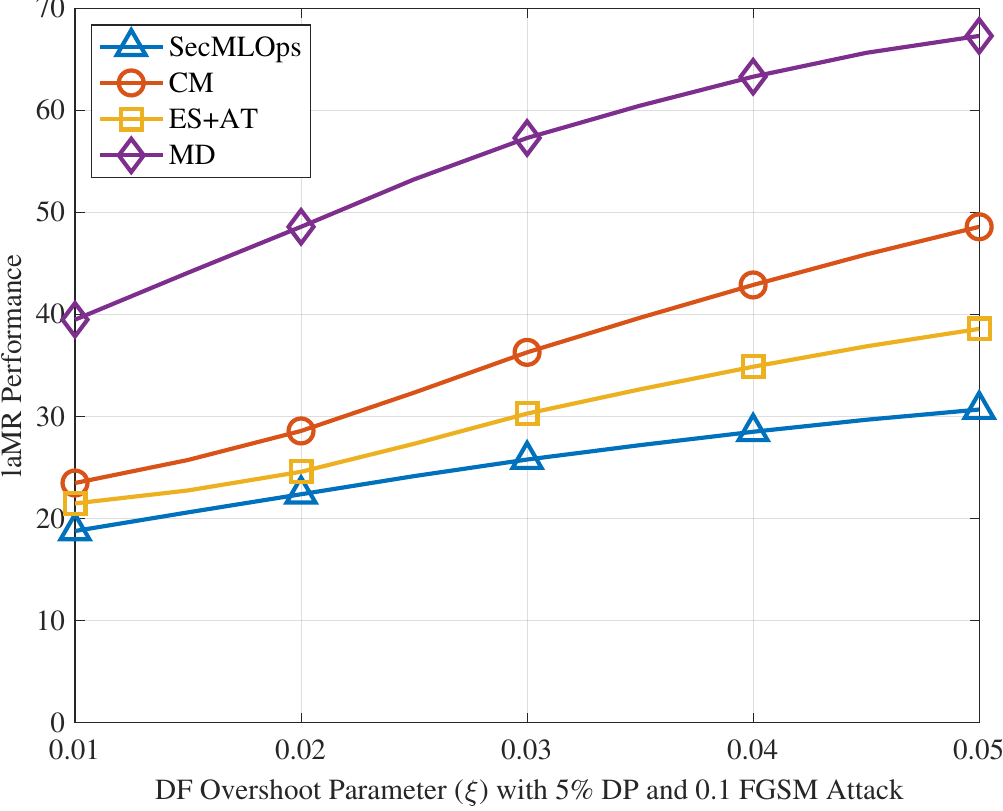}
        \end{minipage}}
    \subfigure[]{
        \begin{minipage}{3.1cm}
            \label{fig.All.3}
            \includegraphics[width=3.1cm]{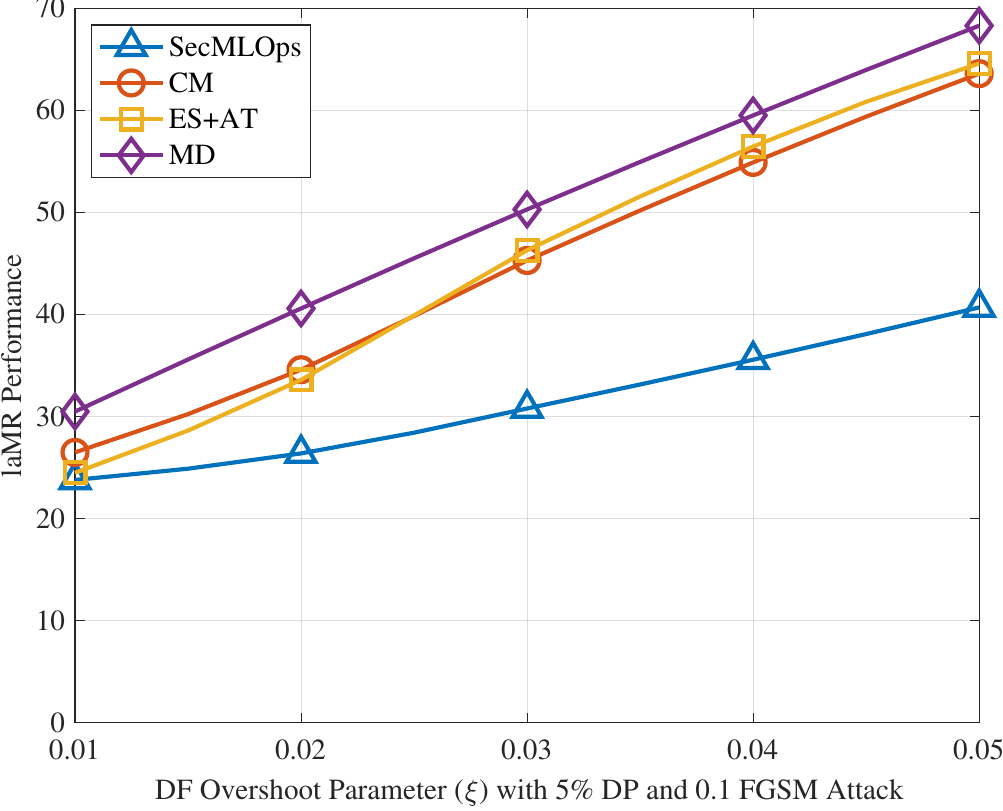}
        \end{minipage}}
    \subfigure[]{
        \begin{minipage}{3.1cm}
            \label{fig.All.4}
            \includegraphics[width=3.1cm]{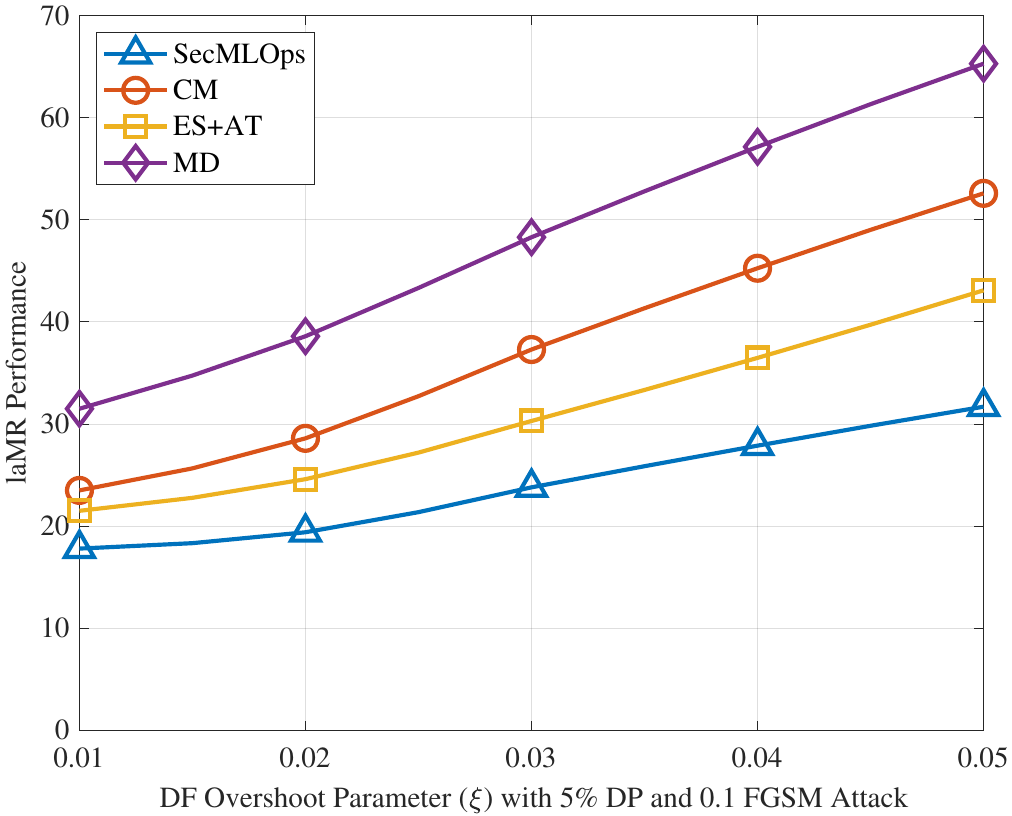}
        \end{minipage}}
    \caption{laMR performance of SecMLOps under combined attack scenarios. (a-d) DP+FGSM, (e-h) DP+DF, (i-l) FGSM+DF, (m-p) DP+FGSM+DF. Results demonstrate SecMLOps' robustness in complex adversarial environments, maintaining lower laMR compared to baseline methods (CM, ES+AT, MD). The defense parameters for the first column figures include: CM probability $P=0.5$, ES patience $p=10$, AT perturbation budget $\epsilon^{AT}$ = 0.01, MD smoothing factor $\alpha = 0.2$; The defense parameters for the second column figures include: $P=0.2$, $p=20$, $\epsilon^{AT}$ = 0.02, $\alpha = 0.2$; The defense parameters for the third column figures include: $P=0.2$, $p=10$ \$ $\epsilon^{AT}$ = 0.01, $\alpha = 0.4$; The defense parameters for the fourth column figures include: $P=0.5$, $p=20$, $\epsilon^{AT}$ = 0.02, $\alpha = 0.2$.}
      \label{fig:combinedattacks}

\end{figure}

\subsubsection{Ablation Study of SecMLOps}
To evaluate the effectiveness of individual components in our SecMLOps framework, we conducted an extensive ablation study comparing our approach with three baseline methods: CutMix (CM), Early Stopping plus Adversarial Training (ES+AT), and Model Distillation (MD). Figures~\ref{fig:individualattack} and~\ref{fig:combinedattacks} present the laMR performance under various attack scenarios and strengths, demonstrating the robustness of SecMLOps across a wide range of adversarial conditions. In Figure~\ref{fig:individualattack}, we observe that SecMLOps consistently outperforms the baseline methods across different types of attacks. For instance, in the case of DP attacks (Figures~\ref{fig.dp.1}-\ref{fig.dp.4}), SecMLOps maintains the lowest laMR even as the poisoning percentage increases from 5\% to 15\%. At 15\% poisoning, SecMLOps achieves an laMR of approximately 22\%, compared to 35\% for CM, 30\% for ES+AT, and 40\% for MD, demonstrating its superior robustness to training-time attacks. Similarly, for FGSM attacks (Figures~\ref{fig.FGSM.1}-\ref{fig.FGSM1.4}), SecMLOps maintains an laMR below 25\% even at the highest attack strength ($\epsilon$ = 0.05), while the best-performing baseline (ES+AT) reaches an laMR of about 35\%. In the case of DF attacks (Figures~\ref{fig.DF.1}-\ref{fig.DF.4}), SecMLOps shows remarkable resilience, with an laMR staying below 30\% across all overshoot parameters, compared to the next best method (ES+AT) which exceeds 40\% laMR at the highest attack strength. The effectiveness of SecMLOps is particularly evident in more challenging scenarios, such as combined attacks, where it maintains a significant performance advantage over the baseline methods, often by margins of 10-20\% in terms of laMR.

Figure~\ref{fig:combinedattacks} further illustrates the resilience of SecMLOps under more complex attack combinations, providing a comprehensive view of its performance in highly adversarial environments. In scenarios combining DP with FGSM (Figures~\ref{fig.DPandFGSM.1}-\ref{fig.DPandFGSM.4}) or DF (Figures~\ref{fig.DPandDF.1}-\ref{fig.DPandDF.4}), SecMLOps consistently achieves the lowest laMR across various attack strengths. For example, under the combined DP (5\%) and FGSM attack, SecMLOps maintains an laMR below 30\% even at the highest FGSM strength ($\epsilon$ = 0.05), while the best baseline method (ES+AT) approaches 45\% laMR. Similarly, for the DP (5\%) and DF combination, SecMLOps keeps the laMR under 35\% across all DF overshoot parameters $\xi$, outperforming the nearest competitor by at least 10\%. The framework's robustness is most pronounced in the most challenging scenarios, such as the combination of DP, FGSM, and DF attacks (Figures~\ref{fig.All.1}-\ref{fig.All.4}), where SecMLOps maintains a substantial performance lead over all baseline methods. In this extreme case, SecMLOps achieves an laMR of approximately 40\% at the highest attack strengths, while the best baseline method (ES+AT) exceeds 55\% laMR, and the worst-performing baseline (MD) approaches 70\% laMR. This comprehensive evaluation demonstrates that the integrated approach of SecMLOps, which combines multiple defense strategies and continuous monitoring, provides superior protection against a wide range of adversarial threats compared to individual defense mechanisms. The consistent performance advantage of SecMLOps across diverse attack scenarios, often maintaining a 15\%-25\% lower laMR than the best baseline method in the most severe attack conditions, underscores the importance of a holistic, multi-layered approach to securing ML systems, validating the core principles of our framework.

Notably, the in-training defense mechanism (ES+AT) emerged as a crucial component, sometimes performing comparably to, or even outperforming, the full SecMLOps framework in certain cases. This observation highlights the significance of in-training defenses in enhancing model robustness. To further optimize our approach and understand the trade-offs between security and performance, we conducted additional experiments exploring various defense parameters and their impacts. These investigations, detailed in the following results, aim to identify the most effective defense strategies and provide insights into balancing security measures with model efficiency.

\subsubsection{Optimization of Defense Parameters and Trade-off}
\begin{figure}[pt]
    \centering
    \subfigure[]{
        \begin{minipage}{4.2cm}
        \label{fig.heatmap.1}
        \includegraphics[width=4.2cm]{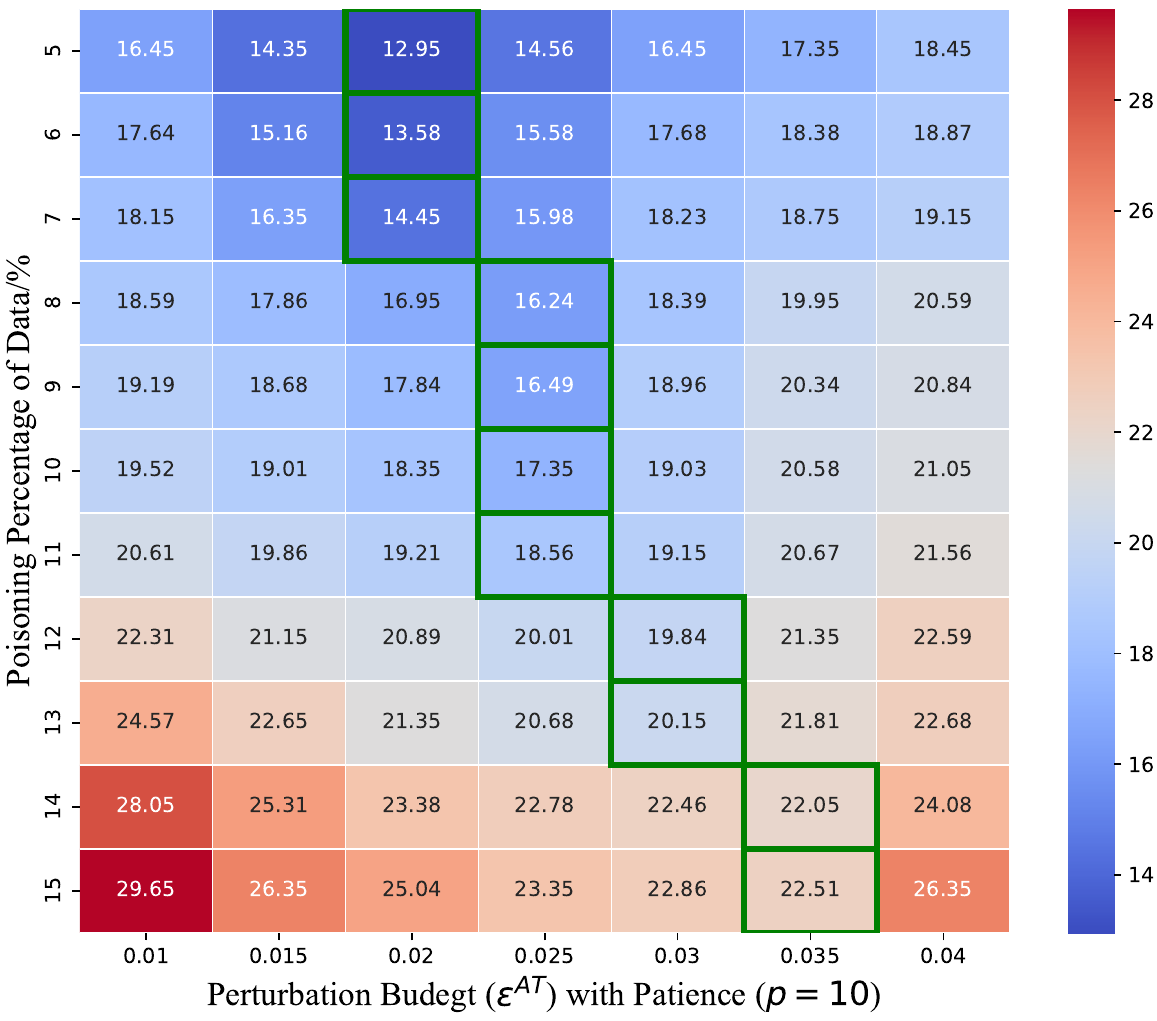}
      \end{minipage}}
      \subfigure[]{
       \begin{minipage}{4.2cm}
       \label{fig.heatmap.2}
      \includegraphics[width=4.2cm]{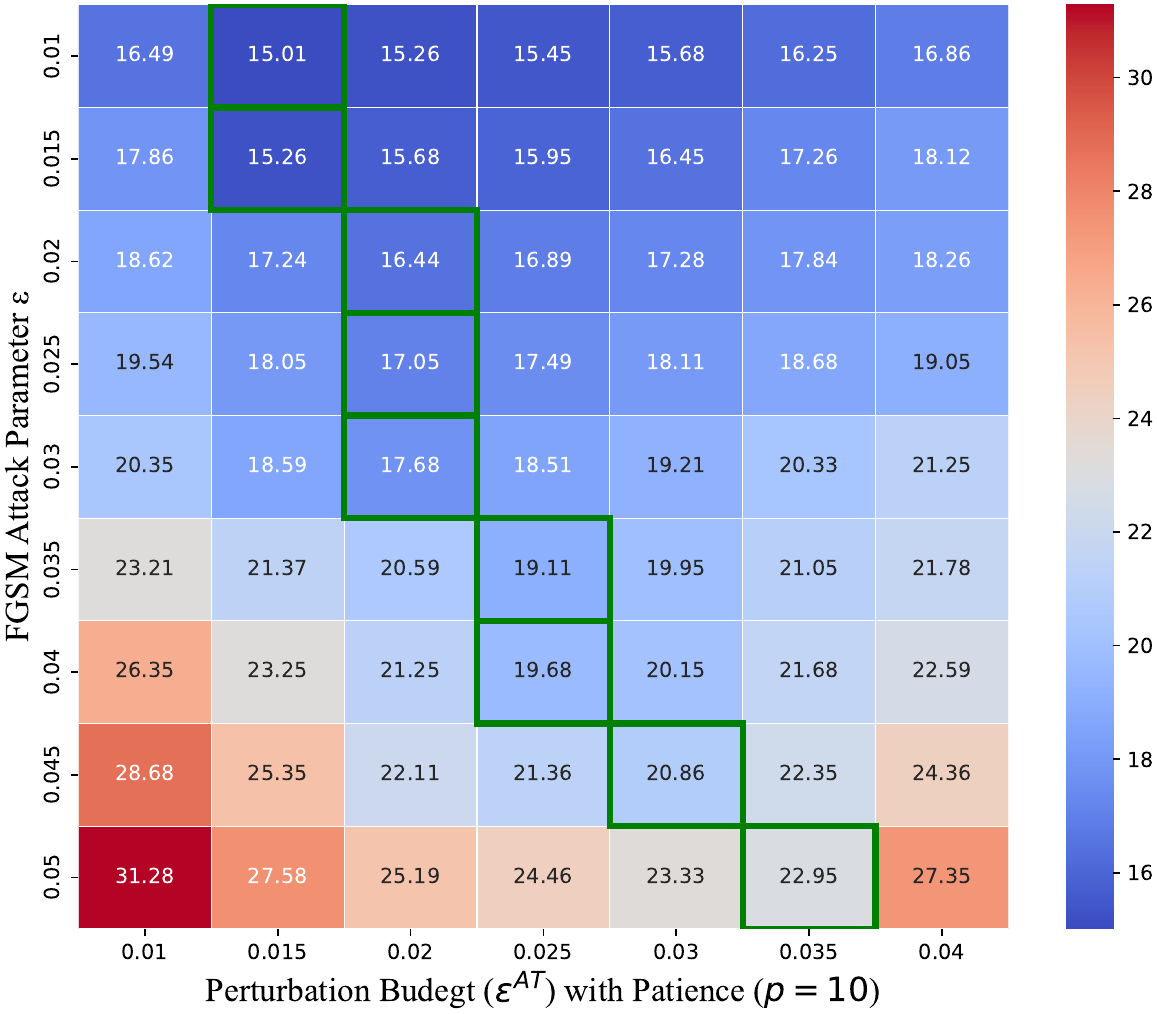}
      \end{minipage}}
        \subfigure[]{
       \begin{minipage}{4.2cm}
       \label{fig.heatmap.3}
      \includegraphics[width=4.2cm]{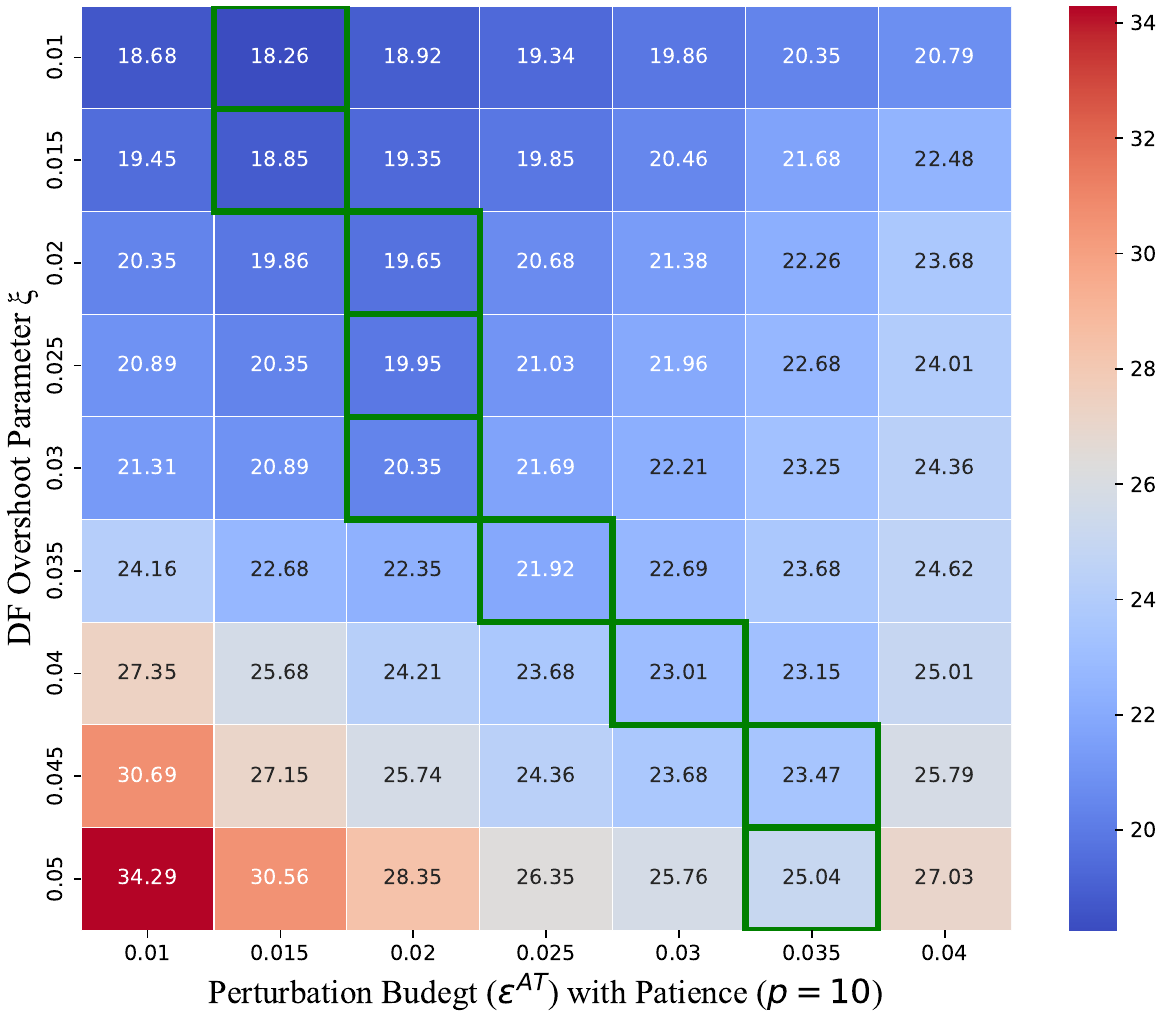}
      \end{minipage}}
    \caption{Performance comparison against different attack parameters (strengths) under different defense strategy. a) poisoning percentage of data (for DP attack); b) FGSM attack parameter; c) DF overshoot parameter.}
      \label{fig:heatmaps}
  
\end{figure}
The heatmaps in Figure~\ref{fig:heatmaps} illustrate the laMR performance of our pedestrian detection model under varying in-training defense mechanisms against DP, FGSM, and DF attacks, respectively. In Figure~\ref{fig.heatmap.1}, the model's laMR under DP attacks is minimized at a defense level parameter (perturbation budget $\epsilon^{AT}$) of 0.02 to 0.03, particularly effective at lower poisoning percentages, demonstrating laMR as low as 12.95\%. This indicates that our method's ability to adapt the defense strategy is crucial for mitigating the effects of poisoned data. Figure~\ref{fig.heatmap.2} shows a significant reduction in laMR under FGSM attacks, with optimal defense achieved at a $\epsilon^{AT}$ of 0.02 to 0.035, leading to a minimum miss rate of 15.01\% at an attack budget of 0.015. This suggests that our approach effectively strengthens the model against gradient-based adversarial perturbations. In Figure~\ref{fig.heatmap.3}, under DF attacks, adjusting the $\epsilon^{AT}$ from 0.025 to 0.03 yields the best performance, with laMR as low as 19.65\% for overshoot parameters around 0.02 to 0.03. These results demonstrate that our proposed method significantly improves laMR performance by dynamically optimizing defense strategies against varying attack intensities. Additionally, by providing clear guidelines on how to adjust defense levels in response to different types of attacks, our SecMLOps framework not only enhances model robustness but also offers valuable insights for developers to maintain and adapt ML models more effectively in the face of evolving threats.

\begin{figure}[t!]
    \centering
    \includegraphics[width=0.6\linewidth]{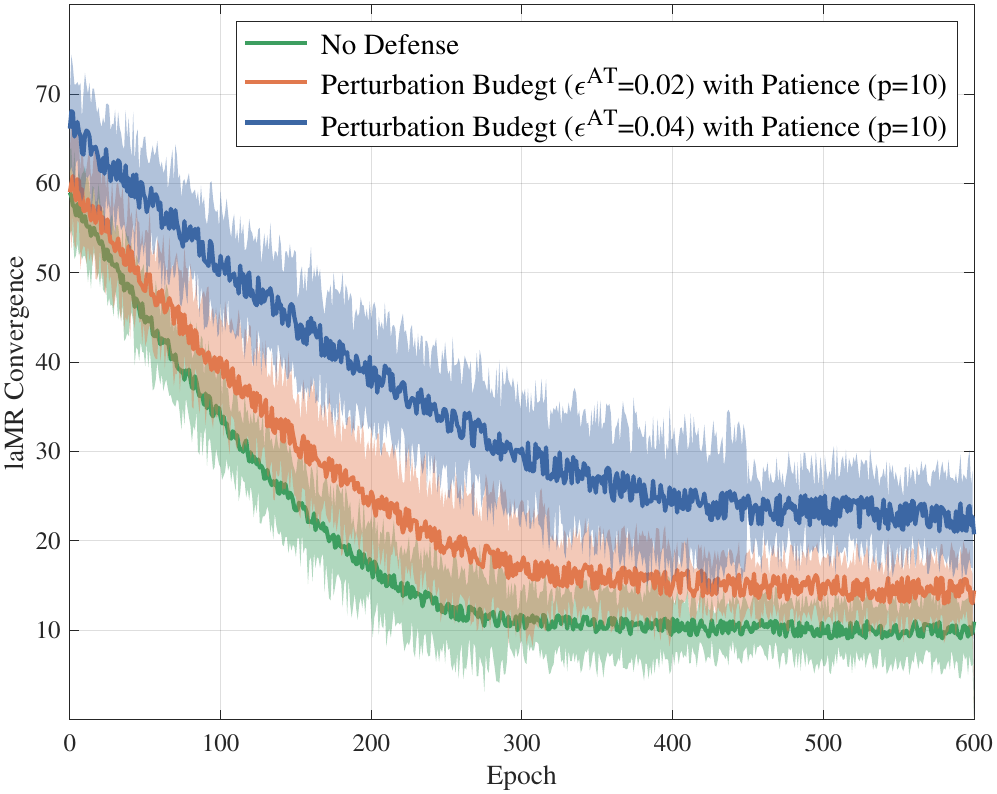}
    \caption{Convergence performance comparison under different defense strategy, illustrating that there exists tradeoff between security and model efficiency.}
    \label{fig:convergence}

\end{figure}

Figure~\ref{fig:convergence} illustrates the convergence performance of our pedestrian detection model under three different security defense strategies: No Defense, perturbation budget ($\epsilon^{AT}$=0.02) with patience $p=10$, and perturbation budget ($\epsilon^{AT}$=0.04) with patience $p=10$. No Defense strategy achieves convergence at the 290th epoch with a laMR of 11.3\%. This suggests a rapid stabilization in performance but without any security assurances against adversarial threats. In contrast, the model with defense strategy ($\epsilon^{AT}$=0.02) converges at the 320th epoch with a higher laMR of 16.35\%, indicating a moderate delay in reaching stability. The increased convergence time reflects the model's enhanced ability to withstand attacks, though at the cost of higher initial miss rates. The model with the highest defense level, $\epsilon^{AT}$=0.04, converges much later, at the 450th epoch, with an even higher laMR of 23.65\%. This significant delay and higher laMR underscore the trade-off between robust security and convergence efficiency. The defense mechanisms enhance the model's resilience, offering better protection against poisoning and adversarial attacks, but they also slow down the convergence process and increase the laMR. These findings highlight the importance of balancing defense strength with model performance. Our SecMLOps framework facilitates this balance, guiding developers to optimize defense settings effectively, ensuring that the model is both secure and capable of maintaining acceptable performance levels over time. This result provides crucial insights for maintaining and adapting ML models in the face of evolving security challenges, ensuring they remain robust and effective in real-world applications.

\subsection{Discussion}
The use case of implementing SecMLOps in the VLPD system provides valuable insights into the practical application, benefits, and challenges of integrating comprehensive security measures throughout the MLOps lifecycle. This section delves into a thorough analysis of these insights, discussing the implications for the broader adoption of SecMLOps in various domains and highlighting the importance of striking a balance between security and operational efficiency.

\subsubsection{Enhanced Security and Reliability}
Our findings underscore the significant enhancements in security and reliability achieved by implementing SecMLOps in the VLPD system. By integrating robust security measures at every stage of the MLOps lifecycle, from data ingestion to model deployment and monitoring, SecMLOps effectively safeguards the system against prevalent threats such as AE and data poisoning. This comprehensive approach to security ensures that the VLPD system maintains its integrity and robustness even in the face of evolving adversarial tactics, thereby instilling greater confidence in its performance and reliability.
The successful application of SecMLOps in the VLPD system serves as a compelling case study for the potential benefits of adopting this framework across diverse domains. From autonomous vehicles to healthcare and finance, numerous industries rely on the accurate and secure functioning of ML systems to drive critical decision-making processes. By demonstrating the efficacy of SecMLOps in enhancing the security and reliability of the VLPD system, this use case provides a valuable template for organizations seeking to fortify their own ML pipelines against the ever-growing landscape of cybersecurity threats.

\subsubsection{Trade-offs and Optimization}
While the implementation of SecMLOps yields substantial security benefits, it is crucial to acknowledge and address the potential trade-offs in terms of computational complexity and system performance. The addition of multiple layers of security controls, such as adversarial training, input validation, and encrypted data storage, inevitably introduces overhead that can impact the real-time processing capabilities essential for time-sensitive applications like autonomous driving.
Balancing these trade-offs requires a thoughtful and strategic approach to optimizing SecMLOps configurations based on the specific operational requirements and risk tolerance of each application. For instance, in safety-critical systems where the consequences of a security breach could be catastrophic, the emphasis on robust security measures may justify a certain degree of performance trade-off. Conversely, in applications where real-time responsiveness is paramount, streamlining the security controls and focusing on the most critical threat vectors may be necessary to maintain optimal performance.
The insights gained from the VLPD use case highlight the importance of conducting thorough performance evaluations and iterative optimizations when implementing SecMLOps. By carefully monitoring system metrics and continuously refining the balance between security and efficiency, organizations can ensure that their ML systems remain both highly secure and operationally viable.

The computational overhead introduced by SecMLOps security measures represents a critical dimension of these trade-offs, particularly for real-time applications like autonomous driving. Our empirical measurements reveal that input validation and sanitization (SC-4) adds approximately 5-8ms of latency per frame, while adversarially trained models exhibit 15-20\% longer inference times compared to unprotected models~\cite{wong2020fastbetterfreerevisiting}. In our experiments, this translates to 47ms versus 39ms average inference time on NVIDIA A100 hardware. Resource utilization also increases, with security monitoring consuming an additional 2.3GB of memory and 12-15\% more CPU cycles for continuous statistical analysis and drift detection.

To mitigate these performance impacts, practitioners can employ adaptive security levels that adjust based on operational context, like reducing validation frequency during highway driving while maintaining comprehensive monitoring in complex urban environments. Hardware acceleration through NPUs can parallelize security checks with primary inference, potentially reducing overhead to under 3ms. Additionally, model optimization techniques such as quantization can recover much of the performance loss while maintaining robustness; our quantized models show only 8\% inference overhead while preserving 94\% of adversarial robustness. Organizations must establish clear policies for these trade-offs based on their specific latency requirements and risk tolerance.

\subsubsection{Compliance and Regulatory Alignment}
Ensuring compliance with the ever-evolving landscape of data privacy and security regulations is a critical aspect of implementing SecMLOps. As demonstrated in the VLPD use case, aligning the system with stringent standards such as GDPR and ISO 26262 requires a proactive and comprehensive approach to data protection and secure system design.
The SecMLOps framework provides a structured methodology for incorporating compliance requirements into every phase of the MLOps lifecycle. By establishing clear security policies, access controls, and auditing mechanisms, SecMLOps enables organizations to maintain a high level of regulatory compliance while still leveraging the power of ML technologies.
However, the rapid pace of regulatory evolution presents an ongoing challenge for SecMLOps implementations. As new data privacy laws and industry-specific standards emerge, organizations must remain agile in adapting their SecMLOps practices to ensure continued compliance. This underscores the importance of regular policy reviews, staff training, and collaboration with legal and compliance teams to stay ahead of the regulatory curve.

\subsubsection{Continuous Monitoring and Adaptation}
The VLPD use case emphasizes the crucial role of continuous monitoring and adaptation in maintaining the long-term effectiveness of SecMLOps implementations. In the face of an ever-evolving threat landscape, it is essential to establish robust monitoring mechanisms that can detect anomalies, performance deviations, and potential security breaches in real-time.
By leveraging advanced monitoring tools and automated alert systems, organizations can proactively identify and respond to emerging threats before they escalate into full-blown security incidents. This real-time visibility into system behavior is particularly critical in the context of ML systems, where subtle manipulations of data or models can lead to significant performance degradations or biased outputs.
Moreover, the insights gained from continuous monitoring should be leveraged to iteratively refine and adapt the SecMLOps framework itself. As new attack vectors and defensive techniques emerge, organizations must be prepared to update their security controls, policies, and incident response plans accordingly. This culture of continuous improvement ensures that the SecMLOps implementation remains effective and relevant in the face of evolving challenges.

\subsubsection{Application and Future Directions}
\textcolor{red}{The successful application of SecMLOps in the VLPD system serves as a demonstrative example of how the paradigmatic framework can be instantiated for specific contexts, rather than as a universal solution to be directly replicated. While the security controls and optimization strategies employed, such as our adversarial training budget ($\epsilon^{AT}$=0.02) and data poisoning thresholds, were carefully calibrated for pedestrian detection, these same parameters would be inappropriate or even counterproductive in other domains. The framework's true value lies in guiding organizations to develop their own context-appropriate security measures based on their unique threat landscapes, compliance requirements, and operational constraints.}

As the field of ML continues to advance at a rapid pace, it is crucial that the SecMLOps framework evolves concurrently. Future research should focus on refining the efficiency and scalability of SecMLOps practices, exploring new techniques for detecting and mitigating emerging threats, and developing more granular metrics for quantifying the security and performance of ML systems.
Additionally, there is a need for greater collaboration and knowledge-sharing among researchers, practitioners, and policymakers to establish industry-wide best practices and standards for secure ML development and deployment. By fostering a vibrant ecosystem of SecMLOps research and practical implementation, we can collectively work towards a future where the transformative potential of ML is fully realized without compromising on the fundamental principles of security, reliability, and trust.

The emergence of Large Language Models (LLMs) has introduced a paradigm shift in ML security, presenting novel attack vectors that extend beyond traditional ML threats~\cite{greshake2023not}. The framework's principles are equally critical when applied to LLM deployments. LLMs face unique vulnerabilities including prompt injection attacks, where malicious inputs manipulate model behavior through carefully crafted prompts; jailbreaking attempts that bypass safety alignments~\cite{zou2023universal}; and training data extraction attacks that can recover sensitive information from model parameters~\cite{carlini2021extracting}.

The SecMLOps framework adapts to these LLM-specific challenges through several key extensions. First, the security analysis phase must expand beyond traditional STRIDE modeling to include prompt-based attack surfaces. This involves analyzing the entire prompt pipeline, from user input to prompt templates to model responses, identifying injection points where malicious content could compromise system behavior~\cite{yao2024survey}. Second, the security controls for LLMs require fundamental reconceptualization. While traditional ML models benefit from adversarial training with perturbed inputs, LLMs demand semantic-level defenses~\cite{carlini2021extracting}. This includes implementing prompt filtering mechanisms that detect and neutralize injection attempts, output validation systems that ensure responses align with safety policies, and dynamic monitoring that tracks conversation patterns for signs of jailbreaking attempts. Third, the governance and compliance components of SecMLOps must address the unique ethical and legal challenges of LLMs. These models' ability to generate human-like text raises concerns about misinformation, bias amplification, and privacy violations through memorized training data~\cite{9798870}. 

Looking forward, SecMLOps for LLMs will likely require integration with emerging technologies such as constitutional AI for maintaining behavioral bounds, watermarking techniques for generated content attribution, and federated learning approaches that enhance privacy while maintaining model capability~\cite{yao2024survey}. The framework must also evolve to address the supply chain risks of foundation models, where organizations deploy pre-trained LLMs whose training data and procedures remain opaque. This necessitates new validation protocols and security controls specifically designed for transfer learning scenarios where the base model itself may contain hidden vulnerabilities or backdoors. 

\subsubsection{Generalizability to Other Domains}
While our empirical evaluation focuses on pedestrian detection in autonomous driving, the SecMLOps framework is designed to generalize across diverse ML domains through its abstract security principles and adaptable implementation strategies. The framework's PTPGC structure provides a domain-agnostic foundation where security considerations integrate systematically into any MLOps pipeline, regardless of the specific application area or data modality.

In healthcare applications, SecMLOps adapts to address domain-specific security requirements while maintaining its core structure. The threat model extends to include patient privacy concerns under HIPAA regulations, where data poisoning attacks might target diagnostic models to cause misdiagnosis~\cite{finlayson2019adversarial}, and model inversion attacks could potentially expose sensitive patient information. The framework's data validation controls (SC-4) would emphasize differential privacy techniques and federated learning approaches to protect patient confidentiality during model training~\cite{kaissis2020secure}. Healthcare-specific adaptations would include mandatory audit trails for all model decisions affecting patient care, enhanced access controls reflecting medical hierarchies, and continuous monitoring for distributional shifts that might indicate demographic biases in diagnostic predictions~\cite{chen2021ethical}. However, the fundamental SecMLOps processes, including systematic threat analysis, role-based security responsibilities, and lifecycle-integrated controls, remain unchanged.

Financial services present different challenges where adversarial attacks might attempt to manipulate fraud detection or credit scoring models for economic gain~\cite{cartella2021adversarial}. In this context, SecMLOps would prioritize integrity and non-repudiation, with enhanced focus on model explainability for regulatory compliance~\cite{bhatt2020explainable}. The framework would adapt to include real-time anomaly detection for unusual transaction patterns that might indicate ongoing attacks~\cite{weber2019anti}, cryptographic proof of model versions used for specific decisions, and segregation of duties between model developers and deployment teams to prevent insider threats. The STRIDE analysis would emphasize tampering and repudiation threats specific to financial transactions, while maintaining the same systematic approach to threat identification and mitigation.

For natural language processing (NLP) applications, particularly LLMs, SecMLOps addresses fundamentally different attack vectors including prompt injection, jailbreaking attempts, and training data memorization~\cite{papernot2018sok}. The framework's security controls would expand to include semantic-level input validation, output filtering for harmful content generation, and specialized monitoring for conversation pattern anomalies~\cite{yao2024survey}. This shows that while specific threats vary, the SecMLOps framework remains flexible and can be systematically extended to secure different data modalities and application settings.

The framework's generalizability derives from its abstraction of security concerns into CIAAAA that apply regardless of domain. While specific threats vary, the systematic approach to identifying, analyzing, and mitigating these threats through the PTPGC model remains consistent. Organizations can instantiate SecMLOps using domain-appropriate tools and controls while following the same fundamental security integration methodology.

This domain adaptability validates our design choice to present SecMLOps as a paradigm rather than a rigid implementation~\cite{kumar2020adversarial}. By demonstrating detailed application to pedestrian detection while discussing adaptation to other domains, we show that the framework provides practical value across the diverse landscape of ML applications without requiring domain-specific reimplementation of core security principles. 

\subsubsection{Limitations}
While SecMLOps provides comprehensive security integration for ML systems, several limitations warrant consideration. Edge devices with constrained computational resources present particular challenges for implementing the full SecMLOps framework~\cite{zhou2019edge}. Devices with limited memory may struggle to maintain the comprehensive audit logs and statistical baselines required for drift detection, while low-power processors may not support real-time adversarial validation without unacceptable latency~\cite{ABUABED2023103391,kaissis2020secure}. In such scenarios, practitioners must implement lightweight variants of SecMLOps, potentially focusing on critical security controls while accepting reduced monitoring capabilities.

The framework also assumes organizational maturity in both MLOps and security practices. Organizations lacking established ML pipelines or security expertise may find the eight specialized roles and comprehensive processes overwhelming~\cite{ruf2021demystifying} to implement simultaneously. Small teams may need to consolidate roles, potentially reducing the separation of duties that strengthens security~\cite{warnett2024understandability}. Additionally, the framework's emphasis on continuous monitoring and validation assumes infrastructure for data collection and analysis that may not exist in all deployment contexts.

Furthermore, SecMLOps cannot fully eliminate the fundamental tension between model utility and security. While our evaluation demonstrates acceptable trade-offs for pedestrian detection, other domains may face scenarios where security requirements make models unusable for their intended purpose~\cite{10062371,wazir2023mlops}. Highly adversarial environments might require security measures so stringent that model performance degrades below acceptable thresholds, forcing organizations to reconsider deployment altogether~\cite{wong2020fastbetterfreerevisiting}.

In conclusion, SecMLOps implementation requires careful adaptation to specific contexts rather than rigid application of all components. Organizations must assess their resources, threat models, and operational requirements to determine appropriate SecMLOps configurations.

\section{Threats to Validity}
\label{Sec.VI_Threats}

\subsection{Internal Validity}
One potential threat to internal validity concerns the selection and implementation of adversarial attacks in our evaluation. We focused on specific attacks (DP, FGSM, and DF) which, while widely recognized, may not represent the full spectrum of potential threats to ML systems. However, these attacks are highly representative of common threats in the industry~\cite{10057473}, covering both training-time and inference-time vulnerabilities. To mitigate this limitation, we carefully selected these attacks to cover different aspects of ML vulnerabilities: DP for training-time attacks, and FGSM and DF for inference-time perturbations. Furthermore, we conducted experiments with various attack strengths and combinations (e.g., DP+FGSM, DP+DF, FGSM+DF) to provide a more comprehensive assessment of our framework's robustness. This approach allows us to demonstrate the effectiveness of SecMLOps against a diverse range of threat scenarios, enhancing the internal validity of our findings.

Another potential threat relates to the implementation of defense strategies within the SecMLOps framework. The effectiveness of techniques such as adversarial training and model distillation can be sensitive to specific parameter settings and implementation details. To address this concern, we conducted extensive experiments with varying defense level parameters, as illustrated in Figure~\ref{fig:heatmaps}. By systematically exploring different defense strengths (e.g., defense level parameters from 0 to 0.5), we were able to identify optimal configurations for different attack scenarios. This thorough approach not only enhances the reliability of our results but also provides valuable insights into the trade-offs between security and model performance, strengthening the internal validity of our conclusions.

Lastly, the generalizability of our findings from the VLPD use case to other ML applications could be questioned. To address this, we designed our SecMLOps framework with modularity and adaptability in mind. While our experiments focused on pedestrian detection, the core principles of our approach – including comprehensive threat modeling, adaptive defense strategies, and continuous monitoring – are applicable across various ML domains. By providing detailed descriptions of our methodology and implementation, we enable other researchers and practitioners to adapt and validate our approach in different contexts, thereby strengthening the internal validity and broader applicability of our work.

\subsection{External Validity}

A threat to the external validity of our findings concerns VLPD system, which represents a specific application within the broader field of machine learning. While VLPD systems are crucial for applications like autonomous driving and urban surveillance, they may not fully represent the diverse range of ML applications in other domains such as healthcare, finance, or NLP. The unique characteristics of the VLPD system, including its use of vision-language semantic segmentation and prototypical semantic contrastive learning, may have influenced the effectiveness of our SecMLOps framework in ways that might not directly translate to other ML architectures or problem domains. However, we believe that the core principles of our approach - such as comprehensive threat modeling, adaptive defense strategies, and continuous monitoring - are broadly applicable. To address this limitation, we have provided detailed descriptions of our methodology and implementation in Sections~\ref{Sec.III_SecMLOps} and~\ref{Sec.IV_CaseStudy}, enabling researchers and practitioners to adapt our framework to their specific contexts.

In addition, our experiments were conducted using the CityPersons dataset, which, although widely recognized and challenging, represents a specific urban environment scenario. The dataset's focus on images captured across 27 cities in Germany may not fully encompass the diversity of pedestrian detection scenarios globally. This geographic limitation could impact the framework's performance in significantly different urban layouts or cultural contexts. To mitigate this concern, we evaluated our system under various subsets of the data, including ``Reasonable,'' ``Small,'' ``Heavy occlusion,'' and ``All,'' as detailed in Table~\ref{tab:evaluation_settings}. This approach helps to demonstrate the framework's effectiveness across different detection conditions, enhancing its potential applicability to diverse real-world scenarios.

\subsection{Construct Validity}
A potential threat to construct validity in our study relates to the metrics used to evaluate the effectiveness of our SecMLOps framework. While we primarily relied on the laMR as our performance metric, which is standard in the field of pedestrian detection, it may not capture all aspects of security and robustness in ML systems. The laMR (as shown in Equation~\ref{equ:lamr}) focuses on detection accuracy but may not fully represent the system's resilience against all types of adversarial attacks or its ability to maintain performance under various security measures. To mitigate this, we evaluated our framework under multiple attack scenarios and defense levels, providing a more comprehensive view of its effectiveness.

To further address construct validity concerns, we incorporated additional analyses, such as convergence performance comparisons and trade-off evaluations between security measures and system performance. These supplementary assessments help to provide a more holistic understanding of the SecMLOps framework's impact on both security and operational efficiency in ML systems.

Additionally, the construct validity of our study is supported by the systematic approach used to develop the SecMLOps framework. However, the framework's foundations in existing literature and established practices may potentially limit its ability to capture emerging or unconventional security challenges in rapidly evolving ML ecosystems. Future work should continuously refine the framework's constructs to accommodate new attack vectors and defense mechanisms as they emerge in the field of ML security.

\section{Conclusion}
\label{Sec.VII_Conclusion}
In conclusion, this paper presents a pioneering SecMLOps framework, designed to embed comprehensive security measures throughout the MLOps lifecycle. By adopting a holistic approach to security, SecMLOps significantly enhances the robustness, reliability, and compliance of ML systems, addressing the growing concerns surrounding the vulnerabilities of these systems in an increasingly complex and interconnected world.

A detailed advanced PDS use case demonstrates the practical application of the SecMLOps framework. Empirical results confirm the effectiveness of this framework in mitigating risks such as DP and AE, thereby bolstering system resilience and reliability. Our study underscores the importance of integrating security from the onset of system development, offering a systematic strategy that addresses the predominant challenges in ML security. This approach sets a new benchmark for developing dependable and secure ML systems tailored to complex real-world applications.

\textcolor{red}{However, our findings also draw attention to the inherent trade-offs between implementing advanced security measures and maintaining system performance, trade-offs that must be navigated differently in each deployment context. Therefore, future work should focus on developing domain-specific instantiations of the SecMLOps paradigm rather than seeking universal implementations, as the security requirements of a medical imaging system differ fundamentally from those of financial fraud detection or autonomous navigation. Future research should also explore how the paradigmatic principles of SecMLOps can guide organizations in developing their own security measures while maintaining the flexibility to adapt to emerging threats and evolving regulatory requirements.}

\section{Declarations}

\subsection{Funding}
This research is supported by the Natural Sciences and Engineering Research Council of Canada (NSERC) grant RGPIN-2019-06306.

\subsection{Ethical Approval}
Not applicable - this study did not require ethical approval.

\subsection{Informed Consent}
The authors consent to the use of this work in the Journal.

\subsection{Author Contributions}
\textbf{Xinrui Zhang:} Conceptualization, investigation, methodology, validation, writing - original draft. \\
\textbf{Pincan Zhao:} Data analysis, validation, writing - review \& editing. \\
\textbf{Jason Jaskolka:} Supervision, funding acquisition, writing - review \& editing.\\
\textbf{Heng Li:} Writing - review \& editing.\\
\textbf{Rongxing Lu:} Writing - review \& editing.\\

\subsection{Data Availability Statement}
The data that support the findings of this study are available upon reasonable request.

\subsection{Conflict of Interest}
The authors declare that they have no conflict of interests.

\subsection{Clinical Trial Number}
Clinical trial number: not applicable.

\bibliography{ReferencesR2}

\end{document}